%% file: symmultiplescf-arxiv.tex
\def\pdfoutput{1}
\def\notikzex{1}
\def\twocolumnmode{1}

\documentclass[%
	journal=jctcce,%
	manuscript=article,%
	layout=twocolumn%
]{achemso}

\usepackage[version=4]{mhchem} % Further chemistry typesetting support

\usepackage[T1]{fontenc} % Use modern font encodings

\usepackage{bm}

\usepackage{siunitx}

\usepackage{amsmath, amssymb, mathtools, mathrsfs, braket, amsthm} % packages for further maths support

\usepackage{mleftright} %loading package for ensuring correct spacing before brackets
\mleftright

\usepackage[eulergreek]{sansmath}

\usepackage{booktabs, multirow, tabularx, bigdelim}

\usepackage{lscape}

\usepackage[svgnames]{xcolor}

\usepackage{caption, subcaption}

\usepackage{pgfplots} % pgfplots loads tikz automatically

\usetikzlibrary{external}

\newif\iftikzex
\ifdefined\notikzex
\tikzexfalse
\else
\tikzextrue
\fi

\iftikzex
\pgfplotsset{compat=1.16}

\usetikzlibrary{calc, luamath, positioning, math, pgfplots.groupplots, decorations.pathreplacing}

\pgfplotscreateplotcyclelist{coloronly}{%
	{red},%
	{blue},%
	{black!60!green},%
	{black!20!orange},%
	{green!30!brown},%
	{blue!40!red},%
	{black!60!blue},%
	{black!40!yellow},%
	{red!50!pink},%
	{green!70!blue},%
}
\fi

\usepackage{mciteplus}

\makeatletter
\newcommand*{\useexternalfile}[4]{%
	\iftikzex
	\tikzsetnextfilename{tikzoutput/#4-output}%
	\scalebox{1}{\input{\tikzexternal@filenameprefix#4.tikz}}
	\else
	\includegraphics[scale=#1, trim=#2 0 #3 0]{\tikzexternal@filenameprefix tikzoutput/#4-output.pdf}
	\fi
}
\makeatother

\newcommand*{\D}{\mathrm{d}}

\newtheorem{theorem}{Theorem}

\author{Bang C. Huynh}
\email{cbh31@cam.ac.uk}
\author{Alex J. W. Thom}
\affiliation{University Chemical Laboratory, Lensfield Road, Cambridge CB2 1EW, United Kingdom}

\title{Symmetry in Multiple Self-Consistent-Field Solutions of Transition-Metal Complexes}

\begin{document}
	
	\input{abstract/abstract}
	\input{introduction/introduction}
	\input{theory/theory}
	\input{comp-details/comp-details}
	\input{results-d1/results-d1}
	\input{results-d2/results-d2-model}
	\input{results-d2/results-d2-full}
	\input{results-JT/results-JT}
	\input{conclusion/conclusion}
	
	\appendix
	\input{apps/grouptheory}
	\input{apps/symanalysis}
	\input{apps/symtransformation}
	\input{apps/nocinatorbs}

	\section*{Supporting Information}
	
	\paragraph{Supporting Information.}
	Detailed \textsc{Q-Chem} inputs for searches of SCF solutions in \ce{[TiF6]^{3-}} and \ce{[VF6]^{3-}} using metadynamics.
	Spin-orbital isosurfaces of SCF solutions in \ce{(B^{3+})(q^-)6}.
	Detailed SCF and NOCI results of \ce{[TiF6]^{3-}} and \ce{[VF6]^{3-}} listing overlap matrices, energy and $\langle \hat{S}^2 \rangle$ values.
	
	This information is available free of charge via the Internet at \url{http://pubs.acs.org}.

	\bibliography{symmultiplescf-arxiv}
	
	\input{TOC/TOC}

\end{document}

\tableofcontents
	\clearpage
	\input{suppinfo/compdetails/compdetails}
	\input{suppinfo/results-d1/results-d1-supp}
	\input{suppinfo/results-d2/results-d2-supp}

%	\bibliography{../bib/symmultiplescf}

%% file: abstract/abstract.tex
\begin{abstract}
  We use a method based on metadynamics to locate multiple low-energy Unrestricted Hartree--Fock (UHF) self-consistent-field (SCF) solutions of two model octahedral $d^1$ and $d^2$ transition-metal complexes, \ce{[MF6]^{3-}} (\ce{M = Ti, V}).
  By giving a group-theoretical definition of symmetry breaking, we classify these solutions in the framework of representation theory and observe that a number of them break spin or spatial symmetry, if not both.
  These solutions seem unphysical at first, but we show that they can be used as bases for Non-Orthogonal Configuration Interaction (NOCI) to yield multi-determinantal wavefunctions that have the right symmetry to be assigned to electronic terms.
  Furthermore, by examining the natural orbitals and occupation numbers of these NOCI wavefunctions, we gain insight into the amount of static correlation that they incorporate.
	We then investigate the behaviors of the most low-lying UHF and NOCI wavefunctions when the octahedral symmetry of the complexes is lowered and deduce that the symmetry-broken UHF solutions must first have their symmetry restored by NOCI before they can describe any vibronic stabilization effects dictated by the Jahn--Teller theorem.
\end{abstract}

%% file: introduction/introduction.tex
\tikzsetexternalprefix{./introduction/tikz/}

\section{Introduction}

	Strongly correlated systems such as mono- or multi-nuclear transition-metal (TM) complexes present an exciting, yet daunting, arena for electronic structure theory and computation.
	The interaction between electrons in the partially filled $d$ shells of the TM centers gives rise to many low-energy states\cite{book:Griffith1961} which exhibit various degrees of degeneracy when placed in a symmetric or nearly symmetric molecular geometry\cite{book:Mabbs1973,book:Sugano1970,article:Tanabe1954,*article:Tanabe1954a,*article:Tanabe1956}.
	As the spacing between these states ranges from \textit{ca.} \SI{200}{cm^{-1}} (\SI{2.48e-2}{eV}, $\sim k_\mathrm{B}T$ at room temperature) to \textit{ca.} \SI{50000}{cm^{-1}} (\SI{6.20}{eV}, $\sim \SI{200}{nm}$, in the ultraviolet region), they play a vital role in determining the spectroscopic and magnetic behaviors of these complexes\cite{book:Griffith1961,book:Condon1959}.
	Therefore, if we wish to predict or understand such properties on an \textit{ab initio} level successfully, we must first develop methods that are capable of describing these states reliably.

	Most conventional electronic structure theories and methods such as Hartree--Fock (HF) theory or Density Functional Theory (DFT) were formulated with the aim of calculating the electronic ground state of molecular systems to great precision\cite{article:Dreuw2005}.
	Both of the aforementioned methods are inherently variational: the HF equations are derived by minimizing the energy expression for a single Slater determinant\cite{article:Hartree1928,article:Fock1930,article:Hartree1935,article:Roothaan1951,book:Szabo1996}, whereas DFT is based upon the two famous theorems first proven by Hohenberg and Kohn, the second of which asserts that the ground-state energy and density can be found by minimizing the exact energy functional of the system\cite{article:Hohenberg1964}.
	Unfortunately, the variational principle is not applicable to most excited states in general because its formulation only allows for an upper bound to the exact ground-state energy to be determined\cite{book:Szabo1996,book:Bransden2000,book:Piela2013}.
	As a consequence, variational techniques often struggle to ascertain if their calculations of excited states and excitation energies are sensible.
	Even though the above-mentioned ground-state methods can be exploited to give variationally optimized upper bounds for certain electronic excited states by imposing orthogonality to the ground state via appropriate spin or spatial irreducible representation constraints\cite{article:Dreuw2005,book:Bransden2000}, their applicability can be rather limited if the irreducible representation of the exact ground state is not known or if the low-lying states span many repeated irreducible representations due to, for example, the low molecular symmetry of the system.
	
	In the last few decades, several methods have been developed to remove the need for such \textit{a priori} constraints in the calculation of excited states.
	The conceptually simplest wavefunction-based approach is Configuration Interaction Singles (CIS) in which an ansatz for excited states is written as a linear expansion in terms of single-replacement Slater determinants formed from a reference HF determinant, and the expansion coefficients can be determined using Rayleigh--Ritz variational principle\cite{article:Bene1971,article:Dreuw2005}.
	As CIS excited-state wavefunctions are simply linear combinations of Slater determinants, they are well defined and can be interpreted relatively easily: the expansion coefficients give an indication of the contribution of the constituting Slater determinants whose spin-orbitals provide chemical understanding.
	However, the construction of CIS is severely rigid: the inclusion of only single-replacement Slater determinants prevents it from describing any excited states that contain considerable double or higher excitation character, which can be abundant in strongly correlated systems. Consequently, CIS commonly produces vertical excitation energies with errors of the order of \SI{1}{eV}\cite{article:Stanton1995}.

	There exist more sophisticated methods still.
	Most notable are a host of techniques that attempt to calculate poles in the linear response of a ground-state reference to a small time-dependent external electric field perturbation: the locations of these poles and the corresponding residues in the frequency domain give vertical excitation energies and oscillator strengths, respectively.
	The single-reference wavefunction implementation of this idea is known as Time-Dependent Hartree--Fock (TDHF)\cite{article:McLachlan1964} and the density functional analogue is Time-Dependent Density Functional Theory (TDDFT)\cite{booksection:Casida1996}.
	Whereas TDHF has only seen very limited applications in the quantum chemistry community\cite{article:Dreuw2005}, TDDFT has enjoyed a much more successful growth owing to its ability to produce vertical excitation energies and excited-state properties for large systems at low computational cost\cite{article:Burke2005}.
	Even though TDDFT is formally an exact theory, approximations to the time-varying exchange-correlation potential must be made since its exact functional form is not known, therefore many TDDFT results are very sensitive to the choice of exchange-correlation functional\cite{article:Dreuw2005,booksection:Elliott2009}.
	Nevertheless, much effort has been put into assessing and improving the validity and quality of the large number of currently available ground-state exchange-correlation functionals used in TDDFT under the adiabatic approximation\cite{booksection:Elliott2009,article:Maier2016}.
	
	Our main concern, however, lies with the inherent dependence of these methods on the ground-state reference.
	For if there are errors in this reference such that it does not form a good description of the ground state, any excited-state calculation that relies on it cannot be expected to give accurate results.
	This property is wittily referred to by \citeauthor{article:Burke2005} as \emph{the sin of the ground state}\cite{article:Burke2005}.
	In fact, \citeauthor{article:Thom2008} demonstrate that the non-linear self-consistent-field (SCF) HF equations exhibit multiple solutions and also remark that the same behavior can be expected for the SCF Kohn--Sham (KS) equations\cite{article:Thom2008}.
	As such, choosing a suitable HF or DFT reference becomes non-trivial, especially if the SCF equations yield multiple low-energy solutions that are degenerate or nearly degenerate in strongly correlated systems.
	Indeed, a number of prior studies on the electronic structures of systems ranging from small radicals such as \ce{F2+} and \ce{O4+} \cite{article:Sherrill1999,article:Cohen2001} to several TM organometallic complexes such as ferrocene\cite{article:Cook1992}, bis($\eta^4$-cyclobutadiene)nickel\cite{article:Jaworska1995}, \ce{Fe(CH)2}\cite{article:Jaworska1999}, a triply bridged chromium dimer\cite{article:Pantazis2019}, and various iron--sulfur clusters\cite{article:Noodleman1981,*article:Noodleman1988} show that there exist several low-energy SCF HF or DFT solutions, some of which are not exact eigenfunctions of $\hat{S}^2$, or contain spin-orbitals that do not respect the molecular symmetry of the structure, or both.
	Therefore, it can be difficult to assign these solutions to actual electronic states or give them meaningful physical interpretations.
	
	In this work, we look for multiple low-energy SCF HF solutions in two representative octahedral hexafluoridometallate(III) complexes, \ce{[MF6]^{3-}} (\ce{M = Ti, V}), using an approach inspired by metadynamics\cite{article:Thom2008}.
	We then carefully analyze these solutions in terms of symmetry and degeneracy so as to classify them systematically according to the irreducible representations of the underlying molecular point group.
	This allows for a rigorous but straightforward definition of \textit{symmetry breaking} and \textit{symmetry conserving} in SCF HF solutions.
	Afterwards, we carry out Non-Orthogonal Configuration Interaction (NOCI)\cite{article:Thom2009b} between different, possibly symmetry-broken, HF solutions to obtain multi-determinantal wavefunctions that conserve symmetry.
	The choice of the model octahedral complexes is thus deliberate: the group-theoretical description of their low-energy electronic terms is well known\cite{book:Sugano1970} while their high molecular symmetry and the presence of one or two unpaired $d$ electrons result in various symmetry-breaking effects that are illuminating yet sufficiently uncomplicated for a tractable in-depth examination.
	The purpose of this paper is therefore twofold: firstly, to demonstrate that symmetry-broken HF solutions are not unphysical and should not be discarded as they can be assigned definitively to actual electronic terms after their symmetry has been restored; and secondly, to show that NOCI using appropriate selections of symmetry-broken HF solutions yields wavefunction descriptions of both ground and excited electronic states that recover a decent amount of static correlation missed out by HF single determinants\cite{article:Benavides-Riveros2017}.
	Even though the use of NOCI on symmetry-broken determinants to form symmetry-conserved multi-determinantal wavefunctions describing ground and excited states has already been formalized and explored under the scope of Projected HF Theory\cite{article:Jimenez-Hoyos2012,article:Jimenez-Hoyos2013,article:Jimenez-Hoyos2013b},
	we think that the detailed analysis presented here adds an extra layer of understanding on the nature and properties of multiple SCF solutions in TM complexes and other strongly correlated systems involving unpaired electrons, and that this understanding can better inform the choice of reference for ground- and excited-state calculations, especially when many of these states are low-lying and degenerate or nearly degenerate.
	We also note that, even though a comparison between NOCI and other standard multi-configurational methods such as Complete Active Space Self-Consistent Field (CASSCF) is highly desirable for a quantification of the amount of static correlation recovered by NOCI, the symmetry breaking nature of many determinants reported in this paper complicates both the CASSCF calculation and the interpretation of the results.
	We therefore defer this comparison to a more comprehensive study in the future.

	The present paper is organized as follows.
	In Section~\ref{sec:theory}, we formalize the notion of symmetry breaking and symmetry conserving in the language of group and representation theory and present a concrete computational method for this classification.
	We outline the computational details for the SCF calculations, symmetry analysis and symmetry restoration of our model systems \ce{[MF6]^{3-}} (\ce{M = Ti, V}) in Section~\ref{sec:comp-details}.
	We then discuss the results of the simpler $d^1$ system \ce{[TiF6]^{3-}} first in Section~\ref{sec:results-d1} to showcase some of the features of symmetry-broken SCF solutions and set the scene for what will follow.
	In Section~\ref{sec:results-d2}, we aim to analyze the results of the more complicated $d^2$ system \ce{[VF6]^{3-}}.
	However, as these can be rather confusing, we start out in Section~\ref{subsec:results-d2-model} with a toy system that only contains two $d$ electrons in an octahedral electrostatic field without any core or ligand electrons. Here, we compare the expected electronic terms derived purely from a symmetry perspective with those obtained using a combination of HF and NOCI.
	We then use this understanding to investigate the SCF and NOCI wavefunctions in the full \ce{[VF6]^{3-}} anion in Section~\ref{subsec:results-d2-full}, paying particular attention to the nature of the correlation recovered via symmetry restoration.
	With a decent insight into the nature of the various SCF and NOCI wavefunctions obtained for \ce{[MF6]^{3-}} (\ce{M = Ti, V}), we study their behaviors as the molecular symmetry descends from $\mathcal{O}_h$ to $\mathcal{D}_{4h}$ or $\mathcal{D}_{2h}$ in Section~\ref{sec:results-JT} and see how the symmetry restoration of SCF solutions is required to correctly describe the stabilization effects necessitated by the Jahn--Teller theorem.

%% file: theory/theory.tex
\tikzsetexternalprefix{./theory/tikz/}

\section{Theory}
\label{sec:theory}

	We briefly review the HF formulation and revisit a fundamental theorem relating group theory to quantum mechanics to define symmetry breaking and symmetry conserving for SCF HF wavefunctions.
	We then give expressions for the representation matrix of a symmetry operation in the single-determinant basis and the corresponding NOCI basis.

	In what follows, we define $N_\mathrm{e}$ and $N_\mathrm{n}$ as the numbers of electrons and nuclei in the system, respectively, and refer to the non-relativistic time-independent electronic Schr\"{o}dinger equation as
		\begin{equation}
			\label{eq:electronicSchroedinger}
			\hat{\mathscr{H}}\Psi =  E\Psi
		\end{equation}
	where the spinless electronic Hamiltonian $\hat{\mathscr{H}}$ takes the following form in atomic units:\cite{book:Griffith1961, book:Szabo1996}
		\ifdefined\twocolumnmode
			\begin{multline}
				\hat{\mathscr{H}} =%
					- \sum_{i}^{N_\mathrm{e}} \frac{1}{2}\nabla_{i}^{2}%
					- \sum_{i}^{N_\mathrm{e}} \: \sum_{A}^{N_\mathrm{n}} \frac{Z_{A}}{\left| \boldsymbol{r}_{i\vphantom{j}} - \boldsymbol{R}_A \right|} \\%
					+ \sum_{i}^{N_\mathrm{e}} \: \sum_{j>i}^{N_\mathrm{e}} \frac{1}{\left| \boldsymbol{r}_i - \boldsymbol{r}_j \right|}
				\label{eq:genhamil}
			\end{multline}
		\else
			\begin{equation}
				\hat{\mathscr{H}} =%
					- \sum_{i}^{N_\mathrm{e}} \frac{1}{2}\nabla_{i}^{2}%
					- \sum_{i}^{N_\mathrm{e}} \: \sum_{A}^{N_\mathrm{n}} \frac{Z_{A}}{\left| \boldsymbol{r}_{i\vphantom{j}} - \boldsymbol{R}_A \right|}%
					+ \sum_{i}^{N_\mathrm{e}} \: \sum_{j>i}^{N_\mathrm{e}} \frac{1}{\left| \boldsymbol{r}_i - \boldsymbol{r}_j \right|}
				\label{eq:genhamil}
			\end{equation}
		\fi
	We also write Slater determinants as\cite{article:Slater1929}
		\ifdefined\twocolumnmode
			\begin{align}
				\Psi_\mathrm{HF}%
					&= |\chi_{1} \ldots \chi_{i} \ldots \chi_{N_\mathrm{e}}| \nonumber \\%
					&= \frac{1}{\sqrt{N_\mathrm{e}!}}%
					\sum_{\sigma}
						(-1)^\sigma
						\hat{\mathscr{P}}_\sigma
						\left[%
							\prod_{i}^{N_\mathrm{e}}
							\chi_{i}(\boldsymbol{x}_i)
						\right]%
				\label{eq:singledet}
			\end{align}
		\else
			\begin{equation}
				\Psi_\mathrm{HF}%
					= |\chi_{1} \ldots \chi_{i} \ldots \chi_{N_\mathrm{e}}| %
					= \frac{1}{\sqrt{N_\mathrm{e}!}}%
						\sum_{\sigma}
							(-1)^\sigma
							\hat{\mathscr{P}}_\sigma
							\left[%
								\prod_{i}^{N_\mathrm{e}}
								\chi_{i}(\boldsymbol{x}_i)
							\right]%
				\label{eq:singledet}
			\end{equation}
		\fi
		where $\chi\left(\boldsymbol{x}\right)$ is a one-electron spin-orbital which can be written most generally as
		\begin{equation}
		\label{eq:spinorb}
		\chi(\boldsymbol{x}) =%
		\omega_{\cdot \delta}(s) \varphi_{\cdot \mu}(\boldsymbol{r}) G^{\delta\mu,\cdot}
		\end{equation}
		where $\omega$ and $\varphi$ are the spin and spatial basis functions and $G^{\delta\mu,\cdot}$ a contravariant component of the generalized spin-orbital coefficient vector labeled by the double index $\delta\mu$ in the covariant direct-product basis,
		\begin{equation*}
		\label{eq:directproductbasis}
		\left\lbrace\omega_{\cdot \delta},\omega_{\cdot \epsilon},\ldots \vphantom{\varphi_{\cdot \mu}}\right\rbrace \otimes \left\lbrace\varphi_{\cdot \mu},\varphi_{\cdot \nu},\ldots\right\rbrace
		\end{equation*}
		Any twice-occurring Greek indices are implicitly contracted over\cite{article:Head-Gordon1998}.
		Furthermore, we use $\boldsymbol{x}$ to denote composite spin-spatial coordinates in overall spin-orbitals, but $s$ and $\boldsymbol{r}$ to indicate separate spin and spatial coordinates in basis functions, respectively.

	\subsection{Symmetry Breaking in the HF Approximation}
	\label{subsec:symbreakingHF}
	
		The symmetry of the Hamiltonian in Equation~(\ref{eq:electronicSchroedinger}) imposes strict constraints on the symmetry and degeneracy of its eigenfunctions.
		This is formalized by the following theorem relating group theory to quantum mechanics\cite{book:Landau1977,book:Heine1960,book:Tsukerblat2006}:
		\begin{theorem}
			\label{theorem:grouptheorytoqm}
			If a Hamiltonian is invariant under a particular symmetry group $\mathcal{G}$, that is, it commutes with all symmetry operations in $\mathcal{G}$, then the eigenfunctions corresponding to one energy level form a basis for an irreducible representation of $\mathcal{G}$.
		\end{theorem}

		We first consider the implications of this theorem for the spatial symmetry of SCF HF solutions.
		If $\mathcal{B}$ is the largest spatial point group under which the spinless Hamiltonian $\hat{\mathscr{H}}$ defined in (\ref{eq:electronicSchroedinger})~and~(\ref{eq:genhamil}) is invariant, then Theorem~\ref{theorem:grouptheorytoqm} dictates that any exact eigenfunction $\Psi$ of $\hat{\mathscr{H}}$ together with all of its linearly independent degenerate partners must transform according to a single irreducible representation of $\mathcal{B}$.
		However, the wavefunction $\Psi_\mathrm{HF}$ optimized under the SCF HF approximation is not guaranteed to be an exact eigenfunction of $\hat{\mathscr{H}}$ and therefore need not obey Theorem~\ref{theorem:grouptheorytoqm}.
		In fact, this non-restriction of spatial symmetry can be traced back to the form of the Fock operator,
			\ifdefined\twocolumnmode
				\begin{multline}
					\label{eq:fockop}
					\hat{f}\left(\boldsymbol{x}_1\right) = \hat{h}\left(\boldsymbol{r}_1\right) \\%
					+ \sum_j^{N_\mathrm{e}} \int \D \boldsymbol{x}_2 \; \chi^*_j\left(\boldsymbol{x}_2\right) \frac{1 - \hat{\mathscr{P}}_{12}}{\left| \boldsymbol{r}_1 - \boldsymbol{r}_2 \right|} \chi_j\left(\boldsymbol{x}_2\right)
				\end{multline}
			\else
				\begin{equation}
					\label{eq:fockop}
					\hat{f}\left(\boldsymbol{x}_1\right) =%
						\hat{h}\left(\boldsymbol{r}_1\right)%
						+ \sum_j^{N_\mathrm{e}} \int \D \boldsymbol{x}_2 \; \chi^*_j\left(\boldsymbol{x}_2\right) \frac{1 - \hat{\mathscr{P}}_{12}}{\left| \boldsymbol{r}_1 - \boldsymbol{r}_2 \right|} \chi_j\left(\boldsymbol{x}_2\right)
				\end{equation}
			\fi
		with $\hat{h}$ being the one-electron Hamiltonian operator.
		Here, the explicit dependence on the bare coordinates of an electron via all the spin-orbitals $\chi_j$ in the second term implies that $\hat{f}$ is not necessarily invariant under $\mathcal{B}$ because there are really no \textit{a priori} symmetry constraints placed on the $\chi_j$ \cite{book:Cook2005}.
		Consequently, the overall SCF wavefunction $\Psi_\mathrm{HF}$ assembled from these spin-orbitals together with its linearly independent degenerate partners, if any, does not have to transform as any single irreducible representation of $\mathcal{B}$ but may form a basis for a reducible representation instead.
		
		Theorem~\ref{theorem:grouptheorytoqm} is also applicable to groups containing spin rotation and time reversal operations, the most general of which is the direct-product group $\mathcal{S} \otimes \mathcal{T}$ where $\mathcal{S}$ is the full spin rotation group and $\mathcal{T}$ the time reversal group.
		$\mathcal{S}$ is identical to $\mathsf{SU}(2)$ (the two-dimensional special unitary group) by definition\cite{article:Fukutome1981} whereas $\mathcal{T}$ is isomorphic to $\mathcal{C}_4$ (the cyclic group of order 4) (Appendix~\ref{appsubsec:timerevgroup}).
		The irreducible representations and characters of $\mathcal{S} \otimes \mathcal{T}$ can thus be obtained from those of $\mathsf{SU}(2)$ and $\mathcal{C}_4$, which are well known\cite{book:Wigner1959,book:James2001}.
		The invariance of $\hat{\mathscr{H}}$ under $\mathcal{S} \otimes \mathcal{T}$ ensures that any eigenfunction $\Psi$ and its degenerate partners must also transform as one of the irreducible representations of $\mathcal{S} \otimes \mathcal{T}$.
		However, once again the approximate nature of the HF method does not guarantee that the SCF wavefunctions $\Psi_\mathrm{HF}$ always satisfy this condition.
		Nevertheless, with suitable constraints imposed on the spin and reality of the spin-orbitals, $\Psi_\mathrm{HF}$ and its degenerate partners can be forced to transform as a single irreducible representation of one of the subgroups of $\mathcal{S} \otimes \mathcal{T}$ and thus be eigenfunctions of appropriate operators constructed from the generators of the subgroup (Appendix~\ref{appsubsec:symbreakingTSgroup}).		
		The hierarchy of these subgroups was originally deduced by Fukutome\cite{article:Fukutome1981} but has since been presented in connection with the various HF regimes using more modern electronic-structure nomenclatures\cite{book:Stuber2003,article:Jimenez-Hoyos2011,article:Henderson2018}.
		We will be particularly interested in $\hat{\Theta}$ (the time reversal operator), $\hat{S}^2$ (the operator representing the square of the magnitude of the spin angular momentum), and  $\hat{S}_3$ (the operator representing the projection of the spin angular momentum onto the third Cartesian axis, which is typically the $z$-axis).
		In atomic units, the eigenvalues of $\hat{S}^2$ are well known\cite{book:Bransden2000} to be $S(S+1)$ where $S$ can take integer or half-integer values, and the corresponding eigenvalues of $\hat{S}_3$ are $M_S = -S, -S+1, \ldots, S$.
		Whenever necessary, we will use upper-case notations for overall many-electron electronic states and lower-case equivalents for individual spin-orbitals.

		In light of the above discussion, we propose the following definition.
		If a set of degenerate wavefunctions is found to form a basis for a single irreducible representation of a group $\mathcal{G}$, be it a spatial point group $\mathcal{B}$ or one of the subgroups of $\mathcal{S} \otimes \mathcal{T}$, they are \emph{symmetry-conserved} in $\mathcal{G}$.
		On the other hand, if they span a representation that can be reduced to multiple irreducible representations of $\mathcal{G}$, they are \emph{symmetry-broken} in $\mathcal{G}$.
		Theorem~\ref{theorem:grouptheorytoqm} immediately rules out symmetry-broken wavefunctions as eigenfunctions of $\hat{\mathscr{H}}$, whereas it places no such restriction on symmetry-conserved wavefunctions.
		
	\subsection{Representation Matrices of Symmetry Operations in SCF and NOCI Bases}
		\label{subsec:repmats}
	
		Consider a set of $N_\mathrm{det}$ degenerate Slater determinants that are possibly non-orthogonal and not all linearly independent,
			\begin{equation}
				\label{eq:Slaterdetset}
				\left\lbrace
					\prescript{w}{}{\Psi} %
					\ | \ %
					w = 1, 2, \ldots, N_\mathrm{det} %
				\right\rbrace 
			\end{equation}
		The overlap matrix $\boldsymbol{S}$ in this basis is defined as
			\begin{equation}
				\label{eq:scfov}
				\prescript{wx}{}{S} = \left(\boldsymbol{S}\right)_{wx} = \Braket{\prescript{w}{}{\Psi}|\prescript{x}{}{\Psi}}
			\end{equation}
		which is a square matrix with dimensions $N_\mathrm{det} \times N_\mathrm{det}$ and rank $N_\mathrm{indept} \leq N_\mathrm{det}$.
		It is possible to construct an $N_\mathrm{det} \times N_\mathrm{indept}$ matrix $\boldsymbol{X}$ that transforms the above basis into a linearly independent one (henceforth denoted by a tilde) in which the overlap matrix
			\begin{equation}
				\label{eq:canonicalorthogonalizationofS}
				\boldsymbol{\tilde{S}} = \boldsymbol{X}^\dagger \boldsymbol{SX}
			\end{equation}
		is of full rank.
		By means of the non-orthogonal projection operator defined by \citeauthor{article:Soriano2014}\cite{article:Soriano2014}, we show in Appendix~\ref{app:symanalysis} that the representation matrix of a symmetry operation $\hat{R}$ in the linearly independent basis takes the form
			\begin{equation}
				\label{eq:DtildeR}
				\boldsymbol{\tilde{D}}(\hat{R}) =%
					\boldsymbol{\tilde{S}}^{-1} %
					\boldsymbol{X}^\dagger %
					\boldsymbol{T}(\hat{R}) %
					\boldsymbol{X}
			\end{equation}
		where the elements of $\boldsymbol{T}(\hat{R})$ are given by
			\begin{equation}
				\label{eq:TwxR}
				T_{wx}(\hat{R}) = \braket{\prescript{w}{}{\Psi} | \hat{R} \prescript{x}{}{\Psi}}
			\end{equation}
			
		We now let
			\begin{equation*}
				\left\lbrace
					\prescript{m}{}{\Phi} %
					\ | \ %
					m = 1, 2, \ldots, N_\mathrm{indept} %
				\right\rbrace 
			\end{equation*}
		be the set of $N_\mathrm{indept}$ linearly independent NOCI wavefunctions obtained from the above $N_\mathrm{det}$ degenerate Slater determinants as
			\begin{equation}
				\label{eq:nociwavefunction}
				\prescript{m}{}{\Phi} = \sum_{w}^{N_\mathrm{det}} \prescript{w}{}{\Psi} A_{wm}
			\end{equation}
		where $A_{wm}$ are elements of the $N_\mathrm{det} \times N_\mathrm{indept}$ matrix $\boldsymbol{A}$ that solves the secular equation\cite{article:Thom2009b}
			\begin{equation}
				\label{eq:nocisecular}
				\boldsymbol{HA} = \boldsymbol{SAE}
			\end{equation}
		with $\boldsymbol{H}$ given by
			\begin{equation}
				\prescript{wx}{}{H} = \left(\boldsymbol{H}\right)_{wx} =%
					\braket{\prescript{w}{}{\Psi}|\hat{\mathscr{H}}|\prescript{x}{}{\Psi}}
			\end{equation}
		and the eigenvalues collected in $\boldsymbol{E} = \mathrm{diag}\left(E_1,\ldots,E_{N_\mathrm{indept}}\right)$ give the energies of the NOCI wavefunctions.
		The representation matrix of $\hat{R}$ in the NOCI basis can be shown (Appendix~\ref{app:symanalysis}) to take the form
			\begin{align}
				\boldsymbol{D}^\mathrm{NOCI}(\hat{R})%
				&= \boldsymbol{\tilde{D}}^\mathrm{NOCI}(\hat{R}) \nonumber \\%
				&= (\boldsymbol{X}^\dagger \boldsymbol{SA})^{-1} %
				\boldsymbol{X}^\dagger %
				\boldsymbol{T}(\hat{R}) %
				\boldsymbol{A}
				\label{eq:Dnoci}
			\end{align}
		The matrix elements of $\boldsymbol{H}$, $\boldsymbol{S}$, and $\boldsymbol{T}(\hat{R})$ involving non-orthogonal determinants can be easily calculated using L\"{o}wdin's paired orbitals\cite{article:Amos1961a,article:Thom2009b,article:Mayhall2014,article:Sundstrom2014b}.
		
		By solving the NOCI secular equation (\ref{eq:nocisecular}), we effectively allow the different irreducible representation components spanned by the symmetry-broken set $\left\lbrace \prescript{w}{}{\Psi} \right\rbrace$ to interact via the totally symmetric $\hat{\mathscr{H}}$ and form linear combinations that transform as single irreducible representations and hence conserve symmetry.
		$\boldsymbol{D}^\mathrm{NOCI}(\hat{R})$ must therefore have a block-diagonal structure in which each little block is an irreducible matrix representation of $\hat{R}$ in the basis of the corresponding NOCI wavefunctions.%
		\bibnote{Repeated representations are not a problem. For example, if both $\left\lbrace \prescript{m}{}{\Phi} \right\rbrace$ and $\left\lbrace \prescript{n}{}{\Phi} \right\rbrace$ form two bases for the same irreducible representation $\Gamma$, NOCI must already take into account their symmetry-allowed mutual interaction and lift their degeneracy with respect to each other so that the $\left\lbrace \prescript{m}{}{\Phi} \right\rbrace$ wavefunctions only transform amongst themselves under any symmetry operation of the group, as do the $\left\lbrace \prescript{n}{}{\Phi} \right\rbrace$ wavefunctions.}
		Furthermore, we show in Appendix~\ref{app:nocinatorbs} that, from each NOCI wavefunction $\prescript{m}{}{\Phi}$, a one-particle density matrix can be constructed which can be diagonalized to give natural orbitals and the associated occupation numbers.
		In addition to giving us a way to visualize the NOCI wavefunctions, these provide further insight into the correlation that they recover by restoring broken symmetry.

%% file: comp-details/comp-details.tex
\tikzsetexternalprefix{./comp-details/tikz/}

\section{Computational Details}
\label{sec:comp-details}
	
	\begin{figure}
		\centering
		\includegraphics{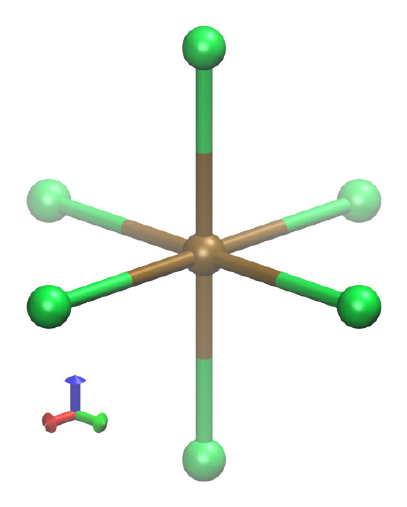}
		\caption{Generic structure of octahedral \ce{[MF6]^{3-}} (\ce{M = Ti, V}) visualized with VMD\cite{article:HUMP96}. Fluoride ligands (green) lie on the Cartesian axes equidistant to the metal center (ocher) at the origin of the coordinate system.}
		\label{fig:MF63-}
	\end{figure}

\begin{figure}
			\centering
			% \useexternalfile{scale}{trimleft}{trimright}{name}
			% Figure compiled with tikzexternalize
			% Pre-compiled figure located at ./comp-details/tikz/tikzoutput/workflow-output.pdf
			\input{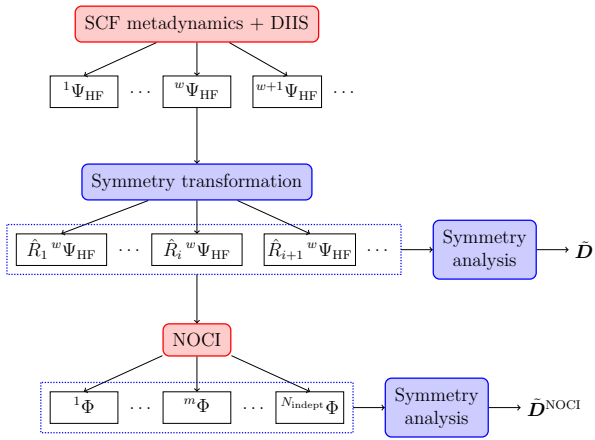}{0}{0}{workflow}
			\caption{Computational procedure for generating and running NOCI on symmetry-equivalent SCF HF solutions. Boxes with solid red boundaries denote calculations run in {\scshape Q-Chem}. Boxes with solid blue boundaries denote routines run in our Python codes. Wavefunctions enclosed in black boxes are passed between {\scshape Q-Chem} and our Python codes.}
			\label{fig:workflow}
		\end{figure}

	In both model \ce{[MF6]^{3-}} (\ce{M = Ti, V}) complexes, the fluoride ligands lie on the Cartesian axes and are equidistant to the metal center at the origin (Figure~\ref{fig:MF63-}).
	The procedure for obtaining and restoring the symmetry of multiple SCF HF solutions in these complexes is sketched in Figure~\ref{fig:workflow}.
	These solutions were located within the Unrestricted Hartree--Fock (UHF) space using the Direct Inversion in the Iterative Subspace (DIIS) algorithm\cite{article:Pulay1980} in conjunction with SCF metadynamics\cite{article:Thom2008} in {\scshape Q-Chem} 5.1\cite{article:Shao2015}.
	Pople's double-zeta split-valence 6-31G* basis sets were employed for all atoms with Cartesian forms for $d$ functions and pure forms for $f$ functions.
	Each solution was considered to have reached convergence when the DIIS error fell below \SI{1e-13}{} (unless stated otherwise).
	More detailed {\scshape Q-Chem} input parameters are listed in the Supporting Information.

	Various spatial symmetry operations $\hat{R}$ of the $\mathcal{O}_h$ point group were then applied to the converged SCF HF solutions to generate their symmetry-equivalent partners.
	The effect of $\hat{R}$ on each single-determinantal solution defined in (\ref{eq:singledet}) can be written as
		\ifdefined\twocolumnmode
			\begin{multline}
				\label{eq:Rsingledet}
				\hat{R}\Psi_\mathrm{HF} = \\%
					\frac{1}{\sqrt{N_\mathrm{e}!}}%
					\sum_{\sigma}
					\left(-1\right)^\sigma
					\hat{\mathscr{P}}_\sigma
					\left\lbrace%
					\prod_{i}^{N_\mathrm{e}}
					\left[\hat{R} \chi_{i}\left(\boldsymbol{x}_i\right)\right]
					\right\rbrace
			\end{multline}
		\else
			\begin{equation}
				\label{eq:Rsingledet}
				\hat{R}\Psi_\mathrm{HF} = %
					\frac{1}{\sqrt{N_\mathrm{e}!}}%
					\sum_{\sigma}
					\left(-1\right)^\sigma
					\hat{\mathscr{P}}_\sigma
					\left\lbrace%
					\prod_{i}^{N_\mathrm{e}}
					\left[\hat{R} \chi_{i}\left(\boldsymbol{x}_i\right)\right]
					\right\rbrace
			\end{equation}
		\fi
	We show in Appendix~\ref{app:symtrans} that $\hat{R}$ transforms the spin-orbitals $\chi_{i}\left(\boldsymbol{x}_i\right)$ into
		\ifdefined\twocolumnmode
			\begin{multline}
				\label{eq:Rchi}
				\hat{R} \chi_{i}\left(\boldsymbol{x}_i\right) = \\%
					\omega_{\cdot \delta}\left(s_i\right) \left\lbrace\hat{R} \tilde{\varphi}_{\cdot \mu}\left[\hat{R}(\boldsymbol{r}_i-\boldsymbol{R}_A)\right] \right\rbrace G_i^{\delta\mu,\cdot}
			\end{multline}
		\else
			\begin{equation}
				\label{eq:Rchi}
				\hat{R} \chi_{i}\left(\boldsymbol{x}_i\right) = %
					\omega_{\cdot \delta}\left(s_i\right) \left\lbrace\hat{R} \tilde{\varphi}_{\cdot \mu}\left[\hat{R}(\boldsymbol{r}_i-\boldsymbol{R}_A)\right] \right\rbrace G_i^{\delta\mu,\cdot}
			\end{equation}
		\fi
	where $\hat{R}\tilde{\varphi}_{\cdot \mu}\left[\hat{R}(\boldsymbol{r}_i-\boldsymbol{R}_A)\right]$ denotes a spatial atomic orbital that is originally centered on nucleus $A$ but that has been moved to a possibly different nucleus, say $A'$, before being transformed by the same $\hat{R}$ in its new local coordinate system centered on $A'$.
	Here, $A'$ is the nucleus onto which $A$ is mapped by the action of $\hat{R}$ on the nuclear framework.
	The calculation of $\hat{R} \tilde{\varphi}_{\cdot \mu}$ in terms of unitary transformation matrices in spherical harmonic bases is also formulated in Appendix~\ref{app:symtrans}, which allows for a simple implementation of all symmetry transformation routines within a set of Python 2.7 codes using NumPy\cite{article:vanderWalt2011} developed in-house.
	Each $\hat{R}\Psi_\mathrm{HF}$ determinant generated using symmetry was used as an initial guess for an SCF calculation with the same convergence criteria as described above.
	The converged solution was then compared with $\hat{R}\Psi_\mathrm{HF}$ using the square state distance metric, $d_{wx}^2$, defined by \citeauthor{article:Thom2008}\cite{article:Thom2008} to ensure that they were identical within numerical errors ($d_{wx}^2 \leq \SI{1e-13}{electrons}$), thus confirming that $\hat{R}\Psi_\mathrm{HF}$ is indeed an SCF HF solution.
	
	The volumetric data for the spatial form of all spin-orbitals was generated in \textsc{Q-Chem} and are represented by isosurfaces plotted using the Tachyon ray-tracing library\cite{masterthesis:STON1998} in VMD\cite{article:HUMP96}.
	
	The multiple SCF HF solutions generated by symmetry were reread into {\scshape Q-Chem} 5.1 and a NOCI calculation was performed on them.
	The NOCI eigenvectors, as well as the associated SCF HF solutions, were then fed back into the Python codes so that the representation matrices in the NOCI basis, $\boldsymbol{\tilde{D}}^\mathrm{NOCI}$, could be computed using (\ref{eq:Dnoci}) and checked for block-diagonality.
	The trace of each block in $\boldsymbol{\tilde{D}}^\mathrm{NOCI}$ was obtained trivially and the irreducible representation spanned by the corresponding NOCI states deduced.

%% file: results-d1/results-d1.tex
\tikzsetexternalprefix{./results-d1/tikz/}

\section{\textit{d}\textsuperscript{1} Metal Ground Configuration}
\label{sec:results-d1}

	\subsection{True \textit{d}\textsuperscript{1} Octahedral System}

		For a single electron described by a set of spin-orbitals with $d$ spatial symmetry in an octahedral field of point charges, group theory dictates that the only possible terms of the exact electronic wavefunctions are $\prescript{2}{}{T}_{2g}$ and $\prescript{2}{}{E}_g$ whose forms are shown in Table~\ref{tab:d1terms} using the notations of \citeauthor{book:Sugano1970}\cite{book:Sugano1970}
		Lower-case irreducible representation symbols ($t_{2g}$, $e_g$) are used to denote the spatial symmetry of a single spin-orbital, whereas upper-case irreducible representation symbols ($T_{2g}$, $E_g$) are reserved for the overall spatial symmetry of the wavefunction.
		Unsurprisingly, the two notations coincide for the one-electron $d^1$ system.
		If the spatial parts of these wavefunctions are constructed from a single set of five degenerate hydrogenic $d$ orbitals with principal quantum number $n$, then first-order perturbation theory gives the following real forms for the components of $t_{2g}$ and $e_g$\cite{book:Sugano1970}:
			\ifdefined\twocolumnmode
				\begin{equation}
					\begin{aligned}
						\xi(\boldsymbol{r}) &= d_{yz}(\theta, \phi) R_{nd}(r) \\
						\eta(\boldsymbol{r}) &= d_{xz}(\theta, \phi) R_{nd}(r) \\
						\zeta(\boldsymbol{r}) &= d_{xy}(\theta, \phi) R_{nd}(r) \\
						u(\boldsymbol{r}) &= d_{z^2}(\theta, \phi) R_{nd}(r) \\
						v(\boldsymbol{r}) &= d_{x^2-y^2}(\theta, \phi) R_{nd}(r)
					\end{aligned}
					\label{eq:1e5dforms}
				\end{equation}
			\else
				\begin{equation}
					\begin{gathered}
						\xi(\boldsymbol{r}) = d_{yz}(\theta, \phi) R_{nd}(r), \quad
						\eta(\boldsymbol{r}) = d_{xz}(\theta, \phi) R_{nd}(r), \quad
						\zeta(\boldsymbol{r}) = d_{xy}(\theta, \phi) R_{nd}(r) \\
						u(\boldsymbol{r}) = d_{z^2}(\theta, \phi) R_{nd}(r), \quad
						v(\boldsymbol{r}) = d_{x^2-y^2}(\theta, \phi) R_{nd}(r)
					\end{gathered}
					\label{eq:1e5dforms}
				\end{equation}
			\fi
		where $d_{yz}$, $d_{xz}$ $d_{xy}$, $d_{z^2}$, and $d_{x^2-y^2}$ denote the familiar (real) spherical harmonics of degree 2 and $R_{nd}$ the hydrogenic radial wavefunction of the $nd$ subshell.
		The energy difference between the $\prescript{2}{}{T}_{2g}$ and $\prescript{2}{}{E}_g$ terms is commonly denoted as $\Delta_\mathrm{oct} = 10Dq$\cite{book:Griffith1961,book:Sugano1970,book:Housecroft2012}.
		
		\begin{table*}
			\centering
			\caption{
				All possible $d^1$ terms in an octahedral field.
				$\xi$, $\eta$, and $\zeta$ are the components of $T_{2g}$ while $u$ and $v$ are the components of $E_g$.
				Each spin-orbital is also an eigenfunction of the $\hat{s}_3$ operator with eigenvalue $m_s = \frac{1}{2}$ (without bar) or $-\frac{1}{2}$ (with bar).
			}
			\label{tab:d1terms}
			\begingroup
			\renewcommand\arraystretch{1.35}
			\begin{tabular}[t]{p{1.5cm} p{1.3cm} p{1.5cm} p{2.0cm} p{2.0cm}}
				\toprule
				Config. & Term & Comp. & $M_S = \frac{1}{2}$ & $M_S = -\frac{1}{2}$ \\
				\midrule
				$t_{2g}^1$ & \multirow[t]{3}{*}{$\prescript{2}{}{T}_{2g}$}%
					& $\xi$ & $\xi(\boldsymbol{x})$ & $\bar{\xi}(\boldsymbol{x})$ \\
				&	& $\eta$ & $\eta(\boldsymbol{x})$ & $\bar{\eta}(\boldsymbol{x})$ \\
				&	& $\zeta$ &	$\zeta(\boldsymbol{x})$ & $\bar{\zeta}(\boldsymbol{x})$ \\
				\midrule
				$e_{g}^1$ & \multirow[t]{2}{*}{$\prescript{2}{}{E}_{g}$}%
					& $u$ & $u(\boldsymbol{x})$ & $\bar{u}(\boldsymbol{x})$ \\
				&	& $v$ & $v(\boldsymbol{x})$ & $\bar{v}(\boldsymbol{x})$ \\
				\bottomrule
			\end{tabular}
			\endgroup
		\end{table*}

	\subsection{\ce{[TiF6]^{3-}}}

		The octahedral \ce{[TiF6]^{3-}} anion has a more complicated electronic structure, however. If all six ligands are considered to be closed-shell \ce{F-}, the metal center is then \ce{Ti^{3+}} which has the configuration $\ce{[Ar]}3d^1$.
		The unpaired $3d^1$ electron still lies in an octahedral field of point charges set up by the nuclei but now exhibits electron--electron interaction with the \ce{Ar} core and the \ce{F-} ligands, and the analytic results for a true $d^1$ system presented above no longer rigorously applies.
		However, in most theoretical treatments, the fully filled shells on both the metal center and the ligands are ignored on the grounds that they only contribute a totally symmetric singlet component to the overall wavefunction which shifts the energy of every term derived for the true $d^1$ system by a constant amount (see Chapter 7 of Ref.~\citenum{book:Griffith1961}).
		While this is often a reasonable assumption to simplify the analytic descriptions and provide an excellent starting point for the qualitative understanding of the electronic terms present, it glosses over any electron correlations between the valence $d$ electrons and the other electrons in the system.
		
		\paragraph{Overview of UHF solutions.}
		
\begin{figure*}
					\centering
					% \useexternalfile{scale}{trimleft}{trimright}{name}
					% Figure compiled with tikzexternalize
					% Pre-compiled figure located at ./results-d1/tikz/tikzoutput/d1_nonoci-output.pdf
					\input{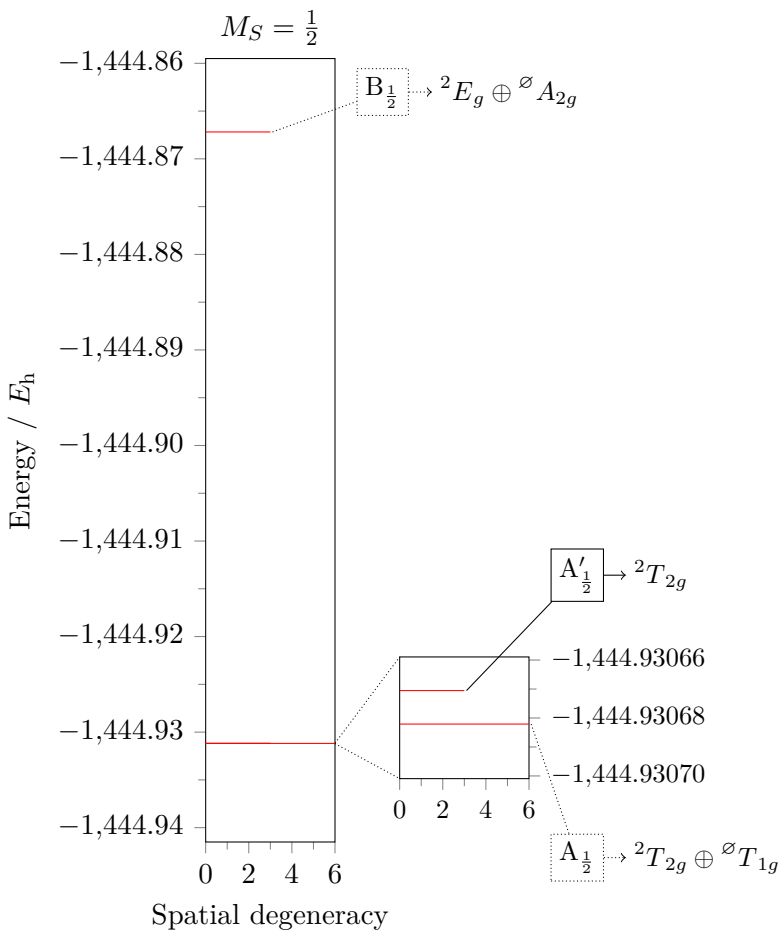}{0}{0}{d1_nonoci}
					\caption{
						Energy and symmetry of low-lying UHF solutions (6-31G* basis) in octahedral \ce{[TiF6]^3-} having Ti--F = \SI{2.0274}{\angstrom}.
						All solutions have DIIS errors smaller than \SI{1e-13}{}.
						All solutions have $N_\alpha - N_\beta = 1$ and are thus eigenfunctions of $\hat{S}_3$ with the same eigenvalue $M_S = \frac{1}{2}$ given as subscripts.
						Solutions are considered degenerate when there energies are at most \SI{e-9}{\hartree} apart and are labeled alphabetically in increasing order of their energies.
						Nearly degenerate solutions share the same letter but are distinguished by dashes.
						Solutions within one degenerate set are distinct and linearly independent.
						Solutions that conserve spatial symmetry are enclosed in solid boxes whereas solutions that break spatial symmetry are enclosed in dotted ones.
						See the main text for a discussion on spin multiplicities.
					}
					\label{fig:d1_nonoci}
				\end{figure*}

			Figure~\ref{fig:d1_nonoci} presents the energy, spatial symmetry, and labels of the lowest-energy $M_S = \frac{1}{2}$ UHF solutions located by SCF metadynamics in octahedral \ce{[TiF6]^3-} at Ti--F bond length of \SI{2.0274}{\angstrom}.
			The representative Pipek--Mezey-localized\cite{article:Pipek1989,article:Thom2009} spatial forms of the highest-occupied $m_s = \frac{1}{2}$  spin-orbitals in these solutions are shown in Table~\ref{tab:d1_nonoci} and detailed solution energies and overlap matrices for the degenerate sets can be found in the Supporting Information.
			There are two groups of solutions approximately $\SI{0.063}{\hartree} \approx \SI{14000}{cm^{-1}}$ apart, which is expected from the two possible $d^1$ terms in an octahedral field presented earlier and from the experimental values for $\Delta_\mathrm{oct}$ in this anionic complex\cite{article:Hatfield1971}.
			However, in order to assign these solutions to the expected terms, their spin and spatial symmetry properties need to be examined.
			
				\begin{table*}
					\centering
					\caption{
						Representative isosurface plots for the Pipek--Mezey-localized spatial parts of the highest-occupied $m_s = \frac{1}{2}$ spin-orbitals, spatial symmetry, and $\langle \hat{S}^2 \rangle$ of the $M_S = \frac{1}{2}$ UHF solutions in \ce{[TiF6]^{3-}}.
						The spin-orbitals of all solutions within one set have similar forms and are related by the symmetry operations of $\mathcal{O}_h$.
						Spatial symmetry lists the irreducible representations of $\mathcal{O}_h$ spanned by the degenerate sets.
						Axis triad: red\textendash $x$; green\textendash $y$; blue\textendash $z$.
					}
					\label{tab:d1_nonoci}
					\footnotesize
					\begin{tabular}[t]{>{\raggedright\arraybackslash}m{0.9cm} >{\centering\arraybackslash}m{1.9cm} >{\centering\arraybackslash}m{1.9cm} S[table-figures-decimal=4, table-figures-integer=1, table-number-alignment=center]}
						\toprule
						$\Psi_\mathrm{UHF}$ & $\chi_{40}$ &
						Spatial symmetry & {$\langle \hat{S}^2 \rangle$} \\
						\midrule
						$\mathrm{A}_{\frac{1}{2}}$ & \includegraphics{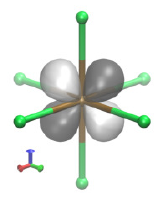} & $T_{1g} \oplus T_{2g}$ & 0.7522 \\
						$\mathrm{A}_{\frac{1}{2}}'$ & \includegraphics{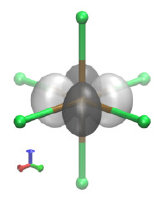} & $T_{2g}$ & 0.7522 \\
						$\mathrm{B}_{\frac{1}{2}}$ & \includegraphics{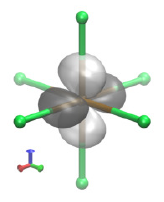} & $A_{2g} \oplus E_g$ & 0.7527 \\
						\bottomrule
					\end{tabular}
				\end{table*}
			
			We turn first to the lower group of solutions.
			There are two degenerate sets within this group, namely $\mathrm{A}_{\frac{1}{2}}$ and $\mathrm{A}_{\frac{1}{2}}'$, whose energies differ by about \SI{1.2e-5}{\hartree}.
			The $\mathrm{A}_{\frac{1}{2}}$ set is six-fold degenerate and forms a basis for the $T_{1g} \oplus T_{2g}$ representation in $\mathcal{O}_h$.
			It is therefore symmetry-broken because it spans more than one irreducible representation.
			A consideration of the spatial forms of the highest-occupied spin-orbitals of the $\mathrm{A}_{\frac{1}{2}}$ solutions (one representative shown in Table~\ref{tab:d1_nonoci}) suggests why this is the case: the valence $3d$ spin-orbital in each solution has the familiar $d_{xy}/d_{yz}/d_{xz}$ form but has been rotated by $\pi/4$ out of the Cartesian planes of the ligands.
			There are thus six equivalent orientations of the valence $3d$ spin-orbital, each of which corresponds to a linearly independent but degenerate UHF solution.
			On the other hand, the $\mathrm{A}_{\frac{1}{2}}'$ set only spans $T_{2g}$ and is therefore symmetry-conserved: the valence $3d$ spin-orbital in each $\mathrm{A}_{\frac{1}{2}}'$ solution lies properly in one of the three Cartesian ligand planes.
			Consequently, the valence $3d$ electron in the $\mathrm{A}_{\frac{1}{2}}'$ solutions experiences a stronger repulsion from the ligand electrons which raises the energy of the $\mathrm{A}_{\frac{1}{2}}'$ set slightly above that of $\mathrm{A}_{\frac{1}{2}}$ (Figure~\ref{fig:d1_nonoci}).
		
			We must now pause for a moment to consider the significance of the existence of the symmetry-broken $\mathrm{A}_{\frac{1}{2}}$ solutions in \ce{[TiF6]^{3-}}.
			In a true $d^1$ system, there \emph{cannot} exist any symmetry-broken one-electron wavefunctions comprising spin-orbitals similar to the valence $3d$ ones in the $\mathrm{A}_{\frac{1}{2}}$ solutions.
			Firstly, any $T_{1g}$ component is strictly forbidden by symmetry.
			Secondly, the $T_{2g}$ space is fully spanned by the degenerate $\xi$, $\eta$, and $\zeta$ components shown in (\ref{eq:1e5dforms}), so any linear combinations of these three functions to give the valence $3d$ spin-orbitals of $\mathrm{A}_{\frac{1}{2}}$ necessarily live in the same space and have the same energy.
			However, in \ce{[TiF6]^{3-}}, the \ce{Ar} core and the \ce{F-} electrons can relax differently in the presence of the different orientations of the valence $3d$ spin-orbitals relative to the ligand exes to give $\mathrm{A}_{\frac{1}{2}}$ and $\mathrm{A}_{\frac{1}{2}}'$ UHF solutions with different energies and symmetries.
			Both $\mathrm{A}_{\frac{1}{2}}$ and $\mathrm{A}_{\frac{1}{2}}'$ solutions contain a $T_{2g}$ component and are thus approximations to an exact wavefunction with $T_{2g}$ spatial symmetry which we assume to be the ground wavefunction%
				\bibnote{Strictly speaking, one cannot be sure that such a $T_{2g}$ wavefunction is indeed the exact ground eigenfunction of the electronic Hamiltonian for \ce{[TiF6]^{3-}} without solving the electronic Schr\"{o}dinger equation exactly, and so one cannot claim with certainty that the UHF solutions reported provide approximations to the ground electronic state of \ce{[TiF6]^{3-}}. Group theory only requires that the exact eigenfunctions of the electronic Hamiltonian for \ce{[TiF6]^{3-}} transform as single irreducible representations of $\mathcal{O}_h$ but is completely silent to their ordering. Nevertheless, if \ce{[TiF6]^{3-}} is assumed to behave qualitatively similar to a true $d^1$ system by virtue of the approximately totally symmetric singlet \ce{Ar} core and \ce{F-} ligands (see Chapter 7 of Ref.~\citenum{book:Griffith1961}), then its true ground and first excited wavefunctions are expected to be described by the $\prescript{2}{}{T}_{2g}$ and $\prescript{2}{}{E}_{g}$ terms, respectively.}%
			, but the $\mathrm{A}_{\frac{1}{2}}$ solutions provide a tighter approximation in light of the variational principle because they incorporate some correlation between the valence $3d$ electron and the remaining electrons to result in symmetry-brokenness and become lower in energy.
			
			We now shift our attention to the $\mathrm{B}_{\frac{1}{2}}$ set.
			This set spans $A_{2g} \oplus E_g$ and is thus considered to be a symmetry-broken approximation of an exact wavefunction with $E_g$ spatial symmetry, but the $d_{x^2-y^2}$-like form of the valence $3d$ spin-orbital (Table~\ref{tab:d1_nonoci}) seems to suggest that this set conserves spatial symmetry, so what is going on?
			It turns out that, for a true $d^1$ system, the spatial orbital $d_{x^2-y^2}(\theta, \phi) R_{nd}(r)$ and its symmetry partners, $d_{y^2-z^2}(\theta, \phi) R_{nd}(r)$ and $d_{x^2-z^2}(\theta, \phi) R_{nd}(r)$, are clearly not all linearly independent and can be shown to indeed span the same $E_g$ space as $u(\boldsymbol{r})$ and $v(\boldsymbol{r})$ defined in (\ref{eq:1e5dforms})\cite{book:Sugano1970}.
			In contrast, the presence of the core and ligand electrons in \ce{[TiF6]^{3-}} removes this linear dependence such that the three $\mathrm{B}_{\frac{1}{2}}$ solutions span a larger space that contains an additional $A_{2g}$ component and become symmetry-broken, effectively incorporating a similar kind of correlation as discussed for the $\mathrm{A}_{\frac{1}{2}}$ solutions.
		
			\begin{figure*}
				\centering
				% \useexternalfile{scale}{trimleft}{trimright}{name}
				% Figure compiled with tikzexternalize
				% Pre-compiled figure located at ./results-d1/tikz/tikzoutput/d1_noci-output.pdf
				\input{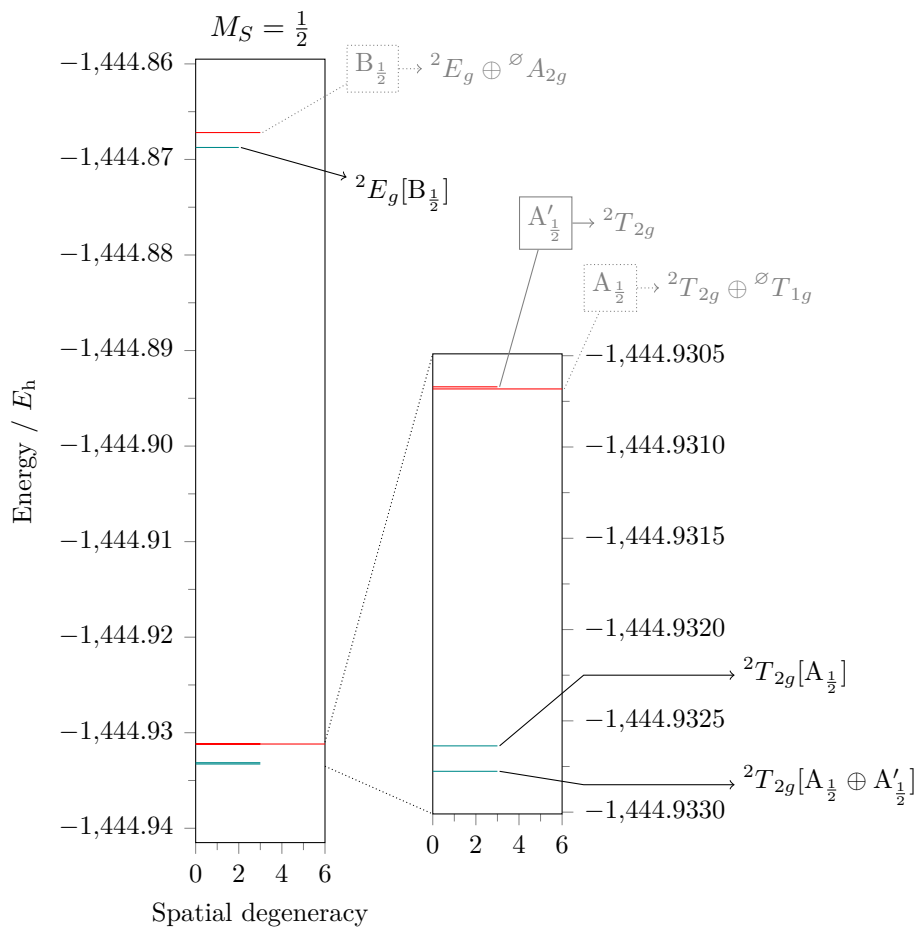}{0}{0}{d1_noci}
				\caption{
					Low-lying NOCI wavefunctions (dark cyan) constructed from the low-lying $M_S = \frac{1}{2}$ UHF solutions in octahedral \ce{[TiF6]^3-} (red, replotted from Figure~\ref{fig:d1_nonoci} for comparison).
					NOCI wavefunctions are considered degenerate when there energies are at most \SI{e-7}{\hartree} apart.
					In generic notations, $\Gamma[\mathrm{A}\oplus\mathrm{B}]$ denotes a specific NOCI set of symmetry $\Gamma$ constructed from both $\mathrm{A}$ and $\mathrm{B}$ solutions.
				}
				\label{fig:d1_noci}
			\end{figure*}
			
			\begin{table*}
				\centering
				\caption{
					Isosurface plots for the Pipek--Mezey-localized spatial parts of the $3d$ natural orbitals and $\langle \hat{S}^2 \rangle$ of the low-energy NOCI wavefunctions $\Phi$ in \ce{[TiF6]^{3-}}.
					Each $3d$ natural orbital shown in each set corresponds to one of the NOCI wavefunctions within the set that transforms as the labeled component.
					Shown in parentheses underneath are the occupation numbers of the natural orbitals.
					The occupation numbers and natural orbitals shown are solely in the $m_s= \frac{1}{2}$ space.	
					Axis triad: red\textendash $x$; green\textendash $y$; blue\textendash $z$.
				}
				\label{tab:d1_noci}
				\footnotesize
				\begin{tabular}[t]{>{\raggedright\arraybackslash}m{2.3cm} >{\centering\arraybackslash}m{2.0cm} >{\centering\arraybackslash}m{2.0cm} >{\centering\arraybackslash}m{2.0cm} S[table-figures-decimal=4, table-figures-integer=1, table-number-alignment=center] >{\raggedright\arraybackslash}m{2.0cm}}
					\toprule
					$\Phi$ & \multicolumn{3}{c}{$3d$ natural orbitals} &
					{$\langle \hat{S}^2 \rangle$} & Term \\
					\midrule
					$\prescript{2}{}{T}_{2g}[\mathrm{A}_{\frac{1}{2}}]$ & \includegraphics{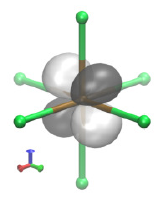} & \includegraphics{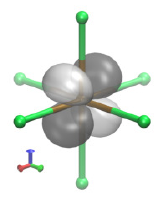} & \includegraphics{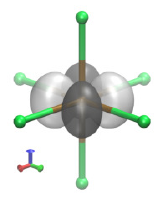} & 0.7512 & $\prescript{2}{}{T}_{2g}(t_{2g}^1)$ \\
					& $\xi$ & $\eta$ & $\zeta$ & & \\
					& (\SI{0.999}{}) & (\SI{0.999}{}) & (\SI{0.999}{}) & & \\[0.3cm]
					$\prescript{2}{}{T}_{2g}[\mathrm{A}_{\frac{1}{2}}\oplus\mathrm{A}_{\frac{1}{2}}']$ & \includegraphics{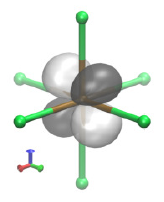} & \includegraphics{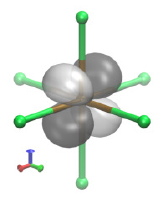} & \includegraphics{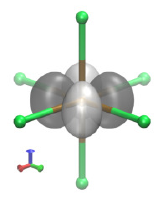} & 0.7512 & $\prescript{2}{}{T}_{2g}(t_{2g}^1)$ \\
					& $\xi$ & $\eta$ & $\zeta$ & & \\
					& (\SI{1.000}{}) & (\SI{1.000}{}) & (\SI{1.000}{}) & & \\[0.3cm]
					$\prescript{2}{}{E}_g[\mathrm{B}_{\frac{1}{2}}]$ & \includegraphics{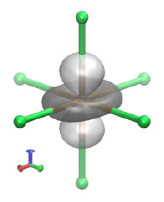} & \includegraphics{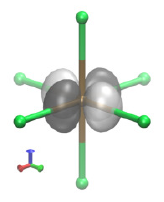} &  & 0.7517 & $\prescript{2}{}{E}_{g}(e_{g}^1)$ \\
					& $u$ & $v$ &  & \\
					& (\SI{1.000}{}) & (\SI{1.000}{}) \\
					\bottomrule
				\end{tabular}
			\end{table*}

		\paragraph{Symmetry restoration of individual symmetry-broken UHF sets.}
		
			Even though the $\mathrm{A}_{\frac{1}{2}}$ and $\mathrm{B}_{\frac{1}{2}}$ solutions include additional valence--core and valence--ligand electron correlation by breaking symmetry, their symmetry-broken nature implies that they cannot be taken as good approximations of the $T_{2g}$ and $E_g$ wavefunctions in \ce{[TiF6]^{3-}} owing to Theorem~\ref{theorem:grouptheorytoqm}.
			This has long been known as the ``symmetry dilemma'' first pointed out by L\"{o}wdin\cite{article:Lykos1963}.
			NOCI can, however, be used to form symmetry-conserved wavefunctions from multiple symmetry-broken SCF solutions.
			Figure~\ref{fig:d1_noci} shows the energy and spatial symmetry of low-lying NOCI wavefunctions constructed from the UHF solutions in octahedral \ce{[TiF6]^3-} discussed so far and Table~\ref{tab:d1_noci} gives the forms of the corresponding $3d$ natural orbitals (see Appendix~\ref{app:nocinatorbs}), also localized using the Pipek--Mezey algorithm.
			Detailed NOCI results can be found in the Supporting Information.
			Arising from the six symmetry-broken $\mathrm{A}_{\frac{1}{2}}$ solutions are a set of triply degenerate multi-determinantal wavefunctions that transform as $T_{2g}$ with the expected $3d_{xy}/3d_{yz}/3d_{xz}$ natural orbitals and another set that transform as $T_{1g}$.
			These two sets are denoted $\prescript{2}{}{T}_{2g}[\mathrm{A}_{\frac{1}{2}}]$ and $\prescript{\varnothing}{}{T}_{1g}[\mathrm{A}_{\frac{1}{2}}]$, respectively.
			The spatial irreducible representation designation in these notations is obvious, but we will come back to the designation of spin multiplicity at the end of this section.
			The $\prescript{2}{}{T}_{2g}[\mathrm{A}_{\frac{1}{2}}]$ set is \SI{1.96e-3}{\hartree} lower in energy than the $\mathrm{A}_{\frac{1}{2}}$ solutions whereas the $\prescript{\varnothing}{}{T}_{1g}[\mathrm{A}_{\frac{1}{2}}]$ set is more than \SI{2}{\hartree} higher and is thus discarded on the basis that it is too high in energy to be of any physical interest.
			Likewise, from the symmetry-broken $\mathrm{B}_{\frac{1}{2}}$ solutions, we obtain a $\prescript{2}{}{E}_g[\mathrm{B}_{\frac{1}{2}}]$ set \SI{1.56e-3}{\hartree} lower in energy with $3d_{x^2-y^2}$ and $3d_{z^2}$ natural orbitals and an unphysical $\prescript{\varnothing}{}{A}_{2g}[\mathrm{B}_{\frac{1}{2}}]$ set approximately \SI{2}{\hartree} higher in energy.
			These results are very encouraging as they illustrate that NOCI has the capability to purify the symmetry component that a set of symmetry-broken SCF wavefunctions are trying to approximate while definitively picking out symmetry components that have been mixed in only as a consequence of symmetry breaking to incorporate electron correlation.

		\paragraph{NOCI between multiple degenerate UHF sets.}
		
			We can, however, take a step further.
			Observing that the $\mathrm{A}_{\frac{1}{2}}$ and $\mathrm{A}_{\frac{1}{2}}'$ sets both contain a spatial $T_{2g}$ component, we expect them to interact with each other via the common $T_{2g}$ space. 
			Indeed, when taking both the $\mathrm{A}_{\frac{1}{2}}$ and $\mathrm{A}_{\frac{1}{2}}'$ solutions as bases for NOCI, we obtain a set denoted by $\prescript{2}{}{T}_{2g}[\mathrm{A}_{\frac{1}{2}} \oplus \mathrm{A}_{\frac{1}{2}}']$ that transforms as $T_{2g}$ and two other high-energy sets denoted by $\prescript{\varnothing}{}{T}_{2g}[\mathrm{A}_{\frac{1}{2}}\oplus \mathrm{A}_{\frac{1}{2}}']$ and  $\prescript{\varnothing}{}{T}_{1g}[\mathrm{A}_{\frac{1}{2}}\oplus \mathrm{A}_{\frac{1}{2}}']$.
			The $\prescript{2}{}{T}_{2g}[\mathrm{A}_{\frac{1}{2}} \oplus \mathrm{A}_{\frac{1}{2}}']$ set has similar $3d$ natural orbitals to $\prescript{2}{}{T}_{2g}[\mathrm{A}_{\frac{1}{2}}]$ but is \SI{1.4e-4}{\hartree} lower in energy and must therefore be a better approximation of the true $\prescript{2}{}{T}_{2g}$ ground-state wavefunctions by the variational principle.

		\paragraph{Spin multiplicity assignments.}
				
			We conclude this section with a few comments on the spin properties of the UHF and NOCI wavefunctions of \ce{[TiF6]^{3-}}.
			All UHF solutions discussed so far have $N_\alpha - N_\beta = 1$ and are thus eigenfunctions of $\hat{S_z}$ with eigenvalue $M_S = \frac{1}{2}$, as expected from Fukutome's classification for UHF\cite{article:Fukutome1981}.
			The NOCI wavefunctions obtained as linear combinations of these UHF solutions must also be eigenfunctions of $\hat{S_z}$ with the same eigenvalue $M_S = \frac{1}{2}$.
			However, the $\langle\hat{S}^2\rangle$ values in Tables~\ref{tab:d1_nonoci}~and~\ref{tab:d1_noci} show that none of these wavefunctions is a strict eigenfunction of $\hat{S}^2$ because a true doublet must have $\langle\hat{S}^2\rangle$ equal to $\frac{3}{4}$ exactly, and the observed deviations on the order of \SI{1e-3}{} must be an indication of spin contamination from the $M_S = \frac{1}{2}$ components of higher spin states.
			Slight but consistent reduction of $\langle\hat{S}^2\rangle$ in the NOCI wavefunctions compared to the basis UHF solutions shows that, even by restoring spatial symmetry, NOCI can improve spin purity as well, although not by much.
			Nevertheless, we can still regard the NOCI wavefunctions $\prescript{2}{}{T}_{2g}[\mathrm{A}_{\frac{1}{2}}]$, $\prescript{2}{}{T}_{2g}[\mathrm{A}_{\frac{1}{2}} \oplus \mathrm{A}_{\frac{1}{2}}']$, and $\prescript{2}{}{E}_g[\mathrm{B}_{\frac{1}{2}}]$ as reasonable approximations of true spin doublets and assign them a spin multiplicity of two as denoted by their pre-superscripts.
			On the other hand, the high-lying NOCI sets that we discarded, $\prescript{\varnothing}{}{T}_{1g}[\mathrm{A}_{\frac{1}{2}}]$, $\prescript{\varnothing}{}{T}_{2g}[\mathrm{A}_{\frac{1}{2}}\oplus \mathrm{A}_{\frac{1}{2}}']$, $\prescript{\varnothing}{}{T}_{1g}[\mathrm{A}_{\frac{1}{2}}\oplus \mathrm{A}_{\frac{1}{2}}']$, and $\prescript{\varnothing}{}{A}_{2g}[\mathrm{B}_{\frac{1}{2}}]$, have $\langle\hat{S}^2\rangle$ values that do not correspond to any integral spin multiplicities (more than \SI{1e-2}{} away from any expected $\hat{S}^2$ eigenvalues, see Supporting Information).
			The symbol ``$\varnothing$'' is therefore used to denote these undefined spin multiplicities.

%% file: results-d2/results-d2-model.tex
\tikzsetexternalprefix{./results-d2/tikz/}

\section{\textit{d}\textsuperscript{2} Metal Ground Configuration}
\label{sec:results-d2}

	\subsection{True \textit{d}\textsuperscript{2} Octahedral System}
	\label{subsec:results-d2-model}

		The main aim of this Section is to discuss the low-energy UHF and NOCI wavefunctions of the octahedral \ce{[VF6]^{3-}} anion.
		However, the interaction between the two $d$ electrons in the presence of the core and ligand electrons complicates the results significantly, so we will first study a toy system where all core and ligand electrons are stripped away.
		We will then see that the results in this system help clarify the correlation nature of the UHF and NOCI wavefunctions in \ce{[VF6]^{3-}}.
		
			\subsubsection{Symmetry-Conserved Wavefunctions in the Strong-Field Coupling Scheme}
			
					%% All d2 terms	
					\begin{table*}
						\centering
						\caption{
							All possible $d^2$ terms in an octahedral field.
							$|\chi_1 \chi_2|$ denotes a normalized Slater determinant constructed from spin-orbitals $\chi_1$ and $\chi_2$.
							$e_1$ is the only component of $A_{1g}$;
							$e_2$ the only component of $A_{2g}$;
							$\alpha$, $\beta$, and $\gamma$ the components of $T_{1g}$;
							$\xi$, $\eta$, and $\zeta$ the components of $T_{2g}$;
							and $u$ and $v$ the components of $E_g$, with $u_x = -\frac{1}{2}u+\frac{\sqrt{3}}{2}v$, $u_y = -\frac{1}{2}u-\frac{\sqrt{3}}{2}v$, $u_z = u$, $v_x = -\frac{\sqrt{3}}{2}u-\frac{1}{2}v$, $v_y = \frac{\sqrt{3}}{2}u-\frac{1}{2}v$, and $v_z = v$.
							Each spin-orbital is also an eigenfunction of the $\hat{s}_3$ operator with eigenvalue $m_s = \frac{1}{2}$ (without bar) or $-\frac{1}{2}$ (with bar).
							Two terms of the same symmetry $\Gamma$ but belonging to two different configurations can interact further to give rise to two new terms $a\Gamma$ and $b\Gamma$ (not shown here) where $E(a\Gamma) < E(b\Gamma)$\cite{book:Sugano1970}.
						}
						\label{tab:d2terms}
						\begingroup
						\renewcommand\arraystretch{1.25}
						\begin{tabular}[t]{p{1.5cm} p{1.3cm} p{1.5cm} p{2.3cm} p{4.5cm} p{2.3cm}}
							\toprule
							Config. & Term & Comp. & $M_S = 1$ & $M_S = 0$ & $M_S = -1$ \\
							\midrule
							\multirow[t]{9}{*}{$t_{2g}^2$}%
							&	\multirow[t]{3}{*}{$\prescript{3}{}{T}_{1g}$}%
								&	$\alpha$ & $|\eta\zeta|$ & $\frac{1}{\sqrt{2}}\left(|\eta\bar{\zeta}| - |\zeta\bar{\eta}|\right)$ & $|\bar{\eta}\bar{\zeta}|$ \\
							&	&	$\beta$ & $|\zeta\xi|$ & $\frac{1}{\sqrt{2}}\left(|\zeta\bar{\xi}| - |\xi\bar{\zeta}|\right)$ & $|\bar{\zeta}\bar{\xi}|$ \\
							&	&	$\gamma$ & $|\xi\eta|$ & $\frac{1}{\sqrt{2}} \left(|\xi\bar{\eta}| - |\eta\bar{\xi}|\right)$ & $|\bar{\xi}\bar{\eta}|$ \\
							&	\multirow[t]{3}{*}{$\prescript{1}{}{T}_{2g}$}%
								&	$\xi$ & -- & $\frac{1}{\sqrt{2}}\left(|\eta\bar{\zeta}| + |\zeta\bar{\eta}|\right)$ & -- \\
							&	&	$\eta$ & -- & $\frac{1}{\sqrt{2}}\left(|\zeta\bar{\xi}| + |\xi\bar{\zeta}|\right)$ & -- \\
							&	&	$\zeta$ & -- & $\frac{1}{\sqrt{2}} \left(|\xi\bar{\eta}| + |\eta\bar{\xi}|\right)$ & -- \\
							&	\multirow[t]{2}{*}{$\prescript{1}{}{E}_g$}%
								&	$u$ & -- & $\frac{1}{\sqrt{6}} \left( -|\xi\bar{\xi}| - |\eta\bar{\eta}| + 2|\zeta\bar{\zeta}| \right)$ & -- \\
							&	&	$v$ & -- & $\frac{1}{\sqrt{2}}\left(|\xi\bar{\xi}| - |\eta\bar{\eta}|\right)$ & -- \\
							& $\prescript{1}{}{A}_{1g}$ & $e_1$ & -- & $\frac{1}{\sqrt{3}} \left( |\xi\bar{\xi}| + |\eta\bar{\eta}| + |\zeta\bar{\zeta}| \right)$ & -- \\
							\midrule
							\multirow[t]{12}{*}{$t_{2g}^1e_g^1$}%
							&	\multirow[t]{3}{*}{$\prescript{3}{}{T}_{1g}$}%
								&	$\alpha$ & $|\xi v_x|$ & $\frac{1}{\sqrt{2}}\left(|\xi\bar{v}_x| - |v_x\bar{\xi}|\right)$ & $|\bar{\xi}\bar{v}_x|$ \\
							&	&	$\beta$ & $|\eta v_y|$ & $\frac{1}{\sqrt{2}}\left(|\eta\bar{v}_y| - |v_y\bar{\eta}|\right)$ & $|\bar{\eta}\bar{v}_y|$ \\
							&	&	$\gamma$ & $|\zeta v_z|$ & $\frac{1}{\sqrt{2}}\left(|\zeta\bar{v}_z| - |v_z\bar{\zeta}|\right)$ & $|\bar{\zeta}\bar{v}_z|$ \\
							&	\multirow[t]{3}{*}{$\prescript{3}{}{T}_{2g}$}%
								&	$\xi$ & $|\xi u_x|$ & $\frac{1}{\sqrt{2}}\left(|\xi\bar{u}_x| - |u_x\bar{\xi}|\right)$ & $|\bar{\xi}\bar{u}_x|$ \\
							&	&	$\eta$ & $|\eta u_y|$ & $\frac{1}{\sqrt{2}}\left(|\eta\bar{u}_y| - |u_y\bar{\eta}|\right)$ & $|\bar{\eta}\bar{u}_y|$ \\
							&	&	$\zeta$ & $|\zeta u_z|$ & $\frac{1}{\sqrt{2}}\left(|\zeta\bar{u}_z| - |u_z\bar{\zeta}|\right)$ & $|\bar{\zeta}\bar{u}_z|$ \\
							&	\multirow[t]{3}{*}{$\prescript{1}{}{T}_{1g}$}%
								&	$\alpha$ & -- & $\frac{1}{\sqrt{2}}\left(|\xi\bar{v}_x| + |v_x\bar{\xi}|\right)$ & -- \\
							&	&	$\beta$ & -- & $\frac{1}{\sqrt{2}}\left(|\eta\bar{v}_y| + |v_y\bar{\eta}|\right)$ & -- \\
							&	&	$\gamma$ & -- & $\frac{1}{\sqrt{2}}\left(|\zeta\bar{v}_z| + |v_z\bar{\zeta}|\right)$ & -- \\
							&	\multirow[t]{3}{*}{$\prescript{1}{}{T}_{2g}$}%
								&	$\xi$ & -- & $\frac{1}{\sqrt{2}}\left(|\xi\bar{u}_x| + |u_x\bar{\xi}|\right)$ & -- \\
								&	&	$\eta$ & -- & $\frac{1}{\sqrt{2}}\left(|\eta\bar{u}_y| + |u_y\bar{\eta}|\right)$ & -- \\
								&	&	$\zeta$ & -- & $\frac{1}{\sqrt{2}}\left(|\zeta\bar{u}_z| + |u_z\bar{\zeta}|\right)$ & -- \\
							\midrule
							\multirow[t]{6}{*}{$e_g^2$}%
							&	\multirow[t]{1}{*}{$\prescript{3}{}{A}_{2g}$}%
								&	$e_2$ & $|uv|$ & $\frac{1}{\sqrt{2}}\left(|u\bar{v}| - |v\bar{u}|\right)$ & $|\bar{u}\bar{v}|$ \\
							&	\multirow[t]{2}{*}{$\prescript{1}{}{E}_{g}$}%
								&	$u$ & -- & $\frac{1}{\sqrt{2}}\left(-|u\bar{u}| + |v\bar{v}|\right)$ & -- \\
							&	&	$v$ & -- & $\frac{1}{\sqrt{2}}\left(|u\bar{v}| + |v\bar{u}|\right)$ & -- \\
							&	\multirow[t]{1}{*}{$\prescript{1}{}{A}_{1g}$}%
								&	$e_1$ & -- & $\frac{1}{\sqrt{2}}\left(|u\bar{u}| + |v\bar{v}|\right)$ & -- \\
							\bottomrule
						\end{tabular}
						\endgroup
					\end{table*}

				For a true $d^2$ system in an octahedral field of point charges, the strong-field coupling scheme\cite{book:Griffith1961} describes each electron by an independent set of $m_s = \pm \frac{1}{2}$ spin-orbitals with spatial symmetry of either $t_{2g}$ or $e_g$ and then couples them together using Clebsch--Gordan coefficients for the spatial part and Wigner coefficients for the spin part\cite{book:Sugano1970} such that the resulted wavefunctions satisfy Pauli's antisymmetry, transform as single irreducible representations in $\mathcal{O}_h$, and are eigenfunctions of $\hat{S}^2$.
				There are thus three possible configurations split into 11 allowed terms whose components are summarized in Table~\ref{tab:d2terms} from \citeauthor{book:Sugano1970}\cite{book:Sugano1970}
				Once again, only if each electron is described by a single set of five degenerate hydrogenic $nd$ orbitals are the $t_{2g}$ and $e_g$ components given by (\ref{eq:1e5dforms}).
				In addition, unlike the $d^1$ case, there is now a clear distinction between the lower-case irreducible representation symbols denoting the spatial symmetry of a single electron in the strong-field coupling scheme and the upper-case irreducible representation symbols describing the overall spatial symmetry of the wavefunction.
				Inspection of Table~\ref{tab:d2terms} reveals that each allowed $M_S = \pm1$ wavefunction only needs a single determinant to satisfy spin and spatial symmetry, whereas every $M_S = 0$ wavefunction requires at least two determinants to fulfill the same conditions and thus exhibits strong correlation between the $d$ electrons that is due solely to spin and spatial symmetry demands.
				
				To further complicate matters, there are pairs of terms arising from different strong-field configurations which have identical spin and spatial symmetry and can therefore interact with each other in what is known as \textit{configuration mixing}\cite{book:Sugano1970}.
				One example would be the interaction between $\prescript{3}{}{T}_{1g}(t_{2g}^2)$ and $\prescript{3}{}{T}_{1g}(t_{2g}^1 e_g^1)$ to give rise to two new terms labeled $a\prescript{3}{}{T}_{1g}$ and $b\prescript{3}{}{T}_{1g}$ with the former having a lower energy than the latter (also defined in the caption of Table~\ref{tab:d2terms}).
				When this happens, the number of determinants required to describe the mixed-configuration wavefunctions must increase, thereby signifying even more correlation effects.
				As a consequence, the true $\prescript{3}{}{T}_{1g}$ ground term cannot be described by any single determinants, even within the $M_S = \pm1$ spaces.
				This behavior turns out to have important implications for the nature of the UHF solutions in both the $d^2$ octahedral toy system and the full \ce{[VF6]^{3-}} anion.
				
			\subsubsection{UHF and NOCI Wavefunctions in a \textit{d}\textsuperscript{2} Octahedral Toy System}
			\label{subsubsec:results-d2-model-UHFNOCI}
			
				The strong-field coupling scheme enables wavefunctions that respect all symmetry requirements to be constructed from spin-orbitals.
				However, this scheme quickly becomes tedious when more interacting electrons are introduced (see, for example, Chapters 3 and 4 of Ref.~\citenum{book:Sugano1970}).
				Furthermore, as the scheme relies on symmetry alone, it cannot provide any functional descriptions of the constituting spin-orbitals \textit{ab initio}.
				Fortunately, the HF method combined with NOCI for symmetry restoration gives us a plausible way around this.
				To investigate how this works for a true $d^2$ octahedral system, we carried out SCF metadynamics calculations on a toy complex consisting of a \ce{B^{3+}} center at the origin surrounded by six \SI{-1}{} point charges placed along the Cartesian axes at a distance of \SI{1.9896}{\angstrom} from the origin.
				We denote this system as \ce{[(B^{3+})(q^-)6]}.
				There are only two electrons in this system, each of which is described by a basis set composed only of five pure $d$ functions contracted from the three primitive Gaussian functions in the first D shell of vanadium's 6-31G*.%
				\bibnote{The D shell in vanadium's 6-31G* is commonly constructed with six Cartesian functions of degree $2$ for the angular part. There is, however, an $s$ component contained in this (see Appendix~\ref{app:symtrans}) which is of no interest to the model system where the two electrons are constrained to occupy only the five $d$ orbitals. We thus use the pure (and real) $d$ form instead for the angular part of the basis functions in the D shell to eliminate the unnecessary totally symmetric $s$ component.}
				The boron nucleus is chosen so that the two electrons experience a nuclear charge of \SI[retain-explicit-plus=true]{+5}{}, which is also the effective nuclear charge experienced by the $3d^2$ electrons in \ce{[VF6]^{3-}} if we assume that the \ce{[Ar]} core shields the vanadium nucleus perfectly and that the \ce{F-} ligands do not shield the vanadium nucleus at all.
				As the sole purpose of the toy system is to help us understand the symmetry of the UHF solutions, we are not too concerned with the accuracy of the above assumptions, nor with the fact that all solutions found have positive energy (Table~\ref{tab:2e5d_d2}).

					%% 2e5d toy system energies
					\begin{table*}
						\centering
						\caption{
							UHF and NOCI wavefunctions in the toy system \ce{[(B^{3+})(q^-)6]}.
							All UHF wavefunctions have DIIS errors smaller than \SI{1e-13}{}.
							The symmetry of each set of symmebry-broken solutions lists terms in increasing order of NOCI energy.
							Each term symbol is immediately followed by the predominant associated strong-field configuration in parentheses.
							If configuration mixing is found to be present following an analysis of NOCI natural orbitals and occupation numbers, a second pair of parentheses denotes the minor configuration introduced to the term by this interaction.
							Repeated terms due to configuration mixing are distinguished based on their energy ordering using the prefixes $a$ and $b$ where the $a$ term is lower in energy than the $b$ term.
							If the energy ordering is not known or cannot be ascertained, a hollow diamond ($\diamond$) is used instead.
						}
						\label{tab:2e5d_d2}
						\footnotesize
						%%
						%% M_S = 1
						\begin{subtable}[t]{\textwidth}
							\centering
							\caption{$M_S = 1$}
							\label{subtab:2e5d_d2_MS1}
							\begingroup
							\renewcommand\arraystretch{1.35}
							\scalebox{0.7}{%
								\begin{tabular}[t]{%
										>{\raggedright\arraybackslash}m{0.9cm} %
										>{\raggedright\arraybackslash}m{2.3cm} %
										l %
										>{\raggedright\arraybackslash}m{1.8cm} | %
										>{\raggedright\arraybackslash}m{2.5cm} %
										S[table-format=1.7, table-alignment=left] %
										>{\raggedright\arraybackslash}m{2.7cm}%
									}
									\toprule
									$\Psi_\mathrm{UHF}$ & Symmetry & Energy/\si{\hartree} & Term & $\Phi(S=1)$ & {Energy/\si{\hartree}} & Term \\
									\midrule
									%% a
									$\mathrm{a}_1$ & $\prescript{3}{}{T}_{1g} \oplus \prescript{3}{}{A}_{2g}$ & \tablenum[table-format=1.7]{0.9488982} & -- %
									   & $\prescript{3}{}{T}_{1g}[\mathrm{a}_1]$  & 0.9488346 & $a\prescript{3}{}{T}_{1g}(t_{2g}^2)(t_{2g}^1 e_g^1)$ \\
									&&&& $\prescript{3}{}{A}_{2g}[\mathrm{a}_1]$  & 0.9547970 & $\prescript{3}{}{A}_{2g}(e_g^2)$ \\[7pt]
									%%
									%% a'
									$\mathrm{a}'_1$ & $\prescript{3}{}{T}_{1g} \oplus \prescript{3}{}{T}_{2g}$ & \tablenum[table-format=1.7]{0.9489908} & -- %
										 & $\prescript{3}{}{T}_{1g}[\mathrm{a}'_1]$ & 0.9488348 & $a\prescript{3}{}{T}_{1g}(t_{2g}^2)(t_{2g}^1 e_g^1)$ \\
									&&&& $\prescript{3}{}{T}_{2g}[\mathrm{a}'_1]$ & 0.9514938 & $\prescript{3}{}{T}_{2g}(t_{2g}^1e_g^1)$ \\[7pt]
									%%
									%% b
									$\mathrm{b}_1$ & $\prescript{3}{}{T}_{1g}$ & \tablenum[table-format=1.7]{0.9677637} & $\prescript{3}{}{T}_{1g}(t_{2g}^2)$ \\[7pt]
									%%
									%% c
									$\mathrm{c}_1$ & $\prescript{3}{}{T}_{2g}$ & \tablenum[table-format=1.7]{0.9514938} & $\prescript{3}{}{T}_{2g}(t_{2g}^1e_g^1)$ \\[7pt]
									%%
									%% d
									$\mathrm{d}_1$ & $\prescript{3}{}{T}_{1g} \oplus \prescript{3}{}{T}_{2g}$ & \tablenum[table-format=1.7]{0.9710669} & -- %
									   & $\prescript{3}{}{T}_{2g}[\mathrm{d}_1]$  & 0.9514938 & $\prescript{3}{}{T}_{2g}(t_{2g}^1e_g^1)$ \\
									&&&& $\prescript{3}{}{T}_{1g}[\mathrm{d}_1]$  & 1.0297863 & $\prescript{3}{}{T}_{1g}(t_{2g}^1e_g^1)$ \\[7pt]
									%%
									%% e
									$\mathrm{e}_1$ & $\prescript{3}{}{T}_{1g}$ & \tablenum[table-format=1.7]{1.0297863} & $\prescript{3}{}{T}_{1g}(t_{2g}^1e_g^1)$ \\[7pt]
									%%
									%% f
									$\mathrm{f}_1$ & $\prescript{3}{}{A}_{2g}$ & \tablenum[table-format=1.7]{0.9547970} & $\prescript{3}{}{A}_{2g}(e_g^2)$ \\
									\midrule
									%%
									%% b+e
									&&&& $a\prescript{3}{}{T}_{1g}[\mathrm{b}_1 \oplus \mathrm{e}_1]$  & 0.9488337 & $a\prescript{3}{}{T}_{1g}(t_{2g}^2)(t_{2g}^1 e_g^1)$  \\
									&&&& $b\prescript{3}{}{T}_{1g}[\mathrm{b}_1 \oplus \mathrm{e}_1]$  & 1.0487163 & $b\prescript{3}{}{T}_{1g}(t_{2g}^1 e_g^1)(t_{2g}^2)$ \\
									\bottomrule
								\end{tabular}
							}
							\endgroup
						\end{subtable}
						
						\vspace{0.5cm}
						%%
						%% M_S = 0
						\begin{subtable}[t]{\textwidth}
							\centering
							\captionsetup{justification=centering}
							\caption{$M_S = 0$}
							\label{subtab:2e5d_d2_MS0}
							\begingroup
							\renewcommand\arraystretch{1.35}
							\scalebox{0.7}{%
								\begin{tabular}[t]{
										>{\raggedright\arraybackslash}m{0.9cm} %
										>{\raggedright\arraybackslash}m{2.3cm} %
										l | %
										>{\raggedright\arraybackslash}m{2.5cm} %
										S[table-format=1.7, table-alignment=left] %
										>{\raggedright\arraybackslash}m{2.7cm} | %
										>{\raggedright\arraybackslash}m{2.5cm} %
										S[table-format=1.7, table-alignment=left] %
										>{\raggedright\arraybackslash}m{2.7cm}%
									}
									% Long symmetries
									\multicolumn{9}{c}{$\Gamma_{\mathrm{a}^*_0} = \prescript{3}{}{T}_{1g} \oplus \prescript{3}{}{A}_{2g} \oplus \prescript{1}{}{T}_{2g} \oplus \prescript{1}{}{E}_{g} \oplus \prescript{1}{}{T}_{1g}; \qquad \Gamma_{\mathrm{a}'_0} = \prescript{3}{}{T}_{1g} \oplus \prescript{3}{}{T}_{2g} \oplus \prescript{1}{}{T}_{2g} \oplus \prescript{1}{}{T}_{1g}; \qquad \Gamma_{\mathrm{b}_0} = \prescript{3}{}{T}_{2g} \oplus \prescript{1}{}{E}_{g} \oplus \prescript{1}{}{A}_{1g}.$} \\[7pt]
									\toprule
									$\Psi_\mathrm{UHF}$ & Symmetry & {Energy/\si{\hartree}} & $\Phi(S=1)$ & {Energy/\si{\hartree}} & Term & $\Phi(S=0)$ & {Energy/\si{\hartree}} & Term \\
									\midrule
									%% a
									$\mathrm{a}^*_0$   & $\Gamma_{\mathrm{a}^*_0}$ & \tablenum[table-format=1.7]{1.0004202} %
										& $\prescript{3}{}{T}_{1g}[\mathrm{a}^*_0]$  & 0.9488449 & $a\prescript{3}{}{T}_{1g}(t_{2g}^2)(t_{2g}^1 e_g^1)$
										& $\prescript{1}{}{T}_{2g}[\mathrm{a}^*_0]$  & 1.0415273 & $b\prescript{1}{}{T}_{2g}(t_{2g}^2)(t_{2g}^1 e_g^1)$ \\
									&&& $\prescript{3}{}{A}_{2g}[\mathrm{a}^*_0]$  & 0.9547970 & $\prescript{3}{}{A}_{2g}(e_g^2)$
									& $\prescript{1}{}{E}_{g}[\mathrm{a}^*_0]$  & 1.0661483 & $\diamond\prescript{1}{}{E}_{g}(t_{2g}^2)(e_g^2)$ \\
									&&&&&& $\prescript{1}{}{T}_{1g}[\mathrm{a}^*_0]$  & 1.0806603 & $\prescript{1}{}{T}_{1g}(t_{2g}^1 e_g^1)$ \\[7pt]
									%%
									%% a'
									$\mathrm{a}'_0$  & $\Gamma_{\mathrm{a}'_0}$ & \tablenum[table-format=1.7]{1.0005129} %
										& $\prescript{3}{}{T}_{1g}[\mathrm{a}'_0]$  & 0.9488462 & $a\prescript{3}{}{T}_{1g}(t_{2g}^2)(t_{2g}^1 e_g^1)$
										& $\prescript{1}{}{T}_{2g}[\mathrm{a}'_0]$  & 1.0461426 & $b\prescript{1}{}{T}_{2g}(t_{2g}^2)(t_{2g}^1 e_g^1)$ \\
										&&& $\prescript{3}{}{T}_{2g}[\mathrm{a}'_0]$  & 0.9514938 & $\prescript{3}{}{T}_{2g}(t_{2g}^1 e_g^1)$
										& $\prescript{1}{}{T}_{1g}[\mathrm{a}'_0]$  & 1.0806604 & $\prescript{1}{}{T}_{1g}(t_{2g}^1 e_g^1)$ \\[7pt]
									%%
									%% a''
									$\mathrm{a}''_0$ & $\prescript{3}{}{T}_{1g} \oplus \prescript{1}{}{T}_{2g}$ & \tablenum[table-format=1.7]{1.0127738} & $\prescript{3}{}{T}_{1g}[\mathrm{a}''_0]$  & 0.9677637 & $\prescript{3}{}{T}_{1g}(t_{2g}^2)$ & $\prescript{1}{}{T}_{2g}[\mathrm{a}''_0]$  & 1.0577840 & $\prescript{1}{}{T}_{2g}(t_{2g}^2)$ \\[7pt]
									%%
									%% b
									$\mathrm{b}_0$   & $\Gamma_{\mathrm{b}_0}$ & \tablenum[table-format=1.7]{1.0030018} %
									& $\prescript{3}{}{T}_{2g}[\mathrm{b}_0]$  & 0.9514938 & $\prescript{3}{}{T}_{2g}(t_{2g}^1 e_g^1)$
									& $\prescript{1}{}{E}_{g}[\mathrm{b}_0]$  & 1.0355619 & $\diamond\prescript{1}{}{E}_{g}(t_{2g}^2)(e_{g}^2)$ \\
									&&&&&& $\prescript{1}{}{A}_{1g}[\mathrm{b}_0]$  & 1.0812665 & $\diamond\prescript{1}{}{A}_{1g}(e_{g}^2)(t_{2g}^2)$ \\[7pt]
									%%
									%% c
									$\mathrm{c}_0$ & $\prescript{3}{}{T}_{2g} \oplus \prescript{1}{}{T}_{2g}$ & \tablenum[table-format=1.7]{1.0030283} & $\prescript{3}{}{T}_{2g}[\mathrm{c}_0]$  & 0.9514938 & $\prescript{3}{}{T}_{2g}(t_{2g}^1e_g^1)$ & $\prescript{1}{}{T}_{2g}[\mathrm{c}_0]$  & 1.0545628 & $\prescript{1}{}{T}_{2g}(t_{2g}^1e_g^1)$\\[7pt]
									%%
									%% d
									$\mathrm{d}_0$ & $\prescript{1}{}{E}_{g} \oplus \prescript{1}{}{A}_{1g}$ & \tablenum[table-format=1.7]{1.1027941} %
									  &&&& $\prescript{1}{}{E}_{g}[\mathrm{d}_0]$  & 1.0577840 & $\prescript{1}{}{E}_{g}(t_{2g}^2)$ \\
									&&&&&& $\prescript{1}{}{A}_{1g}[\mathrm{d}_0]$  & 1.1928144 & $\prescript{1}{}{A}_{1g}(t_{2g}^2)$ \\[7pt]
									%%
									%% e
									$\mathrm{e}_0$ & $\prescript{3}{}{T}_{1g} \oplus \prescript{1}{}{T}_{1g}$ & \tablenum[table-format=1.7]{1.0552233} & $\prescript{3}{}{T}_{1g}[\mathrm{e}_0]$  & 1.0297863 & $\prescript{3}{}{T}_{1g}(t_{2g}^1e_g^1)$ & $\prescript{1}{}{T}_{1g}[\mathrm{e}_0]$  & 1.0806603 & $\prescript{1}{}{T}_{1g}(t_{2g}^1e_g^1)$ \\
									\midrule
									%%
									%% a''+e, triplet
									&&& $a\prescript{3}{}{T}_{1g}[\mathrm{a}''_0 \oplus \mathrm{e}_0]$  & 0.9488337 & $a\prescript{3}{}{T}_{1g}(t_{2g}^2)(t_{2g}^1 e_g^1)$ & %
									%% a''+c, singlet
									$a\prescript{1}{}{T}_{2g}[\mathrm{a}''_0 \oplus \mathrm{c}_0]$  & 1.0335150 & $a\prescript{1}{}{T}_{2g}(t_{2g}^1 e_g^1)(t_{2g}^2)$ \\
									%% a''+e, triplet
									&&& $b\prescript{3}{}{T}_{1g}[\mathrm{a}''_0 \oplus \mathrm{e}_0]$  & 1.0487163 & $b\prescript{3}{}{T}_{1g}(t_{2g}^1 e_g^1)(t_{2g}^2)$ & %
									%% a''+c, singlet
									$b\prescript{1}{}{T}_{2g}[\mathrm{a}''_0 \oplus \mathrm{c}_0]$  & 1.0788318 & $b\prescript{1}{}{T}_{2g}(t_{2g}^2)(t_{2g}^1 e_g^1)$\\
									\bottomrule
								\end{tabular}
							}
							\endgroup
						\end{subtable}
					\end{table*}

				\paragraph{{\boldmath $M_S = 1$}.}
				
				Table~\ref{subtab:2e5d_d2_MS1} gives the energy and symmetry of the low-lying $M_S = 1$ UHF solutions in \ce{[(B^{3+})(q^-)6]}, all having DIIS errors smaller than \SI{1e-13}{}.
				As we will point out in Section~\ref{subsubsec:results-d2-full-MS1}, they can all be correlated to the $M_S = 1$ solutions in \ce{[VF6]^{3-}} by the forms of their spin-orbitals (shown in the Supporting Information) and overall symmetry.
				We therefore pre-emptively denote these solutions with lower-case letters that match the upper-case labels of those in \ce{[VF6]^{3-}}.
				Expectedly, the symmetry-conserved solutions $\mathrm{b}_1$, $\mathrm{c}_1$, $\mathrm{e}_1$, and $\mathrm{f}_1$ correspond exactly to the $M_S = 1$ two-electron single determinants of $\prescript{3}{}{T}_{1g}(t_{2g}^2)$, $\prescript{3}{}{T}_{2g}(t_{2g}^1e_g^1)$, $\prescript{3}{}{T}_{1g}(t_{2g}^1e_g^1)$, and $\prescript{3}{}{A}_{2g}(e_g^2)$, respectively.
				Since there are no core or ligand electrons in this toy system, the symmetry breaking observed in $\mathrm{a}_1$, $\mathrm{a}'_1$, and $\mathrm{d}_1$ must come from the correlation between the two $d$ electrons.
				To determine how much of this correlation is already accounted for by the strong-field coupling scheme in Table~\ref{tab:d2terms}, we first restored their spatial symmetry by NOCI, the results of which are also shown in Table~\ref{subtab:2e5d_d2_MS1}.
				The NOCI wavefunctions $\prescript{3}{}{A}_{2g}[\mathrm{a}_1]$, $\prescript{3}{}{T}_{2g}[\mathrm{a}'_1]$, $\prescript{3}{}{T}_{2g}[\mathrm{d}_1]$, and $\prescript{3}{}{T}_{1g}[\mathrm{d}_1]$ have the same energy as the UHF solutions $\mathrm{f}_1$, $\mathrm{c}_1$, $\mathrm{c}_1$, and $\mathrm{e}_1$, respectively, and therefore do not recover any more correlation energy than the corresponding symmetry-conserved single determinants already do.
				On the other hand, the NOCI wavefunctions $\prescript{3}{}{T}_{1g}[\mathrm{a}_1]$ and $\prescript{3}{}{T}_{1g}[\mathrm{a}'_1]$ are significantly lower in energy than $\mathrm{b}_1$ and not degenerate to any other symmetry-conserved single determinants we have found.
				Hence, they must incorporate some additional correlation missed out by the strong-field coupling scheme of independent particles occupying pure $d$ orbitals prior to configuration mixing.
				In fact, it turns out that $\prescript{3}{}{T}_{1g}[\mathrm{a}_1]$ and $\prescript{3}{}{T}_{1g}[\mathrm{a}'_1]$ are both very close in energy to $a\prescript{3}{}{T}_{1g}[\mathrm{b}_1 \oplus \mathrm{e}_1]$ which is the NOCI description of the lower $\prescript{3}{}{T}_{1g}$ term arising from the configuration mixing between $\prescript{3}{}{T}_{1g}(t_{2g}^2)$ and $\prescript{3}{}{T}_{1g}(t_{2g}^1e_g^1)$.
				We can thus reasonably deduce that the symmetry breaking in $\mathrm{a}_1$ and $\mathrm{a}'_1$ is the consequence of an attempt within the $M_S = 1$ single-determinantal space to incorporate this inherently multi-determinantal correlation effect between the two $d$ electrons.

				\paragraph{{\boldmath $M_S = 0$}.}
				
				Table~\ref{subtab:2e5d_d2_MS0} gives the energy and symmetry of the low-lying $M_S = 0$ UHF solutions in \ce{[(B^{3+})(q^-)6]}, all having DIIS errors smaller than \SI{1e-13}{}.
				These solutions are also denoted with lower-case letters matching the upper-case labels of the $M_S = 0$ solutions in \ce{[VF6]^{3-}}, their spin-orbitals are shown in the Supporting Information, and the reason for the asterisk in the $\mathrm{a}^*_0$ solutions will be explained in Section~\ref{subsubsec:results-d2-full-MS0}.
				We note that the $S = 1$ NOCI wavefunctions $\prescript{3}{}{T}_{1g}[\mathrm{a}''_0]$,
				$\prescript{3}{}{T}_{2g}[\mathrm{b}_0]$, $\prescript{3}{}{T}_{2g}[\mathrm{c}_0]$, and $\prescript{3}{}{T}_{1g}[\mathrm{e}_0]$ are exactly degenerate with the spatial-symmetry-conserved $M_S = 1$ single determinants $\mathrm{b}_1$, $\mathrm{c}_1$, $\mathrm{c}_1$, and $\mathrm{e}_1$, respectively.
				This is not surprising at all: even though the $\mathrm{a}''_0$, $\mathrm{c}_0$, and $\mathrm{e}_0$ solutions are symmetry-broken overall, within the $S = 1$ spin space they conserve spatial symmetry, so that when NOCI restores time-reversal symmetry and subsequently spin symmetry (exactly in this case as $S = 0$ and $S = 1$ are the only possible spin states for two electrons; further explanations will be detailed in Section~\ref{subsubsec:results-d2-full-MS0}), it gives the spatial-symmetry-conserved $M_S = 0$ components in the triplets whose $M_S = 1$ components are described by the $\mathrm{b}_1$, $\mathrm{c}_1$, and $\mathrm{e}_1$ solutions.
				This tells us that the symmetry breaking in $\mathrm{a}''_0$, $\mathrm{b}_0$, $\mathrm{c}_0$, and $\mathrm{e}_0$ indeed comes entirely from the correlation between the two $d$ electrons but does not introduce any more correlation within the $S = 1$ space beyond that already described by the strong-field coupling scheme in Table~\ref{tab:d2terms}.
				A similar observation can also be made for the $\prescript{3}{}{A}_{2g}[\mathrm{a}^*_0]$ and $\prescript{3}{}{T}_{2g}[\mathrm{a}'_0]$ NOCI wavefunctions, despite the fact that the $\mathrm{a}^*_0$ and $\mathrm{a}'_0$ solutions break spatial symmetry within the $S = 1$ space.
				Unfortunately, such comparisons are not possible for the $S = 0$ space because none of the singlets has any single-determinantal components, as shown in Table~\ref{tab:d2terms}.

%% file: results-d2/results-d2-full.tex
\tikzsetexternalprefix{./results-d2/tikz/}

\subsection{\ce{[VF6]^{3-}}}
\label{subsec:results-d2-full}

\begin{figure*}
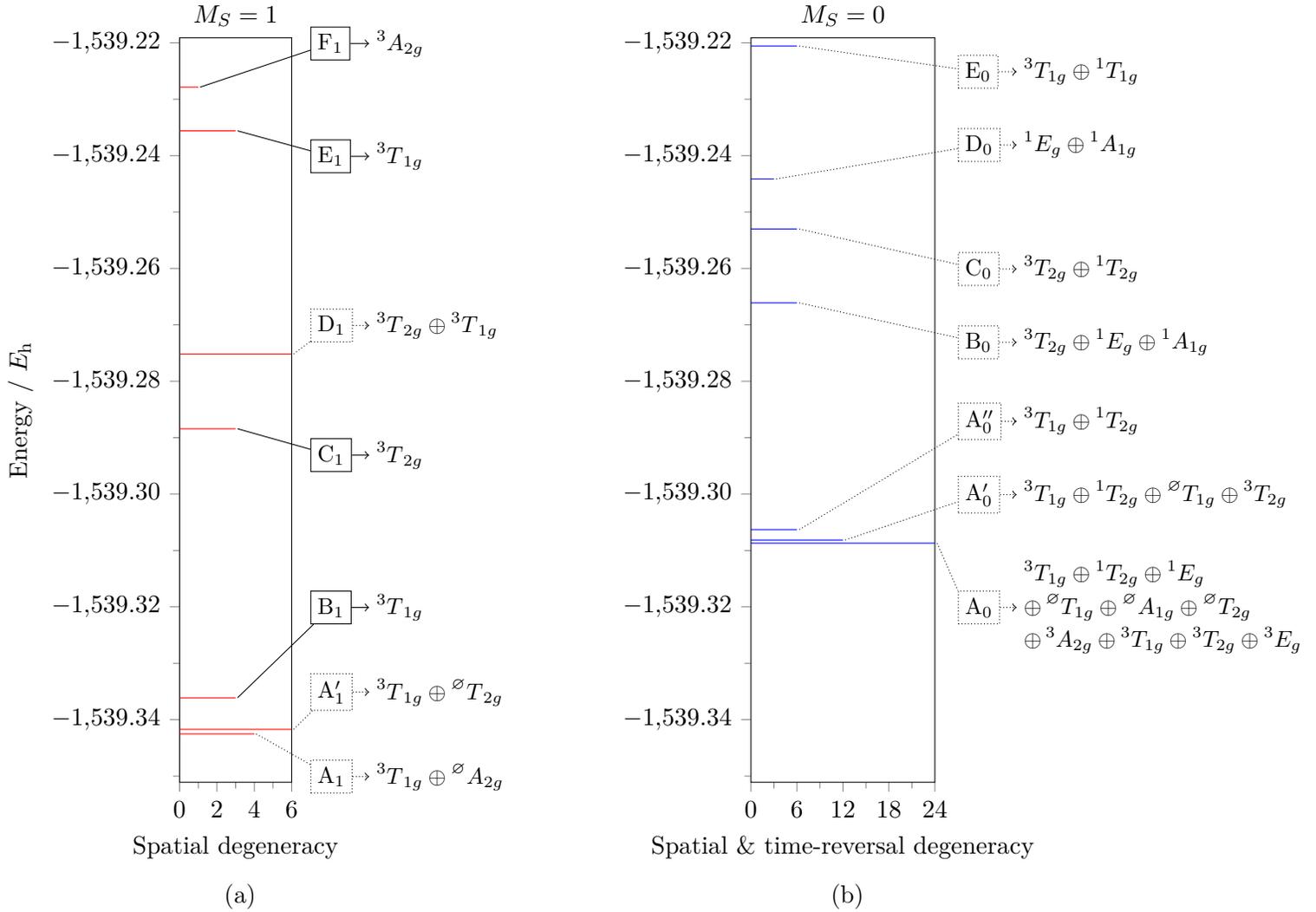

			\centering
			\begin{subfigure}[b]{0.30\textwidth}
				\centering
				% \useexternalfile{scale}{trimleft}{trimright}{name}
				% Figure compiled with tikzexternalize
				% Pre-compiled figure located at ./results-d2/tikz/tikzoutput/d2_MS1_nonoci-output.pdf
				\input{\tikzexternal@filenameprefix1.tikz}{78.73251pt}{95.42513pt}{d2_MS1_nonoci}
				\caption{}
				\label{subfig:d2_MS1_nonoci}
			\end{subfigure}
			\hfill
			\begin{subfigure}[b]{0.65\textwidth}
				\centering
				% \useexternalfile{scale}{trimleft}{trimright}{name}
				% Figure compiled with tikzexternalize
				% Pre-compiled figure located at ./results-d2/tikz/tikzoutput/d2_MS0_nonoci-output.pdf
				\input{\tikzexternal@filenameprefix1.tikz}{59.75038pt}{163.0244pt}{d2_MS0_nonoci}
				\caption{}
				\label{subfig:d2_MS0_nonoci}
			\end{subfigure}
			\caption{
				Energy and symmetry of low-lying UHF solutions (6-31G* basis) in octahedral \ce{[VF6]^3-} having V--F = \SI{1.9896}{\angstrom}.
				(\subref{subfig:d2_MS1_nonoci}) Solutions with $N_\alpha - N_\beta = 2$ (red).
				(\subref{subfig:d2_MS0_nonoci}) Solutions with $N_\alpha = N_\beta$ (blue).
				All solutions have DIIS errors smaller than \SI{1e-13}{} and are eigenfunctions of $\hat{S}_3$ whose $M_S$ eigenvalues are given as subscripts.
				Solutions are considered degenerate when there energies are at most \SI{e-9}{\hartree} apart and are labeled alphabetically in increasing order of their energies.
				Nearly degenerate solutions share the same letter but are distinguished by dashes.
				Solutions within one degenerate set are distinct and linearly independent.
				Solutions that conserve spatial symmetry are enclosed in solid boxes whereas solutions that break spatial symmetry are enclosed in dotted ones.
				The time-reversal partners of the $M_S = 0$ solutions are also eigenfunctions of $\hat{S}_3$ with $M_S = 0$ and are thus included in (b).
				See the main text for a discussion on spin multiplicities.
			}
			\label{fig:d2_nonoci}
		\end{figure*}

	We can now turn to the octahedral \ce{[VF6]^{3-}} anion.
	The $3d^2$ electrons on the \ce{V^{3+}} center are expected to be strongly correlated on grounds of symmetry as discussed in the previous Section.
	However, similar to \ce{[TiF6]^{3-}}, the \ce{Ar} core and the \ce{F-} ligands in \ce{[VF6]^{3-}} also introduce additional correlation which further complicates the electronic structure.
	Figure~\ref{fig:d2_nonoci} presents the energy, spatial symmetry, and labels of the lowest-energy UHF solutions located by SCF metadynamics in octahedral \ce{[VF6]^3-} at V--F bond length of \SI{1.9896}{\angstrom} for both $M_S = 1$ and $M_S = 0$.
	Table~\ref{tab:d2_nonoci} plots the representative spatial forms of the highest-occupied $3d$ spin-orbitals and shows the $\langle \hat{S}^2 \rangle$ values of these solutions.
	Detailed solution energies and overlap matrices for the degenerate sets can be found in the Supporting Information.
	Comparing to Figure~\ref{fig:d1_nonoci}, we immediately see that there are significantly more low-lying UHF solutions in \ce{[VF6]^3-} than in \ce{[TiF6]^3-} as a consequence of the increase in the number of possible terms due to the strong correlation between the two valence $d$ electrons.
	Considering both $M_S = 1$ and $M_S = 0$, these solutions do not appear to segregate into three distinct groups as one would naively expect from the three possible strong-field configurations (Table~\ref{tab:d2terms}).
	It is therefore important that we first understand the symmetry of the UHF solutions, restoring any broken symmetry whenever necessary, before we can assign them to the expected $d^2$ terms.

		%% d2 solution distribution		

		%% d2 isosurface plots
		\begin{table*}
			\centering
			\caption{
				Representative isosurface plots for the Pipek--Mezey-localized spatial parts of the highest-occupied spin-orbitals, spatial symmetry, and $\langle \hat{S}^2 \rangle$ of the UHF solutions in \ce{[VF6]^{3-}}.
				In (\subref{subtab:d2_nonoci_MS1}), both spin-orbitals $\chi_{40}$ and $\chi_{41}$ have $m_s = \frac{1}{2}$.
				In (\subref{subtab:d2_nonoci_MS0}), $\chi_{40}$ has $m_s = \frac{1}{2}$ whereas $\bar{\chi}_{80}$ has $m_s = -\frac{1}{2}$ as indicated by the bar.
				The spin-orbitals of all solutions within one set have similar forms and are related by the symmetry operations of $\mathcal{O}_h$.
				Spatial symmetry lists the irreducible representations of $\mathcal{O}_h$ spanned by the degenerate sets.
				Axis triad: red\textendash $x$; green\textendash $y$; blue\textendash $z$.
			}
			\label{tab:d2_nonoci}
			\footnotesize
			\begin{subtable}[h]{.48\textwidth}
				\centering
				\caption{$M_S = 1$}
				\label{subtab:d2_nonoci_MS1}
				\resizebox{\textwidth}{!}{%
					\begin{tabular}[t]{>{\raggedright\arraybackslash}m{0.9cm} >{\centering\arraybackslash}m{1.9cm} >{\centering\arraybackslash}m{1.9cm} >{\centering\arraybackslash}m{1.9cm} S[table-figures-decimal=4, table-figures-integer=1, table-number-alignment=center]}
						\toprule
						$\Psi_\mathrm{UHF}$ & $\chi_{40}$ & $\chi_{41}$ &
						Spatial symmetry & {$\langle \hat{S}^2 \rangle$} \\
						\midrule
						$\mathrm{A}_1$ & \includegraphics{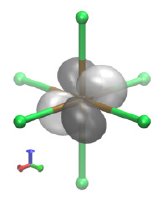} & \includegraphics{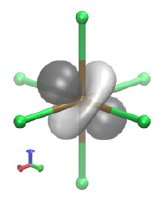} & $T_{1g} \oplus A_{2g}$ & 2.0040 \\
						$\mathrm{A}'_1$ & \includegraphics{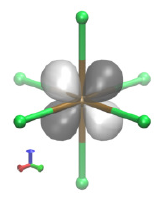} & \includegraphics{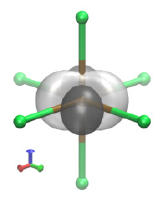}
						& $T_{1g} \oplus T_{2g}$ & 2.0040 \\
						$\mathrm{B}_1$ & \includegraphics{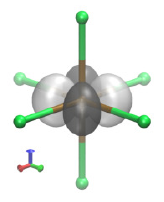} & \includegraphics{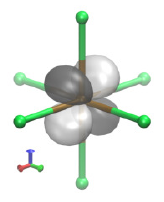} & $T_{1g}$ & 2.0045 \\
						$\mathrm{C}_1$ & \includegraphics{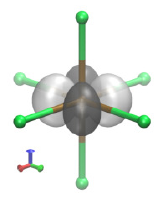} & \includegraphics{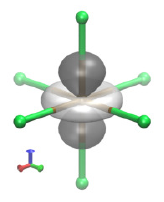} & $T_{2g}$ & 2.0035 \\
						$\mathrm{D}_1$ & \includegraphics{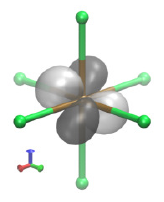} & \includegraphics{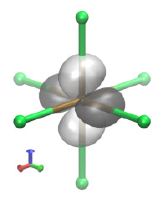} & $T_{1g} \oplus T_{2g}$ & 2.0039 \\
						$\mathrm{E}_1$ & \includegraphics{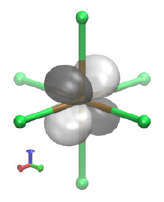} & \includegraphics{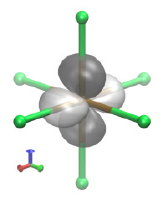}
						& $T_{1g}$ & 2.0049 \\
						$\mathrm{F}_1$ & \includegraphics{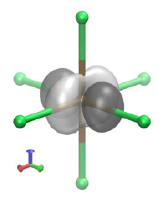} & \includegraphics{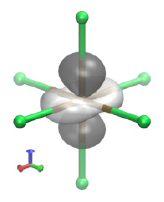} & $A_{2g}$ & 2.0044 \\
						\bottomrule
					\end{tabular}
				}
			\end{subtable}
			\hfill
			\begin{subtable}[h]{.48\textwidth}
				\centering
				\caption{$M_S = 0$}
				\label{subtab:d2_nonoci_MS0}
				\resizebox{\textwidth}{!}{%
					\begin{tabular}[t]{>{\raggedright\arraybackslash}m{0.9cm} >{\centering\arraybackslash}m{1.9cm} >{\centering\arraybackslash}m{1.9cm} >{\centering\arraybackslash}m{1.9cm} S[table-figures-decimal=4, table-figures-integer=1, table-number-alignment=center]}
						\toprule
						$\Psi_\mathrm{UHF}$ & $\chi_{40}$ & $\bar{\chi}_{80}$ &
						Spatial symmetry & {$\langle \hat{S}^2 \rangle$} \\
						\midrule
						$\mathrm{A}_0$ & \includegraphics{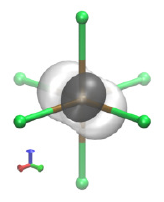} & \includegraphics{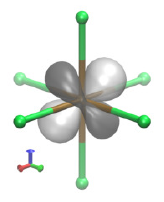} & $A_{1g} \oplus A_{2g}$ \newline $\oplus\ 2E_{g}\ \oplus$ \newline $3T_{1g} \oplus 3T_{2g}$ & 1.0025 \\
						$\mathrm{A}'_0$ & \includegraphics{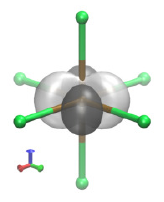} & \includegraphics{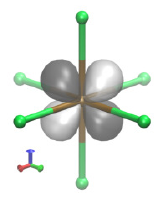}
						& $2T_{1g} \oplus 2T_{2g}$ & 1.0029 \\
						$\mathrm{A}''_0$ & \includegraphics{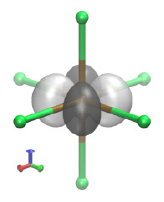} & \includegraphics{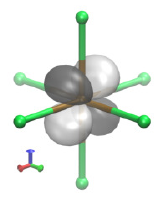} & $T_{1g} \oplus T_{2g}$ & 1.0037 \\
						$\mathrm{B}_0$ & \includegraphics{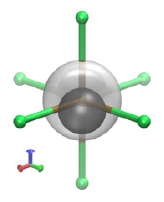} & \includegraphics{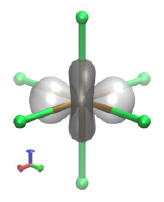} & $A_{1g} \oplus E_{g}$ \newline $\oplus\ T_{2g}$ & 0.8090 \\
						$\mathrm{C}_0$ & \includegraphics{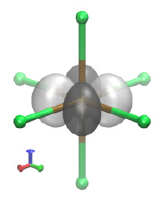} & \includegraphics{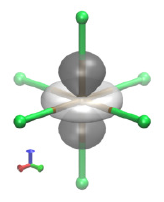} & $2T_{2g}$ & 1.0034 \\
						$\mathrm{D}_0$ & \includegraphics{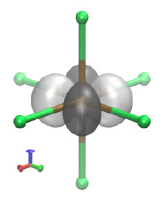} & \includegraphics{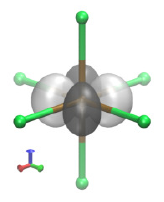}
						& $A_{1g} \oplus E_g$ & 0.0000 \\
						$\mathrm{E}_0$ & \includegraphics{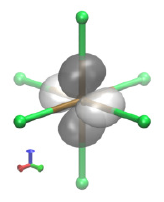} & \includegraphics{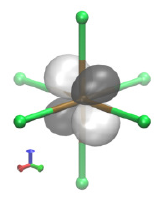} & $2T_{1g}$ & 1.0046 \\
						\bottomrule
					\end{tabular}
				}
			\end{subtable}
		\end{table*}

	\subsubsection{{\boldmath $M_S = 1$}}
	\label{subsubsec:results-d2-full-MS1}
		
		\paragraph{Overview of UHF solutions.}
	
			We focus first on the $M_S = 1$ solutions (Figure~\ref{subfig:d2_MS1_nonoci}) because they are simpler to understand.
			Since the true $d^2$ wavefunctions with $M_S = 1$ are all single-determinantal (Table~\ref{tab:d2terms}), it is without surprise that we have located the spatial-symmetry-conserved $M_S = 1$ solutions $\mathrm{B}_1$, $\mathrm{C}_1$, $\mathrm{E}_1$, and $\mathrm{F}_1$ that have the correct spatial symmetry and reasonable $\langle\hat{S}^2\rangle$ values to describe the $M_S = 1$ components of $\prescript{3}{}{T}_{1g}(t_{2g}^2)$, $\prescript{3}{}{T}_{2g}(t_{2g}^1e_g^1)$, $\prescript{3}{}{T}_{1g}(t_{2g}^1e_g^1)$, and $\prescript{3}{}{A}_{2g}(e_g^2)$, respectively.
			The distinction between $\mathrm{B}_1$ and $\mathrm{E}_1$, both of which transform as $T_{1g}$, is made by referring to the Pipek--Mezey-localized forms of their $\chi_{40}$ and $\chi_{41}$ (Table~\ref{subtab:d2_nonoci_MS1}).
			However, more interesting are the spatial-symmetry-broken solutions $\mathrm{A}_1$, $\mathrm{A}'_1$, and $\mathrm{D}_1$.
			The forms of $\chi_{40}$ and $\chi_{41}$ in these solutions do not let them be assigned to any of the terms listed in Table~\ref{tab:d2terms} immediately: although $\chi_{40}$ and $\chi_{41}$ individually resemble a $d$ orbital, they have been distorted away from the canonical shapes and orientations given in (\ref{eq:1e5dforms}) such that they can no longer be associated with the components of $t_{2g}$ or $e_g$.
			
			\input{results-d2/results-d2-nocifig}

		\paragraph{Symmetry restoration of individual symmetry-broken UHF sets.}
		
			The results of the toy system discussed in Section~\ref{subsubsec:results-d2-model-UHFNOCI} make it clear that symmetry breaking is an indication of electron correlation being incorporated into UHF single determinants and NOCI allows for some of it to be recovered via symmetry restoration.
			Figure~\ref{subfig:d2_MS1_singlenoci} shows the energy of the NOCI wavefunctions constructed from single sets of symmetry-broken $M_S = 1$ UHF solutions, and Table~\ref{subtab:d2_noci_MS1} gives some of their Pipek--Mezey-localized $3d$ natural orbitals which provide a very satisfying visual demonstration of the capability of NOCI to restore spatial symmetry.
			Comparing, for example, the $3d$ orbitals of the $\mathrm{A}_1$ UHF solutions in Table~\ref{subtab:d2_nonoci_MS1} with some of the $3d$ natural orbitals of the $\prescript{3}{}{T}_{1g}[\mathrm{A}_1]$ NOCI wavefunctions in Table~\ref{subtab:d2_noci_MS1}, we see that, unlike the former, the latter are properly symmetrized such that each of them can be attributed to one of the $\xi$, $\eta$, and $\zeta$ components of $t_{2g}$.
			
			However, to assign these NOCI wavefunctions to terms, we need a careful understanding of the correlation they recover.
			By drawing analogies to the $M_S = 1$ UHF and NOCI wavefunctions of the toy model \ce{[(B^{3+})(q^-)6]} analyzed in Section~\ref{subsubsec:results-d2-model-UHFNOCI}, we can infer the correlation nature of the symmetry-broken solutions $\mathrm{A}_1$, $\mathrm{A}'_1$, and $\mathrm{D}_1$ from the corresponding $\mathrm{a}_1$, $\mathrm{a}'_1$, and $\mathrm{d}_1$.
			Firstly, their symmetry breaking is indeed a consequence of the correlation between the two $3d$ electrons, but the symmetry-conserved single determinants $\mathrm{C}_1$, $\mathrm{E}_1$, and $\mathrm{F}_1$ already capture most of this for the corresponding electronic terms.
			However, when the $3d$ electrons break symmetry, they cause the now-present core and ligand electrons to relax accordingly and incorporate correlation to these electrons as a result.
			This is evident by the fact that the NOCI wavefunctions  $\prescript{3}{}{T}_{2g}[\mathrm{D}_1]$ and $\prescript{3}{}{T}_{1g}[\mathrm{D}_1]$ no longer coincide with the symmetry-conserved UHF solutions $\mathrm{C}_1$ and $\mathrm{E}_1$ (Figure~\ref{subfig:d2_MS1_singlenoci}), and that $\prescript{\varnothing}{}{A}_{2g}[\mathrm{A}_1]$ and $\prescript{\varnothing}{}{T}_{2g}[\mathrm{A}'_1]$ do not have well defined spin multiplicities and are so high in energy (well outside the range of Figure~\ref{subfig:d2_MS1_singlenoci}) that they no longer match with $\mathrm{F}_1$ and $\mathrm{C}_1$.
			On the other hand, the symmetry breaking in $\mathrm{A}_1$ and $\mathrm{A}'_1$ arises from the additional configuration-mixing correlation between the $3d$ electrons that is missed out by the symmetry-conserved $\mathrm{B}_1$ solutions.
			This symmetry breaking incidentally also results in the inclusion of some correlation with the core and ligand electrons as previously mentioned.
	
			Based on the above understanding together with the forms and occupation numbers of the natural orbitals (Table~\ref{subtab:d2_noci_MS1}), we can assign the NOCI wavefunctions as follows.
			The predominantly occupied $3d$ natural orbitals of $\prescript{3}{}{T}_{1g}[\mathrm{A}_1]$ (all having occupation numbers of \SI{0.945}{}) suggest that they mainly describe the $M_S = 1$ components of the $\prescript{3}{}{T}_{1g}(t_{2g}^2)$ term, and the minorly occupied $3d$ natural orbitals (all having occupation numbers of \SI{0.055}{}) indicate that they have some $\prescript{3}{}{T}_{1g}(t_{2g}^1e_g^1)$ character mixed in.
			We thus assign $\prescript{3}{}{T}_{1g}[\mathrm{A}_1]$ to the mixed-configuration term $a\prescript{3}{}{T}_{1g}(t_{2g}^2)(t_{2g}^1 e_g^1)$ (see the caption of Table~\ref{tab:2e5d_d2} for the explanation of this notation).
			In a similar manner, we assign $\prescript{3}{}{T}_{1g}[\mathrm{A}'_1]$ to $a\prescript{3}{}{T}_{1g}(t_{2g}^2)(t_{2g}^1 e_g^1)$, $\prescript{3}{}{T}_{2g}[\mathrm{D}_1]$ to $\prescript{3}{}{T}_{2g}(t_{2g}^1 e_g^1)$, and $\prescript{3}{}{T}_{1g}[\mathrm{D}_1]$ to $\prescript{3}{}{T}_{1g}(t_{2g}^1 e_g^1)$, as also listed in Table~\ref{subtab:d2_noci_MS1}.
			It cannot be emphasized enough that, as the symmetry breaking in $\mathrm{A}_1$ and $\mathrm{A}'_1$ is a consequence of configuration mixing, restoring symmetry in these solutions automatically gives symmetry-conserved wavefunctions that also incorporate this effect \textit{without having to explicitly construct or obtain determinants belonging to the various interacting configurations}.
			
			\input{results-d2/results-d2-tab}

		\paragraph{NOCI between multiple degenerate UHF sets.}
		
			The NOCI wavefunctions obtained above can be improved if multiple degenerate UHF sets that span one or more common spaces are to be included in the basis.
			For instance, $\mathrm{A}_1$, $\mathrm{A}'_1$, $\mathrm{B}_1$, $\mathrm{D}_1$, and $\mathrm{E}_1$ all span a common $\prescript{3}{}{T}_{1g}$ space with $M_S = 1$ and we therefore expect them to interact with one another to give rise to various $\prescript{3}{}{T}_{1g}$ NOCI wavefunctions.
			However, while the $\prescript{3}{}{T}_{1g}$ space of $\mathrm{A}_1$ and $\mathrm{A}'_1$ has both $t_{2g}^2$ and $t_{2g}^1 e_g^1$ characters, the $\prescript{3}{}{T}_{1g}$ space of $\mathrm{B}_1$ only has $t_{2g}^2$ character and that of $\mathrm{D}_1$ and $\mathrm{E}_1$ only has $t_{2g}^1 e_g^1$ character.
			Thus, by allowing some or all of $\mathrm{A}_1$, $\mathrm{A}'_1$, and $\mathrm{B}_1$ to interact with none, some, or all of $\mathrm{D}_1$ and $\mathrm{E}_1$, we obtain $\prescript{3}{}{T}_{1g}$ NOCI wavefunctions that have mainly $t_{2g}^2$ character but with some $t_{2g}^1 e_g^1$ mixed in.
			Their energies are labeled collectively as $\prescript{3}{}{T}_{1g}[\mathrm{A}_1, \mathrm{A}'_1, \mathrm{B}_1\ (\mathrm{D}_1, \mathrm{E}_1)]$ in Figure~\ref{subfig:d2_MS1_allnoci}.
			This collection can be considered to consist principally of NOCI wavefunctions constructed from all possible non-trivial combinations of $\mathrm{A}_1$, $\mathrm{A}'_1$, and $\mathrm{B}_1$ and thus incorporating whatever correlation responsible for any symmetry breaking in these UHF solutions.
			These wavefunctions are then further improved by interactions with $\mathrm{D}_1$ and $\mathrm{E}_1$.
			There are 27 non-trivial combinations in this collection: $(2^3-1)\times 2^2$ minus the trivial one that includes only the symmetry-conserved $\mathrm{B}_1$ set.
			It is precisely because this trivial combination is excluded that the NOCI wavefunctions in this collection are all of $a\prescript{3}{}{T}_{1g}(t_{2g}^2)(t_{2g}^1e_g^1)$ character	and spread over a rather small range of \SI{2.36e-3}{\hartree}, or approximately \SI{517}{cm^{-1}} (see Supporting Information for the detailed energies).
			By the variational principle, the best approximation to $a\prescript{3}{}{T}_{1g}(t_{2g}^2)(t_{2g}^1e_g^1)$ must be as low in energy as possible and turns out to be $a\prescript{3}{}{T}_{1g}[\mathrm{A}_1 \oplus \mathrm{A}'_1 \oplus \mathrm{B}_1 \oplus \mathrm{D}_1 \oplus \mathrm{E}_1]$.
			However, from a physical perspective, $a\prescript{3}{}{T}_{1g}[\mathrm{A}_1 \oplus \mathrm{A}'_1 \oplus \mathrm{B}_1 \oplus \mathrm{D}_1 \oplus \mathrm{E}_1]$ does not really recover much more configuration-mixing correlation than $\prescript{3}{}{T}_{1g}[\mathrm{A}_1]$ and $\prescript{3}{}{T}_{1g}[\mathrm{A}'_1]$ already do, so that even if one were not able to locate the higher $\mathrm{D}_1$ and $\mathrm{E}_1$ UHF solutions, one could still get a reasonable description of the ground mixed-configuration $a\prescript{3}{}{T}_{1g}(t_{2g}^2)(t_{2g}^1 e_g^1)$ term from either $\mathrm{A}_1$ or $\mathrm{A}'_1$, as suggested by the small energy range of the $\prescript{3}{}{T}_{1g}[\mathrm{A}_1, \mathrm{A}'_1, \mathrm{B}_1\ (\mathrm{D}_1, \mathrm{E}_1)]$ collection.
			
			Similarly, there are 23 non-trivial NOCI wavefunctions in $\prescript{3}{}{T}_{1g}[\mathrm{D}_1, \mathrm{E}_1\ (\mathrm{A}_1, \mathrm{A}'_1, \mathrm{B}_1)]$ that all contain a major $t_{2g}^1 e_g^1$ character, and those that include at least one of $\mathrm{A}_1$, $\mathrm{A}'_1$, or $\mathrm{B}_1$ also have some $t_{2g}^2$ character mixed in.
			However, since all of these NOCI wavefunctions have the same symmetry as the ground term, and since the exact ground term is unknown, we cannot be sure if any of them is indeed orthogonal to the exact ground term in order for the variational principle to apply \cite{book:Bransden2000}.
			This is rather unfortunate as the collection $\prescript{3}{}{T}_{1g}[\mathrm{D}_1, \mathrm{E}_1\ (\mathrm{A}_1, \mathrm{A}'_1, \mathrm{B}_1)]$ spans a rather large range in energy (about \SI{0.0188}{\hartree} or \SI{4130}{cm^{-1}}) which makes it difficult to give a reasonable estimate for $b\prescript{3}{}{T}_{1g}(t_{2g}^1 e_g^1)(t_{2g}^2)$.

		\paragraph{Spin multiplicity assignments.}
				
			Before moving on, we briefly discuss the spin symmetry of the $M_S = 1$ solutions.
			In a very similar manner to the \ce{[TiF6]^{3-}} case discussed earlier, none of these solutions is an exact eigenfunction of $\hat{S}^2$ due to the slight deviation of their $\langle\hat{S}^2\rangle$ from the exact value of \SI{2}{} expected for a triplet.
			Nevertheless, we still assign them to triplet states on grounds that a discrepancy smaller than \SI{1e-2}{} is negligible for our purposes while acknowledging that there is still some spin symmetry breaking due to contamination from the $M_S = 1$ components of higher spin states.
			Spin purification can be achieved somewhat by spatial symmetry restoration just as noted for \ce{[TiF6]^{3-}}, but ultimately, to fully restore spin symmetry, one needs to include in the NOCI basis all spin symmetry partners generated by the spin rotation operations in $\mathsf{SU}(2)$.

	\subsubsection{{\boldmath $M_S = 0$}}
	\label{subsubsec:results-d2-full-MS0}

		Unlike the $M_S = \pm 1$ wavefunctions, the $M_S = 0$ wavefunctions in a true $d^2$ system are all required by symmetry to be multi-determinantal (Table~\ref{tab:d2terms}).
		It therefore comes as no surprise that the $M_S = 0$ single-determinantal UHF solutions in \ce{[VF6]^{3-}} are all symmetry-broken, be it in $\mathcal{O}_h$, $\mathcal{S} \otimes \mathcal{T}$, or both (Figure~\ref{subfig:d2_MS0_nonoci} and Table~\ref{subtab:d2_nonoci_MS0}).
		In particular, apart from the $\mathrm{D}_0$ solutions which are effectively RHF, have $\langle \hat{S}^2 \rangle = \SI{0.0000}{}$ and can be considered true singlets, these $M_S = 0$ solutions all have $\langle \hat{S}^2 \rangle$ in the vicinity of unity which indicates heavy spin mixing.
		The symmetry breaking in $\mathrm{D}_0$ is therefore purely spatial, whereas the symmetry breaking in the other solutions is much less straightforward to classify without symmetry restoration.
		This is because in the presence of significant spin contamination, the different spin components can break or conserve spatial symmetry independently such that the overall spatial representation obtained from the analysis of symmetry in $\mathcal{O}_h$ (Table~\ref{subtab:d2_nonoci_MS0}) only gives a direct sum of all spatial irreducible representations but does not segregate them into the different spin components.
		Therefore, NOCI must first be carried out so that the $\langle \hat{S}^2 \rangle$ values of the spatial-symmetry-conserved NOCI wavefunctions can be worked out and that the spatial irreducible representations can be assigned to definite spin states.
		The symmetry labels of the UHF solutions in Figure~\ref{subfig:d2_MS0_nonoci} reflect the outcomes of this, and the energy of the resulted low-lying NOCI wavefunctions constructed from single sets of symmetry-broken solutions are plotted in Figure~\ref{subfig:d2_MS0_singlenoci} with some representative Pipek--Mezey-localized natural orbitals shown in Tables~\ref{subtab:d2_noci_MS0_S1}~and~\ref{subtab:d2_noci_MS0_S0}.

		\paragraph{Time-reversal symmetry.}
		
		The $M_S = 0$ space is rather interesting.
		We show in Appendix~\ref{appsubsec:commutativitythetaj3} that the time-reversal operator $\hat{\Theta}$ and the $\hat{S}_3$ operator, which have the following effects on a spin-pure state $\ket{S, M_S}$,\cite{article:Stedman1980}
			\begin{align}
				\hat{\Theta} \ket{S, M_S} &= (-1)^{S-M_S} \ket{S, -M_S} \label{eq:optimerevS} \\
				\hat{S}_3 \ket{S, M_S} &= M_S \ket{S, M_S}
				\label{eq:opS3}
			\end{align}
 		only commute when acting on wavefunctions with $M_S = 0$.
		As such, $M_S = 0$ wavefunctions can be made to be simultaneous eigenfunctions of both $\hat{\Theta}$ and $\hat{S}_3$.
		A given $M_S = 0$ UHF determinant, say $\Psi_0$, is obviously an eigenfunction of $\hat{S}_3$ by construction, but there is no \textit{a priori} restriction that it must also be an eigenfunction of $\hat{\Theta}$.
		However, the commutativity between $\hat{\Theta}$ and $\hat{S}_3$ and the finite cyclic structure of the time-reversal group $\mathcal{T}$ (Section~\ref{subsec:symbreakingHF}, also Appendix~\ref{appsubsec:timerevgroup}) ensure that $\Psi_0$ and all possible linearly independent UHF determinants $(\hat{\Theta}^n) \Psi_0$ where $n = 1,2,3$ can be arranged into linear combinations that have $M_S = 0$ and are also eigenfunctions of $\hat{\Theta}$.
		Since $\mathcal{T}$ is abelian, if the set $\lbrace (\hat{\Theta}^n) \Psi_0\ | \ n=0,\ldots,3 \rbrace$ contains more than one linearly independent member, it is necessarily symmetry-broken in $\mathcal{T}$.
		But as $\hat{\Theta}$ also commutes with the spinless electronic Hamiltonian defined in (\ref{eq:genhamil}) and is antiunitary\cite{book:Wigner1959}, all members of the above set have the same energy which the degeneracy depicted in Figures~\ref{subfig:d2_MS0_nonoci},~\ref{subfig:d2_MS0_singlenoci},~and~\ref{subfig:d2_MS0_allnoci} must take into account.
		
		The above discussion brings to attention the fact that $\hat{\Theta}$ is also a symmetry operator that needs to be considered on top of the spatial symmetry operations $\hat{R}$ in $\mathcal{O}_h$ in order to obtain enough UHF determinants and complete any symmetry-broken set within the $M_S = 0$ space prior to running NOCI.
		However, there is a fortuitous exception to this as demonstrated by the $\mathrm{A}_0$, $\mathrm{A}''_0$, $\mathrm{B}_0$, and $\mathrm{D}_0$ solutions.
		Consider the set of all linearly independent spatial-symmetry-equivalent UHF determinants, $\lbrace \hat{R} \Psi_0 \ | \ \hat{R} \in \mathcal{O}_h \rbrace$, where $\Psi_0$ is a determinant in any one of $\mathrm{A}_0$, $\mathrm{A}''_0$, $\mathrm{B}_0$, or $\mathrm{D}_0$.
		A character analysis shows that this set spans multiple but \textit{complete} irreducible representations in $\mathcal{T}$, and that augmenting this set to $\lbrace (\hat{\Theta}^n \hat{R}) \Psi_0 \ | \ n=0,\ldots, 3;\ \hat{R} \in \mathcal{O}_h \rbrace$ and retaining only the linearly independent elements does not change the representations spanned in $\mathcal{O}_h$ or $\mathcal{T}$.
		There thus exists a subset $T$ in $\mathcal{O}_h$ such that
			\begin{equation*}
				\hat{\Theta} \Psi_0 = \sum_{\hat{R} \in T} c_{\hat{R}} \hat{R}\Psi_0
			\end{equation*}
		Typically, $T$ consists of only one element $\hat{R}'$ with $c_{\hat{R}'} = \pm 1$.
		Consequently, $\mathrm{A}_0$, $\mathrm{A}''_0$, $\mathrm{B}_0$, and $\mathrm{D}_0$ are said to exhibit \textit{time-reversal coincidence} and the origin of this effect can be somewhat understood by noting from Table~\ref{subtab:d2_nonoci_MS0} that the spatial forms of $\chi_{40}$ and $\bar{\chi}_{80}$ in these solutions can be interconverted (up to a phase factor of $\pm 1$) by a symmetry operation in $\mathcal{O}_h$.
		The remaining solutions, however, do not benefit from this coincidence as the set $\lbrace \hat{R} \Psi_0 \ | \ \hat{R} \in \mathcal{O}_h \rbrace$ for these solutions does not span complete irreducible representations in $\mathcal{T}$ and therefore $\hat{\Theta}$ must be involved to complete the set.
		This is also obvious from Table~\ref{subtab:d2_nonoci_MS0} that their $\chi_{40}$ and $\bar{\chi}_{80}$ are not related by any symmetry operations in $\mathcal{O}_h$ at all.
		
		The use of NOCI on time-reversal symmetry partners to restore spin symmetry of the heavily spin-contaminated $M_S = 0$ UHF solutions needs a more careful explanation.
		The effect of $\hat{\Theta}$ on the spin-pure state $\ket{S, 0}$ can be deduced from (\ref{eq:optimerevS}) as
			\begin{equation*}
				\hat{\Theta} \ket{S, 0} = (-1)^{S} \ket{S, 0}
			\end{equation*}
		It is then obvious that an $\ket{S, 0}$ state with even $S$ is totally symmetric in $\mathcal{T}$ whereas an $\ket{S, 0}$ state with odd $S$ spans the $B$ irreducible representation of $\mathcal{T}$ (Table~\ref{tab:Tcharactertable}).
		Thus, by allowing all spin-contaminated $M_S = 0$ determinants and their time-reversal partners to interact via the totally symmetric Hamiltonian, we restore time-reversal symmetry and decouple states with even and odd $S$ as a consequence.
		The $M_S = 0$ NOCI wavefunctions in Figure~\ref{subfig:d2_MS0_singlenoci}, Figure~\ref{subfig:d2_MS0_allnoci}, Table~\ref{subtab:d2_noci_MS0_S1}, and Table~\ref{subtab:d2_noci_MS0_S0} are therefore all symmetry-conserved in $\mathcal{T}$ and contain either odd or even $S$ components, but not both: those with $\langle \hat{S}^2 \rangle \approx 2$ are mainly $\ket{1, 0}$ in character but are contaminated by higher $\ket{S, 0}$ states with odd $S$, while those with $\langle \hat{S}^2 \rangle \approx 0$ contain a main $\ket{0, 0}$ component together with smaller contributions from higher $\ket{S, 0}$ with even $S$.
		As such, by restoring time-reversal symmetry, NOCI also restores spin-rotation symmetry but not completely, unless $S = 0$ and $S = 1$ are the only possible spin states as in the toy system \ce{[(B^{3+})(q^-)6]}.
		The multiplicity labels for these NOCI states therefore only reflect the most dominant $\ket{S, 0}$ component, but this is sufficiently accurate as deviations smaller than \SI{1e-2}{} of $\langle\hat{S}^2\rangle$ from the exact values are negligible for our purposes, as stated before.

		\paragraph{Correlation from symmetry restoration.}
		
		To understand the kind of correlation recovered by NOCI, we once again make reference to the $M_S = 0$ UHF and NOCI wavefunctions of the toy system \ce{[(B^{3+})(q^-)6]} (Section~\ref{subsubsec:results-d2-model-UHFNOCI}).
		We were unfortunately not able to find any solutions in \ce{[VF6]^{3-}} that can be correlated exactly to the symmetry-broken $\mathrm{a}^*_0$ in \ce{[(B^{3+})(q^-)6]} based on their symmetry and degeneracy.
		However, we located a set of symmetry-broken solutions labeled $\mathrm{A}_0$ whose spin-orbitals are similar to those of $\mathrm{a}^*_0$ but that span more irreducible representations than $\mathrm{a}^*_0$.
		This is the reason for the asterisk in the label of $\mathrm{a}^*_0$.
		We posit that the implication of this result is twofold.
		Firstly, the symmetry breaking in $\mathrm{a}^*_0$ must be due to the correlation between the two $d$ electrons, so the space that is common to both $\mathrm{a}^*_0$ and $\mathrm{A}_0$ must contain descriptions of this strong correlation.%
		\bibnote{The irreducible representations spanned by $\mathrm{a}^*_0$ are shown in Table~\ref{subtab:2e5d_d2_MS0} and those spanned by $\mathrm{A}_0$ are shown in Figure~\ref{subfig:d2_MS0_nonoci}. The attentive reader will notice that the $\prescript{1}{}{T}_{1g}$ component in $\mathrm{a}^*_0$ is apparently missing from $\mathrm{A}_0$. This, however, can be attributed to the incomplete spin purification using time reversal, because, as shown in Table S5b in the Supporting Information, $\mathrm{A}_0$ does contain a $T_{1g}$ component denoted by $\prescript{\varnothing}{}{T}_{1g}[\mathrm{A}_0]$ with $\langle\hat{S}^2\rangle \approx 0.1003$. This unfortunately results in it being assigned to an undefined spin multiplicity based on our \SI{1e-2}{} threshold (signified by ``$\varnothing$''). However, $\prescript{\varnothing}{}{T}_{1g}[\mathrm{A}_0]$ still contains a $\prescript{1}{}{T}_{1g}$ component among other even-$S$ contaminations.}
		Secondly, the additional symmetry breaking in $\mathrm{A}_0$, which gives rise to the observed increase in size of the linearly independent space spanned by these solutions compared to that spanned by $\mathrm{a}^*_0$, must be a consequence of the correlation between the two $d$ electrons with the core and ligand electrons which are only present in the full \ce{[VF6]^{3-}} system.
		The severe symmetry breaking in $\mathrm{A}_0$ is therefore the result of both types of correlation.
		
		The symmetry breaking in $\mathrm{A}_0$, $\mathrm{A}'_0$, and $\mathrm{B}_0$ contains configuration mixing information for various electronic terms.
		In fact, an inspection of the $3d$ natural orbitals and occupation numbers of the NOCI wavefunctions constructed from $\mathrm{A}_0$, $\mathrm{A}'_0$, and $\mathrm{B}_0$ (Tables~\ref{subtab:d2_noci_MS0_S1}~and~\ref{subtab:d2_noci_MS0_S0}) reveals that the symmetry breaking of $\mathrm{A}_0$ and $\mathrm{A}'_0$ accounts for some configuration mixing in $a\prescript{3}{}{T}_{1g}(t_{2g}^2)(t_{2g}^1 e_g^1)$ and $a\prescript{1}{}{T}_{2g}(t_{2g}^2)(t_{2g}^1 e_g^1)$, and that the symmetry breaking of $\mathrm{B}_0$ accounts for some configuration mixing in $\diamond\prescript{1}{}{E}_{g}(t_{2g}^2)(e_g^2)$ and $\diamond\prescript{1}{}{A}_{1g}(t_{2g}^2)(e_g^2)$.
		This is further supported by the NOCI results of the corresponding solutions $\mathrm{a}^*_0$, $\mathrm{a}'_0$, and $\mathrm{b}_0$ in the toy system \ce{[(B^{3+})(q^-)6]} as shown in Table~\ref{subtab:2e5d_d2_MS0} where similar configuration mixing effects are also observed.
		
		For the remaining solutions $\mathrm{A}''_0$, $\mathrm{C}_0$, and $\mathrm{E}_0$ that can be corresponded to $\mathrm{a}''_0$, $\mathrm{c}_0$, and $\mathrm{e}_0$ of the toy system \ce{[(B^{3+})(q^-)6]}, we expect from the analysis in Section~\ref{subsubsec:results-d2-model-UHFNOCI} that the most of the correlation in their $S \approx 1$ components is already captured by the symmetry-conserved $M_S = 1$ solutions $\mathrm{B}_1$, $\mathrm{C}_1$, and $\mathrm{E}_1$.
		However, the energies of $\prescript{3}{}{T}_{1g}[\mathrm{A}''_0]$, $\prescript{3}{}{T}_{2g}[\mathrm{C}_0]$, and $\prescript{3}{}{T}_{1g}[\mathrm{E}_0]$ all deviate from those of $\mathrm{B}_1$, $\mathrm{C}_1$, and $\mathrm{E}_1$ by slight amounts (Figure~\ref{fig:d2_singlenoci}, see also the Supporting Information for precise energy values), which must be the result of some correlation with the core and ligand electrons.
		Unfortunately, the NOCI $3d$ natural orbitals and occupation numbers in Tables~\ref{subtab:d2_noci_MS0_S1}~and~\ref{subtab:d2_noci_MS0_S0} show that $\prescript{3}{}{T}_{1g}[\mathrm{A}''_0]$, $\prescript{1}{}{T}_{2g}[\mathrm{A}''_0]$, $\prescript{1}{}{T}_{2g}[\mathrm{C}_0]$ and $\prescript{3}{}{T}_{1g}[\mathrm{E}_0]$ miss out large amounts of configuration mixing correlation between the two $d$ electrons.
		Fortunately, this problem can be overcome by including multiple degenerate sets of UHF solutions describing different configurations of the same term in the NOCI basis, the results of which are shown in Figure~\ref{subfig:d2_MS0_allnoci}.
		Drastic improvements to the NOCI wavefunctions can be most easily seen through the energy distribution of the collections $\prescript{3}{}{T}_{1g}[\mathrm{A}_0, \mathrm{A}'_0, \mathrm{A}''_0\ (\mathrm{E}_0)]$, $\prescript{3}{}{T}_{1g}[\mathrm{E}_0\  (\mathrm{A}_0, \mathrm{A}'_0, \mathrm{A}''_0)]$, and $\prescript{1}{}{T}_{2g}[\mathrm{C}_0\ (\mathrm{A}_0, \mathrm{A}'_0, \mathrm{A}''_0)]$.
		Take the collection $\prescript{3}{}{T}_{1g}[\mathrm{A}_0, \mathrm{A}'_0, \mathrm{A}''_0\ (\mathrm{E}_0)]$ for example: there is a very distinctive gap between the energy of $\prescript{3}{}{T}_{1g}[\mathrm{A}''_0]$ that have no configuration mixing and the energy of the remaining NOCI wavefunctions in the collection that all include configuration mixing to some extent.

%% file: results-d2/results-d2-nocifig.tex
%% !TeX root = ../symmultiplescf.tex
% !TeX program = lualatex

%% d2 single NOCI
\begin{figure*}
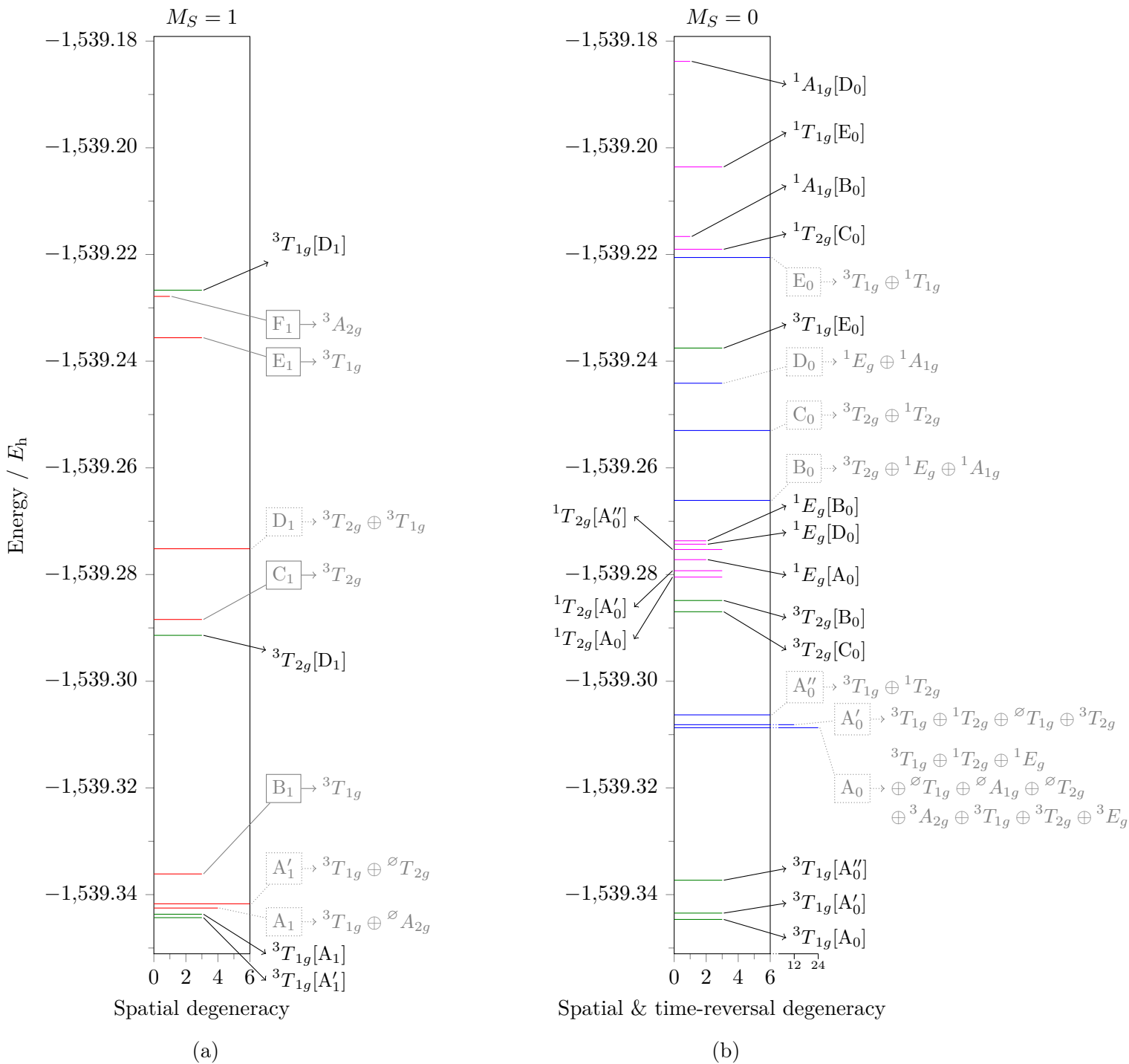

	\centering
	\begin{subfigure}[b]{0.30\textwidth}
		\centering
		% \useexternalfile{scale}{trimleft}{trimright}{name}
		% Figure compiled with tikzexternalize
		% Pre-compiled figure located at ./results-d2/tikz/tikzoutput/d2_MS1_singlenoci-output.pdf
		\input{\tikzexternal@filenameprefix1.tikz}{78.73251pt}{95.42513pt}{d2_MS1_singlenoci}
		\caption{}
		\label{subfig:d2_MS1_singlenoci}
	\end{subfigure}
	\hfill
	\begin{subfigure}[b]{0.65\textwidth}
		\centering
		% \useexternalfile{scale}{trimleft}{trimright}{name}
		% Figure compiled with tikzexternalize
		% Pre-compiled figure located at ./results-d2/tikz/tikzoutput/d2_MS0_singlenoci-output.pdf
		\input{\tikzexternal@filenameprefix1.tikz}{65.09644pt}{185.59612pt}{d2_MS0_singlenoci}
		\caption{}
		\label{subfig:d2_MS0_singlenoci}
	\end{subfigure}
	\caption{
		Low-lying NOCI wavefunctions constructed from single sets of degenerate UHF solutions in octahedral \ce{[VF6]^3-} (replotted from Figure~\ref{fig:d2_nonoci} for comparison).
		The horizontal axis in (b) has been broken at $6$ and a more compact scale is used between $6$ and $24$ so that the degeneracy of most states can be shown clearly.
		Green indicates NOCI wavefunctions with $\langle \hat{S}^2 \rangle \approx 2$ while magenta indicates those with $\langle \hat{S}^2 \rangle \approx 0$.
		NOCI wavefunctions are considered degenerate when there energies are at most \SI{e-7}{\hartree} apart.
	}
	\label{fig:d2_singlenoci}
\end{figure*}

%% d2 all NOCI
\begin{figure*}
	\centering
	\begin{subfigure}[b]{0.30\textwidth}
		\centering
		\input{\tikzexternal@filenameprefix1.tikz}{78.73251pt}{123.66551pt}{d2_MS1_allnoci}
		\caption{}
		\label{subfig:d2_MS1_allnoci}
	\end{subfigure}
	\hfill
	\begin{subfigure}[b]{0.65\textwidth}
		\centering
		\input{\tikzexternal@filenameprefix1.tikz}{63.03935pt}{185.59612pt}{d2_MS0_allnoci}
		\caption{}
		\label{subfig:d2_MS0_allnoci}
	\end{subfigure}
	\caption{
		Low-lying NOCI wavefunctions constructed from all interacting low-lying UHF solutions in octahedral \ce{[VF6]^3-} (replotted from Figure~\ref{fig:d2_nonoci} for comparison).
		The horizontal axis in (b) has been broken at $6$ and a more compact scale is used between $6$ and $24$ so that the degeneracy of most states can be shown clearly.
		Green indicates NOCI wavefunctions with $\langle \hat{S}^2 \rangle \approx 2$ while magenta indicates those with $\langle \hat{S}^2 \rangle \approx 0$.
		NOCI wavefunctions are considered degenerate when there energies are at most \SI{e-7}{\hartree} apart.
		In generic notations, $\Gamma[\mathrm{A}, \mathrm{B}\ (\mathrm{C}, \mathrm{D})]$ denotes multiple NOCI sets of symmetry $\Gamma$ constructed from all non-trivial combinations of \textit{at least one} of $\mathrm{A}$ and $\mathrm{B}$, but of none, some, or all of $\mathrm{C}$ and $\mathrm{D}$.
	}
	\label{fig:d2_allnoci}
\end{figure*}

%% file: results-d2/results-d2-tab.tex
%% !TeX root = ../symmultiplescf.tex
% !TeX program = lualatex

%% d2 NOCI natural orbitals		
\begin{table*}
	\centering
	\caption{
		Isosurface plots for the Pipek--Mezey-localized spatial parts of some $3d$ natural orbitals, $\langle \hat{S}^2 \rangle$, and term assignment of selected NOCI wavefunctions $\Phi$ in \ce{[VF6]^{3-}}.
		Each group of $3d$ natural orbitals shown in each set corresponds to one of the NOCI wavefunctions within the set that transforms as the labeled component.
		From left to right within each group of natural orbitals, shown in the first pair of parentheses are the occupation numbers of the natural orbitals corresponding to the dominant configuration, and whenever applicable, shown in the second pair of parentheses are the occupation numbers of the  natural orbitals corresponding to the minor configuration.
		These configurations are also indicated next to the term symbol in the same order.
		Axis triad: red\textendash $x$; green\textendash $y$; blue\textendash $z$.
	}
	\label{tab:d2_noci}
	\footnotesize
	%%
	%% M_S = 1
	\begin{subtable}[h]{\textwidth}
		\centering
		\caption{
			$M_S = 1$.
			The occupation numbers and natural orbitals are solely in the $m_s= \frac{1}{2}$ space.
		}
		\label{subtab:d2_noci_MS1}
		\scalebox{0.55}{%
			\begin{tabular}[t]{>{\raggedright\arraybackslash}m{1.6cm} >{\centering\arraybackslash}m{7.30cm} >{\centering\arraybackslash}m{7.30cm} >{\centering\arraybackslash}m{7.30cm} S[table-format=1.4, table-number-alignment=center] >{\raggedright\arraybackslash}m{2.7cm}}
				\toprule
				$\Phi$ & \multicolumn{3}{c}{$3d$ natural orbitals} &
				{$\langle \hat{S}^2 \rangle$} & Term \\
				\midrule
				$\prescript{3}{}{T}_{1g}[\mathrm{A}_1]$ &%
				\includegraphics{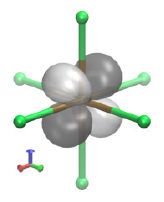} \includegraphics{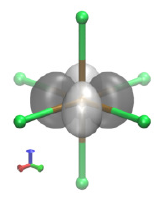}
				\includegraphics{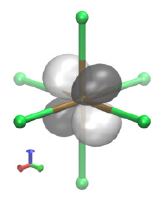} \includegraphics{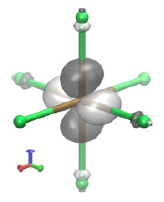} &%
				\includegraphics{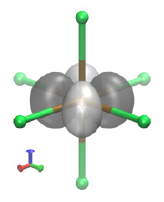} \includegraphics{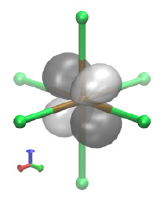}
				\includegraphics{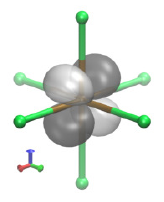}
				\includegraphics{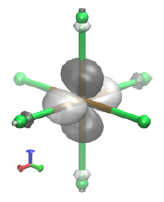} &%
				\includegraphics{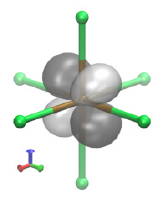} \includegraphics{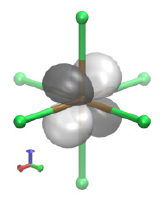}
				\includegraphics{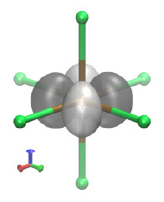} \includegraphics{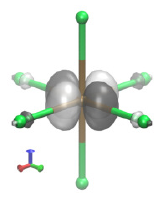} &%
				2.0035 & $a\prescript{3}{}{T}_{1g}(t_{2g}^2)(t_{2g}^1 e_g^1)$ \\
				& $\alpha$ & $\beta$ & $\gamma$ & \\
				& (\SI{0.945}{}, \SI{0.945}{}), (\SI{0.055}{}, \SI{0.055}{}) & (\SI{0.945}{}, \SI{0.945}{}), (\SI{0.055}{}, \SI{0.055}{}) & (\SI{0.945}{}, \SI{0.945}{}), (\SI{0.055}{}, \SI{0.055}{}) \\[0.3cm]
				$\prescript{3}{}{T}_{1g}[\mathrm{A}'_1]$ &%
				\includegraphics{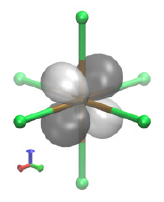} \includegraphics{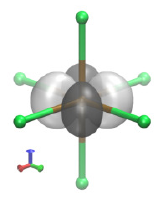}
				\includegraphics{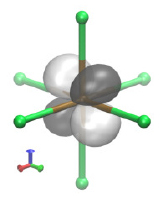} \includegraphics{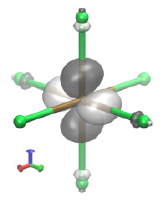} &%
				\includegraphics{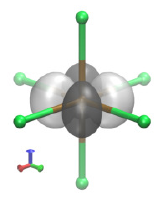} \includegraphics{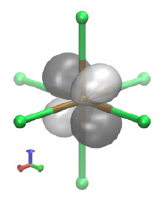}
				\includegraphics{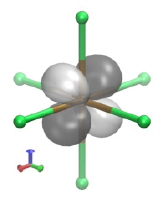} \includegraphics{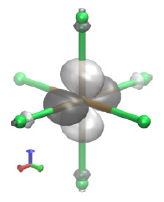} &%
				\includegraphics{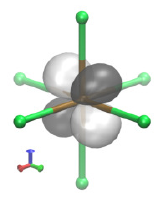} \includegraphics{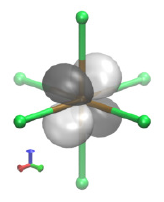}
				\includegraphics{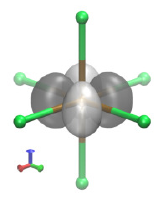} \includegraphics{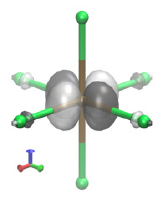} &%
				2.0033 & $a\prescript{3}{}{T}_{1g}(t_{2g}^2)(t_{2g}^1 e_g^1)$ \\
				& $\alpha$ & $\beta$ & $\gamma$ & \\
				& (\SI{0.957}{}, \SI{0.957}{}), (\SI{0.043}{}, \SI{0.043}{}) & (\SI{0.957}{}, \SI{0.957}{}), (\SI{0.043}{}, \SI{0.043}{}) & (\SI{0.957}{}, \SI{0.957}{}), (\SI{0.044}{}, \SI{0.043}{}) \\[0.3cm]
				$\prescript{3}{}{T}_{2g}[\mathrm{D}_1]$ &%
				\includegraphics{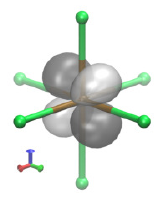} \includegraphics{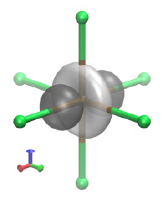} &%
				\includegraphics{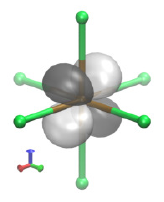} \includegraphics{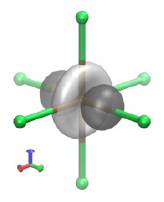} &%
				\includegraphics{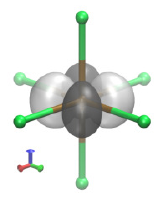} \includegraphics{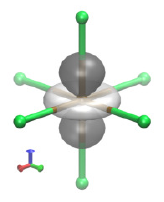} &%
				2.0025 & $\prescript{3}{}{T}_{2g}(t_{2g}^1 e_g^1)$ \\
				& $\xi$ & $\eta$ & $\zeta$ & \\
				& (\SI{0.999}{}, \SI{0.999}{}) & (\SI{0.999}{}, \SI{0.999}{}) & (\SI{0.999}{}, \SI{0.999}{}) \\[0.3cm]
				$\prescript{3}{}{T}_{1g}[\mathrm{D}_1]$ &%
				\includegraphics{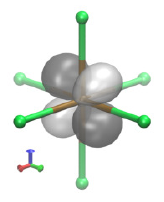} \includegraphics{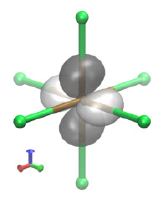} &%
				\includegraphics{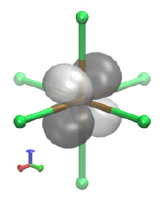} \includegraphics{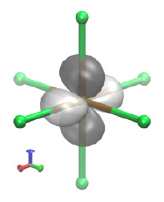} &%
				\includegraphics{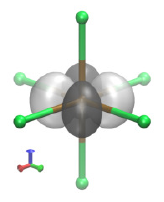} \includegraphics{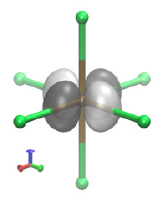} &%
				2.0079 & $\prescript{3}{}{T}_{1g}(t_{2g}^1 e_g^1)$ \\
				& $\alpha$ & $\beta$ & $\gamma$ & \\
				& (\SI{0.983}{}, \SI{0.981}{}) & (\SI{0.990}{}, \SI{0.988}{}) & (\SI{0.998}{}, \SI{0.995}{}) \\
				\bottomrule
			\end{tabular}
		}
	\end{subtable}
\end{table*}

\begin{table*}
	\ContinuedFloat
	\footnotesize
	%%
	%% M_S = 0, S ~ 1
	\begin{subtable}[h]{\textwidth}
		\centering
		\caption{
			$M_S = 0, S \approx 1$.
			The occupation numbers and natural orbitals are identical for both $m_s=\pm \frac{1}{2}$.
		}
		\label{subtab:d2_noci_MS0_S1}
		\scalebox{0.55}{%
			\begin{tabular}[t]{>{\raggedright\arraybackslash}m{1.6cm} >{\centering\arraybackslash}m{7.30cm} >{\centering\arraybackslash}m{7.30cm} >{\centering\arraybackslash}m{7.30cm} S[table-figures-decimal=4, table-figures-integer=1, table-number-alignment=center] >{\raggedright\arraybackslash}m{2.7cm}}
				\toprule
				$\Phi$ & \multicolumn{3}{c}{$3d$ natural orbitals} &
				{$\langle \hat{S}^2 \rangle$} & Term \\
				\midrule
				$\prescript{3}{}{T}_{1g}[\mathrm{A}_0]$ &%
				\includegraphics{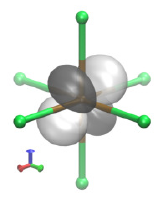} \includegraphics{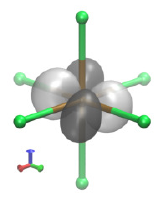}
				\includegraphics{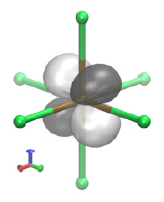} \includegraphics{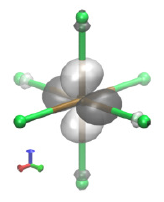} &%
				\includegraphics{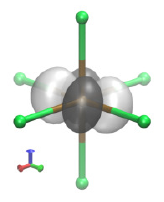} \includegraphics{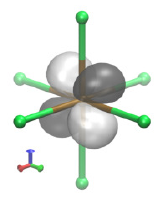}
				\includegraphics{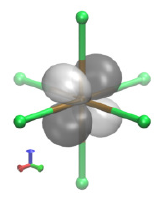} \includegraphics{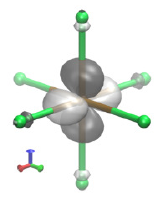} &%
				\includegraphics{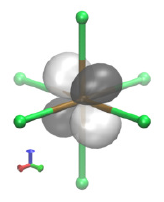} \includegraphics{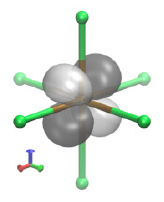}
				\includegraphics{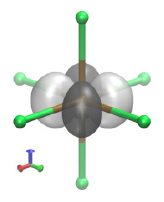} \includegraphics{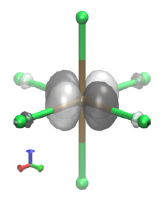} &%
				2.0000 & $a\prescript{3}{}{T}_{1g}(t_{2g}^2)(t_{2g}^1 e_g^1)$ \\
				& $\alpha$ & $\beta$ & $\gamma$ & \\
				& (\SI{0.485}{}, \SI{0.485}{}), (\SI{0.015}{}, \SI{0.015}{}) & (\SI{0.485}{}, \SI{0.485}{}), (\SI{0.015}{}, \SI{0.015}{}) & (\SI{0.485}{}, \SI{0.485}{}), (\SI{0.015}{}, \SI{0.015}{}) \\[0.3cm]
				$\prescript{3}{}{T}_{1g}[\mathrm{A}'_0]$ &%
				\includegraphics{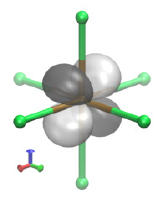} \includegraphics{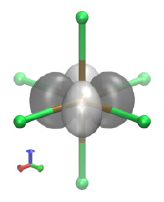}
				\includegraphics{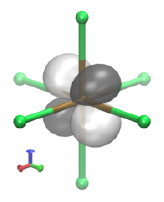} \includegraphics{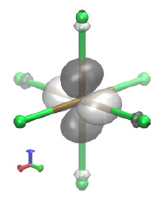} &%
				\includegraphics{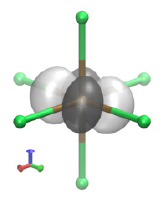} \includegraphics{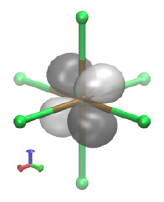}
				\includegraphics{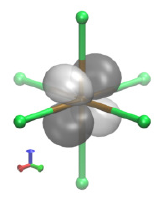} \includegraphics{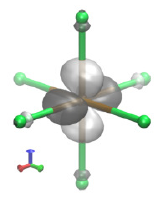} &%
				\includegraphics{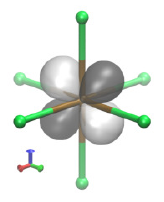} \includegraphics{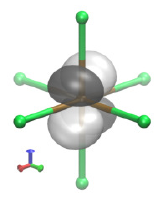}
				\includegraphics{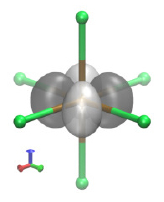} \includegraphics{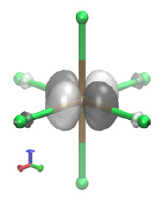} &%
				2.0000 & $a\prescript{3}{}{T}_{1g}(t_{2g}^2)(t_{2g}^1 e_g^1)$ \\
				& $\alpha$ & $\beta$ & $\gamma$ & \\
				& (\SI{0.489}{}, \SI{0.489}{}), (\SI{0.011}{}, \SI{0.011}{}) & (\SI{0.489}{}, \SI{0.489}{}), (\SI{0.011}{}, \SI{0.011}{}) & (\SI{0.489}{}, \SI{0.489}{}), (\SI{0.011}{}, \SI{0.011}{}) \\[0.3cm]
				$\prescript{3}{}{T}_{1g}[\mathrm{A}''_0]$ &%
				\includegraphics{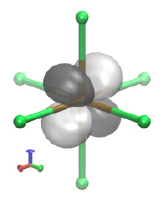} \includegraphics{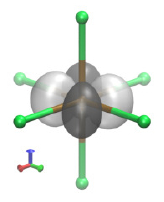} &%
				\includegraphics{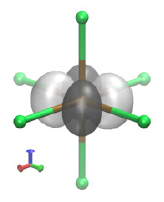} \includegraphics{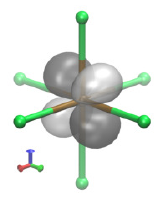} &%
				\includegraphics{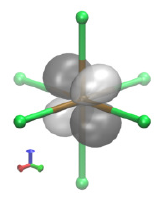} \includegraphics{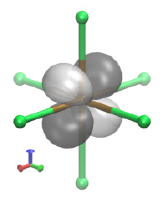} &%
				2.0000 & $\prescript{3}{}{T}_{1g}(t_{2g}^2)$ \\
				& $\alpha$ & $\beta$ & $\gamma$ & \\
				& (\SI{0.500}{}, \SI{0.500}{}) & (\SI{0.500}{}, \SI{0.500}{}) & (\SI{0.500}{}, \SI{0.500}{}) \\[0.3cm]
				$\prescript{3}{}{T}_{2g}[\mathrm{B}_0]$ &%
				\includegraphics{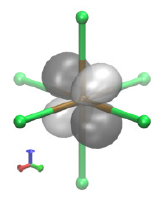} \includegraphics{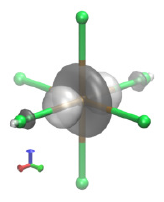} &%
				\includegraphics{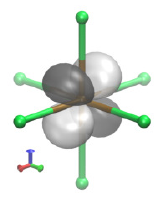} \includegraphics{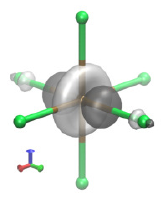} &%
				\includegraphics{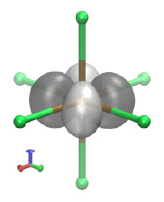} \includegraphics{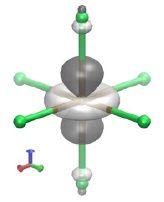} &%
				2.0000 & $\prescript{3}{}{T}_{2g}(t_{2g}^1 e_g^1)$ \\
				& $\xi$ & $\eta$ & $\zeta$ & \\
				& (\SI{0.500}{}, \SI{0.499}{}) & (\SI{0.497}{}, \SI{0.497}{}) & (\SI{0.500}{}, \SI{0.499}{}) \\[0.3cm]
				$\prescript{3}{}{T}_{2g}[\mathrm{C}_0]$ &%
				\includegraphics{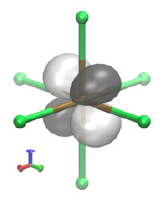} \includegraphics{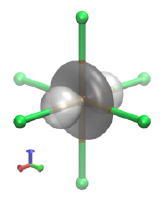} &%
				\includegraphics{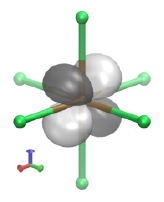} \includegraphics{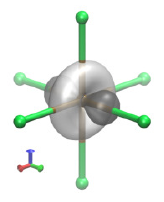} &%
				\includegraphics{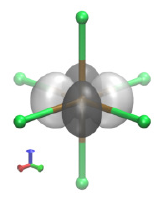} \includegraphics{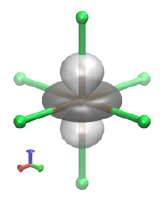} &%
				2.0000 & $\prescript{3}{}{T}_{2g}(t_{2g}^1 e_g^1)$ \\
				& $\xi$ & $\eta$ & $\zeta$ & \\
				& (\SI{0.500}{}, \SI{0.500}{}) & (\SI{0.500}{}, \SI{0.500}{}) & (\SI{0.500}{}, \SI{0.500}{}) \\[0.3cm]
				$\prescript{3}{}{T}_{1g}[\mathrm{E}_0]$ &%
				\includegraphics{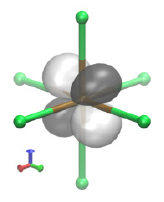} \includegraphics{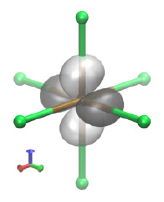} &%
				\includegraphics{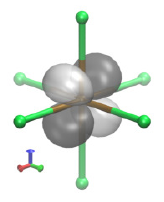} \includegraphics{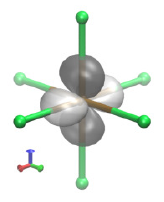} &%
				\includegraphics{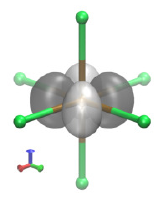} \includegraphics{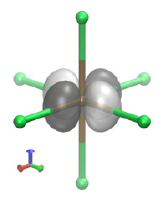} &%
				2.0000 & $\prescript{3}{}{T}_{1g}(t_{2g}^1 e_g^1)$ \\
				& $\alpha$ & $\beta$ & $\gamma$ & \\
				& (\SI{0.500}{}, \SI{0.500}{}) & (\SI{0.500}{}, \SI{0.500}{}) & (\SI{0.500}{}, \SI{0.500}{}) \\
				\bottomrule
			\end{tabular}
		}
	\end{subtable}
\end{table*}

\begin{table*}
	\ContinuedFloat
	\footnotesize
	%%
	%% M_S = 0, S ~ 0
	\begin{subtable}[h]{\textwidth}
		\centering
		\caption{
			$M_S = 0, S \approx 0$.
			The occupation numbers and natural orbitals are identical for both $m_s=\pm \frac{1}{2}$.
		}
		\label{subtab:d2_noci_MS0_S0}
		\scalebox{0.55}{%
			\begin{tabular}[t]{>{\raggedright\arraybackslash}m{1.6cm} >{\centering\arraybackslash}m{7.30cm} >{\centering\arraybackslash}m{7.30cm} >{\centering\arraybackslash}m{7.30cm} S[table-figures-decimal=4, table-figures-integer=1, table-number-alignment=center] >{\raggedright\arraybackslash}m{2.7cm}}
				\toprule
				$\Phi$ & \multicolumn{3}{c}{$3d$ natural orbitals} &
				{$\langle \hat{S}^2 \rangle$} & Term \\
				\midrule
				$\prescript{1}{}{T}_{2g}[\mathrm{A}_0]$ &%
				\includegraphics{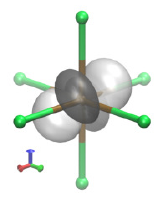} \includegraphics{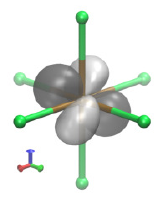}
				\includegraphics{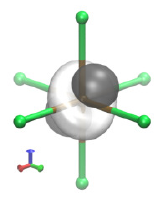} \includegraphics{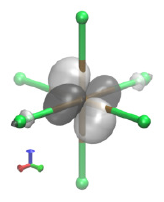} &%
				\includegraphics{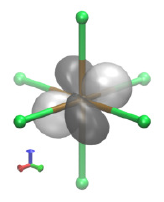} \includegraphics{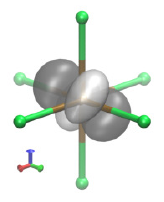}
				\includegraphics{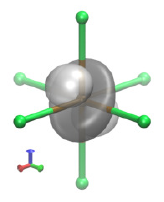} \includegraphics{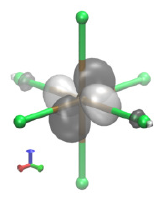} &%
				\includegraphics{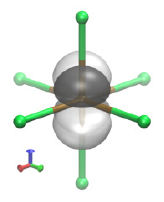} \includegraphics{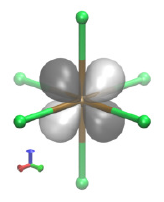}
				\includegraphics{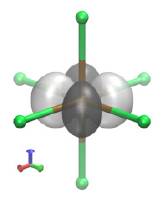} \includegraphics{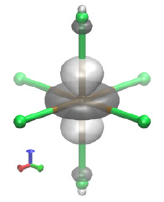} &%
				0.0038 & $a\prescript{1}{}{T}_{2g}(t_{2g}^2)(t_{2g}^1 e_g^1)$ \\
				& $\xi$ & $\eta$ & $\zeta$ & \\
				& (\SI{0.491}{}, \SI{0.491}{}), (\SI{0.009}{}, \SI{0.009}{}) & (\SI{0.496}{}, \SI{0.486}{}), (\SI{0.010}{}, \SI{0.009}{}) & (\SI{0.492}{}, \SI{0.490}{}), (\SI{0.009}{}, \SI{0.009}{}) \\[0.3cm]
				$\prescript{1}{}{T}_{2g}[\mathrm{A}'_0]$ &%
				\includegraphics{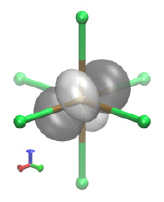} \includegraphics{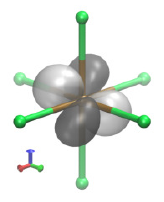}
				\includegraphics{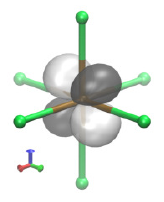} \includegraphics{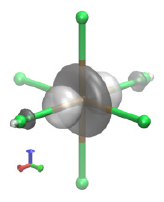} &%
				\includegraphics{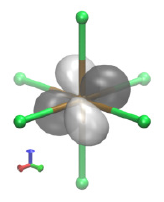} \includegraphics{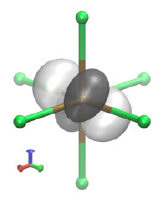}
				\includegraphics{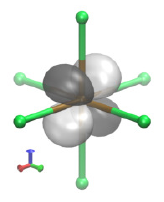} \includegraphics{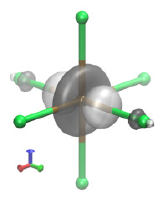} &%
				\includegraphics{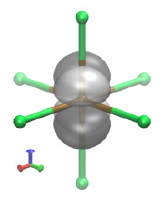} \includegraphics{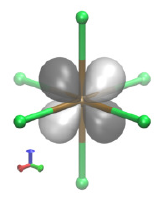}
				\includegraphics{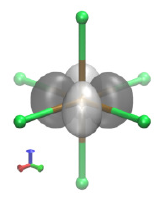} \includegraphics{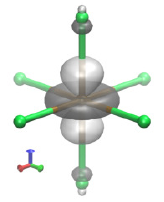} &%
				0.0042 & $a\prescript{1}{}{T}_{2g}(t_{2g}^2)(t_{2g}^1 e_g^1)$ \\
				& $\xi$ & $\eta$ & $\zeta$ & \\
				& (\SI{0.496}{}, \SI{0.496}{}), (\SI{0.004}{}, \SI{0.004}{}) & (\SI{0.496}{}, \SI{0.496}{}), (\SI{0.004}{}, \SI{0.004}{}) & (\SI{0.496}{}, \SI{0.496}{}), (\SI{0.004}{}, \SI{0.004}{}) \\[0.3cm]
				$\prescript{1}{}{T}_{2g}[\mathrm{A}''_0]$ &%
				\includegraphics{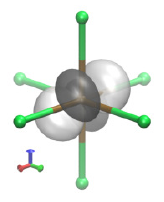} \includegraphics{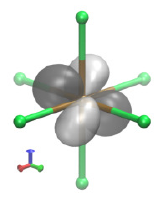} &%
				\includegraphics{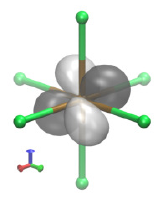} \includegraphics{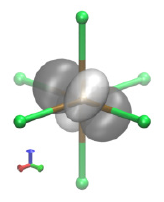} &%
				\includegraphics{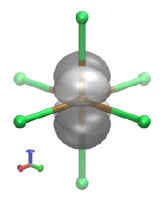} \includegraphics{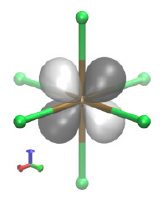} &%
				0.0073 & $\prescript{1}{}{T}_{2g}(t_{2g}^2)$ \\
				& $\xi$ & $\eta$ & $\zeta$ & \\
				& (\SI{0.500}{}, \SI{0.500}{}) & (\SI{0.500}{}, \SI{0.500}{}) & (\SI{0.500}{}, \SI{0.500}{}) \\[0.3cm]
				$\prescript{1}{}{T}_{2g}[\mathrm{C}_0]$ &%
				\includegraphics{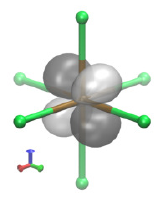} \includegraphics{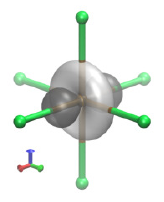} &%
				\includegraphics{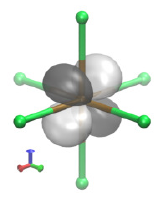} \includegraphics{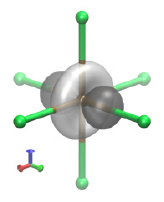} &%
				\includegraphics{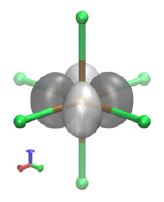} \includegraphics{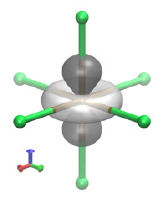} &%
				0.0067 & $\prescript{1}{}{T}_{2g}(t_{2g}^1 e_g^1)$ \\
				& $\xi$ & $\eta$ & $\zeta$ & \\
				& (\SI{0.498}{}, \SI{0.498}{}) & (\SI{0.497}{}, \SI{0.497}{}) & (\SI{0.499}{}, \SI{0.499}{}) \\[0.3cm]
				$\prescript{1}{}{T}_{1g}[\mathrm{E}_0]$ &%
				\includegraphics{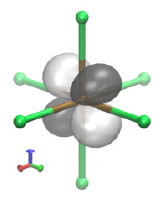} \includegraphics{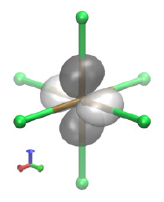} &%
				\includegraphics{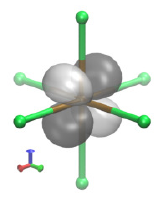} \includegraphics{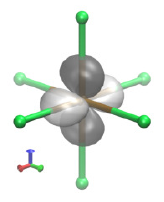} &%
				\includegraphics{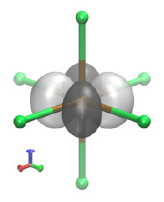} \includegraphics{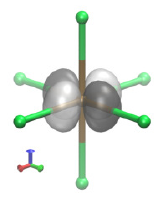} &%
				0.0092 & $\prescript{1}{}{T}_{1g}(t_{2g}^1 e_g^1)$ \\
				& $\alpha$ & $\beta$ & $\gamma$ & \\
				& (\SI{0.500}{}, \SI{0.500}{}) & (\SI{0.499}{}, \SI{0.499}{}) & (\SI{0.500}{}, \SI{0.500}{}) \\
				\bottomrule
			\end{tabular}
		}
		
		\vspace{0.50cm}
		\scalebox{0.55}{%
			\begin{tabular}[t]{>{\raggedright\arraybackslash}m{1.6cm} >{\centering\arraybackslash}m{9.05cm} >{\centering\arraybackslash}m{7.30cm} S[table-figures-decimal=4, table-figures-integer=1, table-number-alignment=center] >{\raggedright\arraybackslash}m{2.7cm}}
				\toprule
				$\Phi$ & \multicolumn{2}{c}{$3d$ natural orbitals} &
				{$\langle \hat{S}^2 \rangle$} & Term \\
				\midrule
				$\prescript{1}{}{E}_{g}[\mathrm{A}_0]$ &%
				\includegraphics{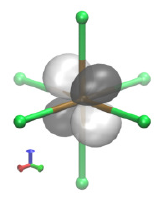} \includegraphics{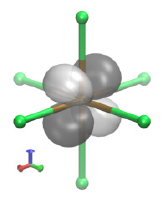}						\includegraphics{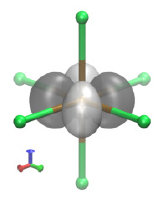} &%
				\includegraphics{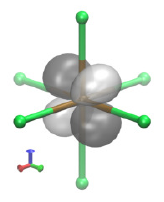} \includegraphics{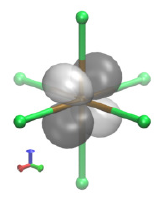} &%
				0.0032 & $\prescript{1}{}{E}_{g}(t_{2g}^2)$ \\
				& $u$ & $v$ & \\
				& (\SI{0.216}{}, \SI{0.122}{}, \SI{0.661}{}) & (\SI{0.433}{}, \SI{0.558}{}) \\[0.3cm]
				$\prescript{1}{}{E}_{g}[\mathrm{B}_0]$ &%
				\includegraphics{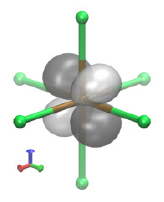} \includegraphics{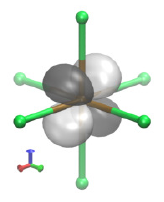}						\includegraphics{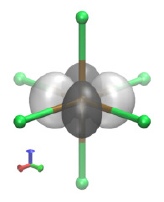}
				\includegraphics{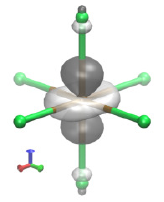}						\includegraphics{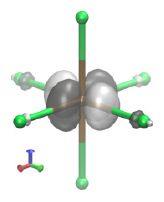} &%
				\includegraphics{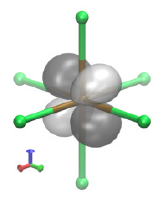} \includegraphics{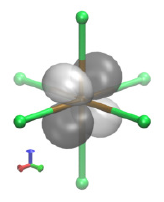}
				\includegraphics{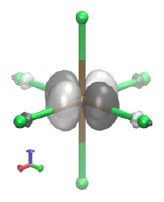} \includegraphics{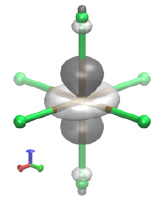} &%
				0.0029 & $\diamond\prescript{1}{}{E}_{g}(t_{2g}^2)(e_g^2)$ \\
				& $u$ & $v$ & \\
				& (\SI{0.188}{}, \SI{0.115}{}, \SI{0.595}{}), (\SI{0.052}{}, \SI{0.051}{}) & (\SI{0.400}{}, \SI{0.493}{}), (\SI{0.052}{}, \SI{0.051}{}) \\[0.3cm]
				$\prescript{1}{}{E}_{g}[\mathrm{D}_0]$ &%
				\includegraphics{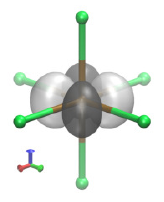} \includegraphics{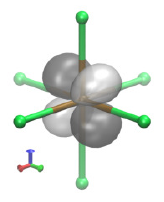}						\includegraphics{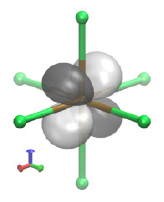} &%
				\includegraphics{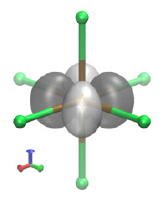} \includegraphics{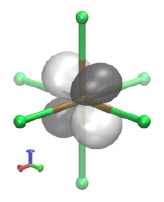} &%
				0.0000 & $\prescript{1}{}{E}_{g}(t_{2g}^2)$ \\
				& $u$ & $v$ & \\
				& (\SI{0.221}{}, \SI{0.118}{}, \SI{0.661}{}) & (\SI{0.446}{}, \SI{0.549}{}) \\[0.3cm]
				$\prescript{1}{}{A}_{1g}[\mathrm{B}_0]$ &%
				\includegraphics{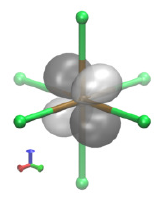} \includegraphics{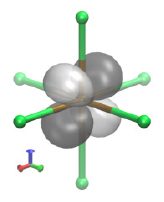}						\includegraphics{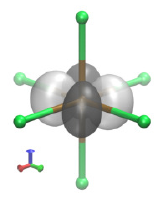}
				\includegraphics{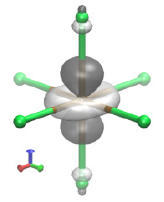}						\includegraphics{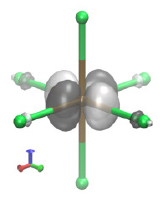} &%
				&%
				0.0027 & $\diamond\prescript{1}{}{A}_{1g}(t_{2g}^2)(e_g^2)$ \\
				& $e_1$ & & \\
				& (\SI{0.272}{}, \SI{0.272}{}, \SI{0.272}{}), (\SI{0.093}{}, \SI{0.093}{}) & \\[0.3cm]
				$\prescript{1}{}{A}_{1g}[\mathrm{D}_0]$ &%
				\includegraphics{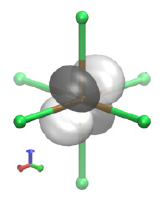} \includegraphics{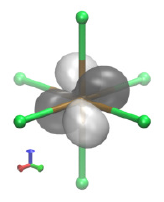}						\includegraphics{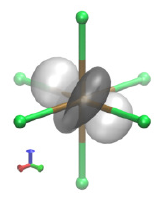} &%
				&%
				0.0000 & $\prescript{1}{}{A}_{1g}(t_{2g}^2)$ \\
				& $e_1$ & & \\
				& (\SI{0.333}{}, \SI{0.333}{}, \SI{0.334}{}) & \\
				\bottomrule
			\end{tabular}
		}
	\end{subtable}
\end{table*}

%% file: results-JT/results-JT.tex
\tikzsetexternalprefix{./results-JT/tikz/}

\section{UHF and NOCI Wavefunctions upon Symmetry Descent}
\label{sec:results-JT}

\begin{figure}
			\centering
			% \useexternalfile{scale}{trimleft}{trimright}{name}
			% Figure compiled with tikzexternalize
			% Pre-compiled figure located at ./results-JT/tikz/tikzoutput/vibmodes-output.pdf
			\input{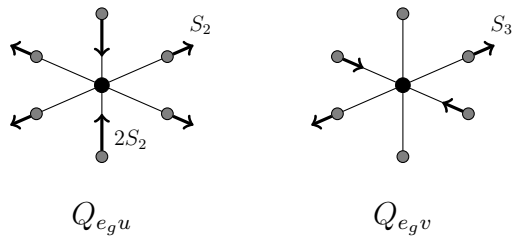}{0}{0}{vibmodes}
			\caption{
				Displacement components of the normal coordinates $Q_{e_gu}$ and $Q_{e_gv}$ of an octahedral \ce{MF6} (\ce{M = Ti, V}) nuclear framework.
				$S_2$ and $S_3$ are magnitudes of displacement vectors as derived by \citeauthor{book:Sugano1970}\cite{book:Sugano1970}
			}
			\label{fig:vibmodes}
		\end{figure}

	The understanding of the symmetry of the various low-lying UHF solutions in octahedral \ce{[TiF6]^{3-}} and \ce{[VF6]^{3-}} obtained so far sheds light on their behaviors when the molecular symmetry of the system descends from the highly symmetric $\mathcal{O}_h$.
	The ground terms for \ce{[TiF6]^{3-}} and \ce{[VF6]^{3-}} are $\prescript{2}{}{T}_{2g}$ and $\prescript{3}{}{T}_{1g}$, respectively.
	With spin--orbit coupling ignored for the current purposes, the Jahn--Teller theorem predicts that both octahedral anions undergo stabilizing and symmetry-lowering distortions via the $e_g$ or $t_{2g}$ vibrational modes since both $E_g$ and $T_{2g}$ are contained in the symmetrized squares of both $T_{1g}$ and $T_{2g}$ \cite{article:Jahn1937,book:Sugano1970}.
	We will distinguish between the symmetry of vibrational modes and the symmetry of electronic terms by the use of lower-case and upper-case symbols, respectively.
	In this work, we consider the $e_gu$ and $e_gv$ normal vibrational coordinates (Figure~\ref{fig:vibmodes}): the $\mathcal{O}_h$ symmetry of \ce{[TiF6]^{3-}} and \ce{[VF6]^{3-}} is lowered to $\mathcal{D}_{4h}$ along $Q_{e_gu}$ and to $\mathcal{D}_{2h}$ along $Q_{e_gv}$.
	We thus expect the octahedral $\prescript{2}{}{T}_{2g}$ and $\prescript{3}{}{T}_{1g}$ ground terms to split as shown in Table~\ref{tab:symdescent}.
	
		\begin{table*}
			\centering
			\caption{
				Splitting of $\prescript{2}{}{T}_{2g}$ and $\prescript{3}{}{T}_{1g}$ upon descending from $\mathcal{O}_h$ to $\mathcal{D}_{4h}$ and $\mathcal{D}_{2h}$.
			}
			\label{tab:symdescent}
			\begingroup
			\renewcommand\arraystretch{1.35}
			\begin{tabular}[t]{p{1.5cm} p{2.5cm} p{3.5cm}}
				\toprule
				$\mathcal{O}_h$ & $\mathcal{D}_{4h}$ & $\mathcal{D}_{2h}$ \\
				\midrule
				$\prescript{2}{}{T}_{2g}$ & $\prescript{2}{}{B}_{2g} \oplus \prescript{2}{}{E}_{g}$ & $\prescript{2}{}{B}_{1g} \oplus \prescript{2}{}{B}_{2g} \oplus \prescript{2}{}{B}_{3g}$ \\
				$\prescript{3}{}{T}_{1g}$ & $\prescript{3}{}{A}_{2g} \oplus \prescript{3}{}{E}_{g}$ & $\prescript{3}{}{B}_{1g} \oplus \prescript{3}{}{B}_{2g} \oplus \prescript{3}{}{B}_{3g}$ \\
				\bottomrule
			\end{tabular}
			\endgroup
		\end{table*}
	
	In Figures~\ref{fig:VF6AAdashJT}~and~\ref{fig:TiF6AAdashJT}, we show the behaviors of the lowest-lying SCF solutions that we have found for \ce{[VF6]^{3-}} and \ce{[TiF6]^{3-}}.
	Immediately obvious is that the symmetry-broken solutions $\mathrm{A}_1$, $\mathrm{A}'_1$, and $\mathrm{A}_\frac{1}{2}$ remain symmetry-broken to various degrees upon symmetry descent.
	Specifically, Figure~\ref{fig:TiF6AAdashJT} reveals that the $\mathrm{A}_\frac{1}{2}$ solutions which span $T_{1g} \oplus T_{2g}$ in $\mathcal{O}_h$ split into a symmetry-conserved set $E_g$ and a symmetry-broken set $A_{2g} \oplus B_{2g} \oplus E_{g}$ in $\mathcal{D}_{4h}$, and into three symmetry-broken sets $B_{1g} \oplus B_{2g}$, $B_{1g} \oplus B_{3g}$, and $B_{2g} \oplus B_{3g}$ in $\mathcal{D}_{2h}$.
	Clearly, the two components $T_{1g}$ and $T_{2g}$ of the $\mathrm{A}_\frac{1}{2}$ set are so entangled together in $\mathcal{O}_h$ that they cannot split independently from each other when descending in molecular symmetry.
	Figure~\ref{fig:VF6AAdashJT} shows similar behaviors for the $\mathrm{A}_1$ and $\mathrm{A}'_1$ solutions in \ce{[VF6]^{3-}}, where we note in particular that the $\mathrm{A}_1$ set remains completely symmetry-broken and quadruply degenerate throughout.
	This seems to suggest that there is some hidden symmetry in the UHF method that persists even when the symmetry of the electronic Hamiltonian is lowered, but it is unfortunately beyond the scope of the current paper and we will have to return to it in a future publication.
	
		\begin{figure*}
			\centering
			% \useexternalfile{scale}{trimleft}{trimright}{name}
			% Figure compiled with tikzexternalize
			% Pre-compiled figure located at ./results-JT/tikz/tikzoutput/VF63-_AAdash_JT_Eg_minimal-output.pdf
			\input{\tikzexternal@filenameprefix1.tikz}{0}{0}{VF63-_AAdash_JT_Eg_minimal}
			\caption{
				$\mathrm{A}_1$ (dashed) and $\mathrm{A}'_1$ (dotted) solutions and the low-lying NOCI wavefunctions constructed from all of them (solid) along the $e_g$ normal vibrational coordinates of octahedral \ce{[VF6]^{3-}}.
				The zero values of $S_2$ and $S_3$ correspond to $\mathcal{O}_h$ while the non-zero values correspond to the lower symmetry groups.
				All symmetry symbols are irreducible representations of the lower-symmetry group unless indicated otherwise.
				Spin multiplicities are only shown for NOCI wavefunctions.
			}
			\label{fig:VF6AAdashJT}
		\end{figure*}
	
		\begin{figure*}
			\centering
			% \useexternalfile{scale}{trimleft}{trimright}{name}
			% Figure compiled with tikzexternalize
			% Pre-compiled figure located at ./results-JT/tikz/tikzoutput/TiF63-_AAdash_JT_Eg_minimal-output.pdf
			\input{\tikzexternal@filenameprefix1.tikz}{0}{0}{TiF63-_AAdash_JT_Eg_minimal}
			\caption{
				$\mathrm{A}_{\frac{1}{2}}$ (dashed) and $\mathrm{A}'_{\frac{1}{2}}$ (dotted) solutions and the low-lying NOCI wavefunctions constructed from all of them (solid) along the $e_g$ normal vibrational coordinates of octahedral \ce{[TiF6]^{3-}}.
				The zero values of $S_2$ and $S_3$ correspond to $\mathcal{O}_h$ while the non-zero values correspond to the lower symmetry groups.
				All symmetry symbols are irreducible representations of the lower symmetry group unless indicated otherwise.
				Spin multiplicities are only shown for NOCI wavefunctions.
				The $\prescript{2}{}{E}_g$ and $\prescript{2}{}{B}_{2g}$ NOCI curves in $\mathcal{D}_{4h}$ lie in the vicinity of \SI{-1444.9328}{\hartree} (barely visible in the $Q_{e_gu}$ graph) for $S_2 \in (\SI{-0.5e-3}{\angstrom},\SI{0.5e-3}{\angstrom})$ but exhibit discontinuities at $S_2 = \pm \SI{0.5e-3}{\angstrom}$.
			}
			\label{fig:TiF6AAdashJT}
		\end{figure*}

	\ifdefined\pdfoutput
	\else
		\clearpage
	\fi
	
	In general, symmetry-broken SCF solutions, although better variationally, do not exhibit the correct splitting consistent with the physical distortions required by the Jahn--Teller theorem.
	This is most easily seen in Figure~\ref{fig:VF6AAdashJT} for \ce{[VF6]^{3-}}: along both $Q_{e_gu}$ and $Q_{e_gv}$, neither $\mathrm{A}_1$ or $\mathrm{A}'_1$ show any appreciable minima that can be attributed to the expected Jahn--Teller stabilization effects.
	However, when symmetry is restored using NOCI, these effects are recovered.
	In particular, the $\prescript{3}{}{T}_{1g}(\mathcal{O}_h)[\mathrm{A}_1 \oplus \mathrm{A}'_1]$ NOCI wavefunctions now exhibit the correct splitting in $\mathcal{D}_{4h}$ and $\mathcal{D}_{2h}$.
	In addition, the $\prescript{3}{}{A}_{2g}[\mathrm{A}_1 \oplus \mathrm{A}'_1]$ component in $\mathcal{D}_{4h}$ has a deeper minimum than $\prescript{3}{}{E}_{g}[\mathrm{A}_1 \oplus \mathrm{A}'_1]$ which suggests that axial elongation is preferred along $Q_{e_gu}$ and which is consistent with the fact that $A_{2g}$ has had all of the degeneracy of $T_{1g}$ lifted while $E_g$ still retains a double degeneracy.
	
	Unfortunately, similar observations for the symmetry-broken $\mathrm{A}_\frac{1}{2}$ solutions in \ce{[TiF6]^{3-}} cannot be made.
	As seen in Figure~\ref{fig:TiF6AAdashJT}, along both $Q_{e_gu}$ and $Q_{e_gv}$, the components arising from $\mathrm{A}_\frac{1}{2}$ quickly disappear upon coalescence with those from $\mathrm{A}'_\frac{1}{2}$ that share one or more irreducible representations.
	In particular, along $Q_{e_gu}$ in $\mathcal{D}_{4h}$, the $A_{2g} \oplus B_{2g} \oplus E_g$ components of the $\mathrm{A}_\frac{1}{2}$ solutions coalesce with both $B_{2g}$ and $E_g$ components of $\mathrm{A}'_\frac{1}{2}$.
	Similarly, along $Q_{e_gv}$ in $\mathcal{D}_{2h}$, the $B_{1g} \oplus B_{2g}$ components of $\mathrm{A}_\frac{1}{2}$ can cross the $B_{3g}$ component of $\mathrm{A}'_\frac{1}{2}$ easily but disappear upon approaching the $B_{1g}$ and $B_{2g}$ components.
	It is possible that the disappearing wavefunctions have become holomorphic beyond the coalescence points\cite{article:Hiscock2014,article:Burton2016}, a behavior we seek to investigate in a future study.
	However, at the moment, the disappearance of these components causes the set of SCF determinants used for NOCI to have varying dimensionality along $Q_{e_gu}$ and $Q_{e_gv}$.
	As a consequence, the NOCI energy curves exhibit discontinuities and are therefore not physically meaningful.
	We demonstrate this for the $Q_{e_gu}$ distortion in Figure~\ref{fig:TiF6AAdashJT}.
	When $S_2$ lies within the open interval $(\SI{-0.5e-3}{\angstrom},\SI{0.5e-3}{\angstrom})$, the $A_{2g} \oplus B_{2g} \oplus E_g$ components of $\mathrm{A}_\frac{1}{2}$ still exist and are linearly independent of all the other components from $\mathrm{A}_\frac{1}{2}$ and $\mathrm{A}'_\frac{1}{2}$.
	The NOCI space $[\mathrm{A}_\frac{1}{2} \oplus \mathrm{A}'_\frac{1}{2}]$ therefore remains nine-dimensional and the lowest $\prescript{2}{}{E}_g$ and $\prescript{2}{}{B}_{2g}$ NOCI wavefunctions lie in the vicinity of \SI{-1444.9328}{\hartree} (barely visible in Figure~\ref{fig:TiF6AAdashJT}).
	However, when $S_2$ lies outside of $(\SI{-0.5e-3}{\angstrom},\SI{0.5e-3}{\angstrom})$, the $A_{2g} \oplus B_{2g} \oplus E_g$ components of $\mathrm{A}_\frac{1}{2}$ no longer exist and the NOCI space $[\mathrm{A}_\frac{1}{2} \oplus \mathrm{A}'_\frac{1}{2}]$ becomes five-dimensional.
	The surviving $E_g$ components of $\mathrm{A}_\frac{1}{2}$ can still interact with the $E_g$ components of $\mathrm{A}'_\frac{1}{2}$ to give the part of the $\prescript{2}{}{E}_g$ NOCI curve outside of $(\SI{-0.5e-3}{\angstrom},\SI{0.5e-3}{\angstrom})$, but the $B_{2g}$ component of $\mathrm{A}'_\frac{1}{2}$ has nothing to interact with and is thus identical to the part of the $\prescript{2}{}{B}_{2g}$ NOCI curve on the same domain.
	The unphysical jumps exhibited by the $\prescript{2}{}{E}_g$ and $\prescript{2}{}{B}_{2g}$ NOCI curves as $S_2$ crosses $\pm \SI{0.5e-3}{\angstrom}$ are clearly an unphysical artifact of the non-inclusion of the $A_{2g} \oplus B_{2g} \oplus E_g$ wavefunctions from the NOCI space.

%% file: conclusion/conclusion.tex
\section{Conclusion}
\label{sec:conclusion}

	In this paper, we demonstrate the existence and analyze the properties of multiple low-energy SCF HF solutions in two representative octahedral hexafluoridometallate(III) complexes, \ce{[MF6]^{3-}} (\ce{M = Ti, V}).
	The presence of $d$ electrons gives rise to many UHF solutions that are degenerate or nearly degenerate, and that can conserve or break symmetry without any apparent predictable patterns.
	We therefore focus particularly on the physical meaning of symmetry-broken solutions and investigate how a careful choice of basis for NOCI can yield wavefunctions that restore symmetry, recover a decent amount of static correlation, and give reasonable descriptions of ground and excited electronic states.
	
	In \ce{[TiF6]^{3-}} where there is only a single $d$ electron in the ground configuration, the only correlation involving this electron is the correlation with the \ce{[Ar]} core electrons on \ce{Ti^{3+}} and the ligand electrons on \ce{F-}.
	This gives rise to one set of symmetry-broken UHF solutions that we label $\mathrm{A}_\frac{1}{2}$, and by restoring the symmetry of these solutions, we obtain NOCI wavefunctions that incorporate this correlation and are thus a variationally better description of the $\prescript{2}{}{T}_{2g}$ ground term.
	
	In \ce{[VF6]^{3-}} where there are two $d$ electrons in the ground configuration, on top of the correlation with the core and ligand electrons which is likely to be weak, there are now strong correlation effects between these two electrons.
	As a consequence, the number of low-lying UHF solutions that can be found increases significantly and many of them turn out to break spin or spatial symmetry, if not both.
	All symmetry breaking due to correlation between the two $d$ electrons also causes the core and ligand electrons to relax accordingly due to the self consistency of the SCF method and thereby inadvertently incorporates some correlation to these electrons.
	More importantly, however, in several cases, symmetry breaking is a direct consequence of the incorporation of inherently multi-determinantal correlation effects due to configuration mixing into single-determinantal wavefunctions, and the restoration of such symmetry allows for configuration mixing to be reasonably described using a single set of degenerate single determinants without having to explicitly construct or obtain determinants belonging to the various interacting configurations.
	
	An investigation of the behaviors of the lowest symmetry-broken UHF solutions in \ce{[VF6]^{3-}} upon molecular distortions from the $\mathcal{O}_h$ symmetry to $\mathcal{D}_{4h}$ and $\mathcal{D}_{2h}$ shows that the symmetry of these solutions must be restored using NOCI in order to obtain wavefunctions that can describe stabilization effects dictated by the Jahn--Teller theorem.
	However, a similar investigation for \ce{[TiF6]^{3-}} runs into difficulties as some of the UHF solutions disappear in $\mathcal{D}_{4h}$ and $\mathcal{D}_{2h}$, rendering the NOCI energy curves discontinuous and unphysical.
	It is likely that these solutions have become holomorphic, and if one can locate them beyond the coalescence points as reported elsewhere for various simpler systems\cite{article:Burton2016}, one can obtain continuous and smooth NOCI energy curves that are physically meaningful.
	
	Even though this paper focuses solely on two simple systems with high molecular symmetry, the insights learned from their results, in particular the use of NOCI to obtain physically meaningful wavefunctions from symmetry-broken SCF solutions and the analysis of correlation using natural orbitals and occupation numbers, are general and can be applied to any system.
	And even though NOCI wavefunctions are by no means the best approximations to electronic states as they are likely to still miss out dynamic correlation, the fact that they have the right symmetry and exhibit physically proper behaviors bolsters their applicability as references for more involved correlation methods.

\section*{Acknowledgment}

	B.C.H. is grateful for the financial support from the Jardine Foundation, Cambridge Commonwealth, European \& International Trust, and Peterhouse during the duration of this work.
	A.J.W.T. thanks the Royal Society for a University Research Fellowship (UF110161).

%% file: apps/grouptheory.tex
\tikzsetexternalprefix{./apps/tikz/}

\section{Group-Theoretical Aspects}
\label{app:gtaspects}

	\subsection{Structure of the Time-Reversal Group}
	\label{appsubsec:timerevgroup}
	
		Let $\mathcal{T}$ be the group generated by the time-reversal operator $\hat{\Theta}$.
		The behavior of any spin-pure state with respect to $\hat{\Theta}$ has long been classified by its eigenvalue under $\hat{\Theta}^2$ which is well known\cite{book:Wigner1959,article:Mead1979,article:Stedman1980} to take on only two possible values, $\pm 1$.
		However, the possibility of $\hat{\Theta}^2$ reversing the sign of certain states means that $\hat{\Theta}^2$ cannot be considered to coincide with the identity $\hat{E}$ but must be treated as a generally distinct operation that satisfies
			\begin{equation*}
				(\hat{\Theta}^2)^2 = \hat{\Theta}^4 = \hat{E}
			\end{equation*}
		$\mathcal{T}$ is therefore isomorphic to the cyclic group of order $4$ which is abelian and has four one-dimensional irreducible representations.
		The characters of the irreducible representations in $\mathcal{T}$ are shown in Table~\ref{tab:Tcharactertable} which is expectedly identical to that of $\mathcal{C}_4$\cite{article:Fukutome1981,book:Sugano1970}.
		The time-reversal group $\mathcal{T}$ that we identify here is the same as that deduced by Fukutome\cite{article:Fukutome1981} but we have explicitly written the square of the time-reversal operator as $\hat{\Theta}^2$ instead of $-\hat{E}$ so as to be completely general and also to reveal its cyclic nature.
		
			\begin{table}
				\centering
				\caption{Character table for $\mathcal{T}$.}
				\label{tab:Tcharactertable}
				\begin{tabular}[t]{
						>{\raggedright\arraybackslash}m{0.3cm} l@{\,} r | *{4}{c}
					}
					\toprule
					\multicolumn{3}{c |}{$\mathcal{T}$} & {$\hat{E}$} & {$\hat{\Theta}$} & {$\hat{\Theta}^2$} & {$\hat{\Theta}^3$} \\
					\midrule
					$A$ && $\Gamma_1$ & $1$ & $\hphantom{-}1$ & $\hphantom{-}1$ & $\hphantom{-}1$ \\
					$B$ && $\Gamma_2$ & $1$ & $-1$ & $\hphantom{-}1$ & $-1$ \\
					\multirow{2}{*}{$E$} & \ldelim\{{2}{*}[] & $\Gamma_3$ & $1$ & $\hphantom{-}i$ & $-1$ & {$-i$} \\
					&& $\Gamma_4$ & $1$ & $-i$ & $-1$ & $\hphantom{-}i$ \\
					\bottomrule
				\end{tabular}
			\end{table}

	\subsection{\boldmath Symmetry Breaking and the Generators of $\mathcal{S} \otimes \mathcal{T}$}
	\label{appsubsec:symbreakingTSgroup}
	
		The following results are well known in the theories of Lie groups, Lie algebras, and representations.
		However, we feel that it is in the interest of rigor and completeness to recapitulate these results in order to bridge the gap between our definition of symmetry breaking (a group representation concept) and the structure of the generators of $\mathcal{S} \otimes \mathcal{T}$ (an algebra representation concept).
		We focus first on the full spin-rotation group $\mathcal{S}$ which is isomorphic to the two-dimensional unitary group $\mathsf{SU}(2)$, a matrix Lie group containing all complex $2 \times 2$ unitary matrices with determinant $1$.
		Let $V_n$ denote the space of homogeneous polynomials with total degree $n$ ($n \geq 0$) in two complex variables $z_1$, $z_2$ that define a column vector $\boldsymbol{z} \in \mathbb{C}^2$,
			\begin{equation*}
				f(\boldsymbol{z}) = f(z_1, z_2) = \sum_{k=0}^{n} a_k z_1^{n-k} z_2^k
			\end{equation*}
		where the $a_k$'s are complex coefficients, then $V_n$ is a complex vector space of dimension $n+1$.
		There exists a Lie group representation $\Pi_n: \mathsf{SU}(2) \rightarrow \mathsf{GL}(V_n)$ which is a Lie group homomorphism acting on this space and given by
			\begin{equation}
				[\Pi_n(\boldsymbol{U})f](\boldsymbol{z}) %
					= f(\boldsymbol{U}^{-1}\boldsymbol{z})
					\label{eq:SU2rep}
			\end{equation}
		that maps any element $\boldsymbol{U}$ of $\mathsf{SU}(2)$ to an $(n+1) \times (n+1)$ invertible complex matrix in $\mathsf{GL}(V_n)$.
		$\Pi_n$ is an irreducible representation of $\mathsf{SU}(2)$\cite{book:Hall2015}.
		Thus, from our perspective, the functions $f(z_1, z_2)$ in $V_n$ are symmetry-conserved in $\mathsf{SU}(2)$ as they form a basis for a single irreducible representation $\Pi_n$ in the group.
		
		We now make the connection to angular momentum operators via the tangent space at the identity of the group.
		Known as the Lie algebra of the group and denoted by $\mathfrak{su}(2)$, this space contains all real-coefficient linear combinations of the traceless skew-Hermitian matrices $\lbrace \boldsymbol{X}_1, \boldsymbol{X}_2, \boldsymbol{X}_3 \rbrace$ satisfying\cite{book:Pfeifer2003,book:Woit2017}
			\begin{equation}
				[\boldsymbol{X}_i, \boldsymbol{X}_j] \equiv \boldsymbol{X}_i\boldsymbol{X}_j - \boldsymbol{X}_j\boldsymbol{X}_i = \sum_{k=1}^{3} \epsilon_{ijk}\boldsymbol{X}_k
				\label{eq:su2liebracket}
			\end{equation}
		where the Lie bracket $[\cdot,\cdot]: \mathfrak{su}(2) \times \mathfrak{su}(2) \rightarrow \mathfrak{su}(2)$ is a map defined in this space as the commutator of square matrices and obeys the properties of a Lie algebra (see Section~3.1 of Ref.~\citenum{book:Hall2015}) by virtue of the Levi-Civita symbol $\epsilon_{ijk}$.
		Elements of $\mathfrak{su}(2)$ are infinitesimal generators of $\mathsf{SU}(2)$.
		By Theorem~3.28 in Ref.~\citenum{book:Hall2015}, every Lie group representation gives rise to a unique Lie algebra representation.
		Therefore, for each of the $\Pi_n$ acting on the space $V_n$, we can find a corresponding Lie algebra representation $\pi_n: \mathfrak{su}(2) \rightarrow \mathfrak{gl}(V_n)$ acting on the same space.
		Physically illuminating forms for $\pi_n$ can be obtained by choosing an appropriate basis for $\mathfrak{su}(2)$.
		One such basis is given in terms of the Pauli spin matrices as
			\begingroup
			\renewcommand\arraystretch{0.65}
			\begin{align*}
					\boldsymbol{X}_1 &= -\frac{i}{2} \begin{pmatrix*}[r] 0 & 1 \\ 1 & 0 \end{pmatrix*} \\
					\boldsymbol{X}_2 &= -\frac{i}{2} \begin{pmatrix*}[r] 0 & -i \\ i & 0 \end{pmatrix*} \\
					\boldsymbol{X}_3 &= -\frac{i}{2} \begin{pmatrix*}[r] 1 & 0 \\ 0 & -1 \end{pmatrix*}
			\end{align*}
			\endgroup
		The Lie algebra representations $\pi_n$ for these elements are easily shown\cite{book:Woit2017} to be
			\begin{align*}
				\pi_n(\boldsymbol{X}_1) &= \frac{i}{2} \left(z_2\frac{\partial}{\partial z_1} + z_1 \frac{\partial}{\partial z_2}\right) \\
				\pi_n(\boldsymbol{X}_2) &= \frac{1}{2} \left(z_2\frac{\partial}{\partial z_1} - z_1 \frac{\partial}{\partial z_2}\right) \\
				\pi_n(\boldsymbol{X}_3) &= \frac{i}{2} \left(z_1\frac{\partial}{\partial z_1} - z_2 \frac{\partial}{\partial z_2}\right)
			\end{align*}
		It is trivial to see that any basis element $z_1^{n-k} z_2^k$ of $V_n$ is an eigenfunction of $\pi_n(\boldsymbol{X}_3)$ with eigenvalue $(i/2)(n-2k)$.
		But as $\boldsymbol{X}_1$, $\boldsymbol{X}_2$, and $\boldsymbol{X}_3$ do not commute with one another (Equation~\ref{eq:su2liebracket}), $\pi_n(\boldsymbol{X}_1)$, $\pi_n(\boldsymbol{X}_2)$, and $\pi_n(\boldsymbol{X}_3)$ do not commute with one another either and therefore $z_1^{n-k} z_2^k$ is not an eigenfunction of either $\pi_n(\boldsymbol{X}_1)$ or $\pi_n(\boldsymbol{X}_2)$.
		However, we can construct the Casimir operator in this representation, $\pi_n(\boldsymbol{C})$, as\cite{book:Hall2015}
			\ifdefined\twocolumnmode
				\begin{align*}
					\pi_n(\boldsymbol{C})%
						&= -\left[\pi_n(\boldsymbol{X}_1)^2 + \pi_n(\boldsymbol{X}_2)^2 + \pi(\boldsymbol{X}_3)^2\right]\\
						&= 
							\begin{multlined}[t]
								\frac{3}{4} \left(z_1\frac{\partial}{\partial z_1} + z_2\frac{\partial}{\partial z_2}\right) +%
								\frac{1}{2}z_1z_2\frac{\partial^2}{\partial z_1 \partial z_2} \\%
								+ \frac{1}{4}\left(z_1^2\frac{\partial^2}{\partial z_1^2} + z_2^2\frac{\partial^2}{\partial z_2^2}\right)
							\end{multlined}
				\end{align*}
			\else
				\begin{align*}
					\pi_n(\boldsymbol{C})%
						&= -\left[\pi_n(\boldsymbol{X}_1)^2 + \pi_n(\boldsymbol{X}_2)^2 + \pi(\boldsymbol{X}_3)^2\right]\\
						&= \frac{3}{4} \left(z_1\frac{\partial}{\partial z_1} + z_2\frac{\partial}{\partial z_2}\right) +%
						   \frac{1}{2}z_1z_2\frac{\partial^2}{\partial z_1 \partial z_2} +%
						   \frac{1}{4}\left(z_1^2\frac{\partial^2}{\partial z_1^2} + z_2^2\frac{\partial^2}{\partial z_2^2}\right)
				\end{align*}
			\fi
		which commutes with $\pi_n(\boldsymbol{X}_1)$, $\pi_n(\boldsymbol{X}_2)$, and $\pi_n(\boldsymbol{X}_3)$.
		Therefore, $z_1^{n-k} z_2^k$ is also an eigenfunction of $\pi_n(\boldsymbol{C})$ with eigenvalue $(n/2)[(n/2)+1]$, independent of $k$.
		
		As $\mathfrak{su}(2)$ contains only skew-Hermitian matrices, any linear combinations of $\boldsymbol{X}_i$ that involve complex coefficients must lie outside of this space.
		However, these combinations are needed to arrive at representations that can be identified to the familiar angular momentum operators.
		It is therefore necessary to consider the complexification of $\mathfrak{su}(2)$, namely $\mathfrak{sl}(2;\mathbb{C})$, the Lie algebra of the two-dimensional special linear group over the complex numbers.
		Proposition~4.6 in Ref.~\citenum{book:Hall2015} allows each Lie algebra representation $\pi_n$ of $\mathfrak{su}(2)$ to be uniquely extended to a Lie algebra representation of $\mathfrak{sl}(2;\mathbb{C})$, also denoted by $\pi_n$.
		By considering the following basis elements in $\mathfrak{sl}(2;\mathbb{C})$:
			\begingroup
			\renewcommand\arraystretch{0.65}
			\begin{alignat*}{3}
				&\boldsymbol{J}_+ &&= i\boldsymbol{X}_1 - \boldsymbol{X}_2 &&= \begin{pmatrix*}[r] 0 & 1 \\ 0 & 0 \end{pmatrix*} \\
				&\boldsymbol{J}_- &&= i\boldsymbol{X}_1 + \boldsymbol{X}_2 &&= \begin{pmatrix*}[r] 0 & 0 \\ 1 & 0 \end{pmatrix*} \\
				&\boldsymbol{J}_3 &&= i\boldsymbol{X}_3 &&= \frac{1}{2}\begin{pmatrix*}[r] 1 & 0 \\ 0 & -1 \end{pmatrix*}
			\end{alignat*}
			\endgroup
		we deduce the corresponding $\pi_n$\cite{book:Hall2015}:
			\ifdefined\twocolumnmode
				\begin{align*}
				\pi_n(\boldsymbol{J}_+)%
					&= i\pi_n(\boldsymbol{X}_1) - \pi_n(\boldsymbol{X}_2)\\
					&= -z_2 \frac{\partial}{\partial z_1} \\
				\pi_n(\boldsymbol{J}_-)%
					&= i\pi_n(\boldsymbol{X}_1) + \pi_n(\boldsymbol{X}_2)\\
					&= -z_1 \frac{\partial}{\partial z_2} \\
				\pi_n(\boldsymbol{J}_3)%
					&= i\pi_n(\boldsymbol{X}_3)\\
					&= -\frac{1}{2} \left(z_1\frac{\partial}{\partial z_1} - z_2 \frac{\partial}{\partial z_2}\right)
				\end{align*}
			\else
				\begin{alignat*}{3}
					&\pi_n(\boldsymbol{J}_+) &&= i\pi_n(\boldsymbol{X}_1) - \pi_n(\boldsymbol{X}_2) &&= -z_2 \frac{\partial}{\partial z_1} \\
					&\pi_n(\boldsymbol{J}_-) &&= i\pi_n(\boldsymbol{X}_1) + \pi_n(\boldsymbol{X}_2) &&= -z_1 \frac{\partial}{\partial z_2} \\
					&\pi_n(\boldsymbol{J}_3) &&= i\pi_n(\boldsymbol{X}_3) &&= -\frac{1}{2} \left(z_1\frac{\partial}{\partial z_1} - z_2 \frac{\partial}{\partial z_2}\right)
				\end{alignat*}
			\fi
		The basis element $z_1^{n-k} z_2^k$ of $V_n$ is now an eigenfunction of $\pi_n(\boldsymbol{J}_3)$ with eigenvalue $(1/2)(2k-n)$ but is transformed by $\pi_n(\boldsymbol{J}_{+})$ and $\pi_n(\boldsymbol{J}_{-})$ into other basis elements within $V_n$ that are also eigenfunctions of $\pi_n(\boldsymbol{J}_3)$ but with eigenvalues raised or lowered by $1$, respectively.
		
		We are now in a suitable position to make the following comments.
		First, we identify $\pi_n(\boldsymbol{C})$ with the $\hat{J}^2$ operator, $\pi_n(\boldsymbol{J}_3)$ with the $\hat{J}_3$ operator, and $\pi_n(\boldsymbol{J}_{\pm})$ with the ladder operators $\hat{J}_{\pm}$.
		The irreducible representation enumerator $n$ can then be equated to $2J$, and within each $V_n$, the basis element label $k$ can be equated to $J + M_J$.
		We noted earlier that the functions $f(z_1, z_2)$ in $V_n$ conserve symmetry in $\mathsf{SU}(2)$ as they form a basis for a single irreducible Lie group representation $\Pi_n$.
		As a consequence, they must also form a basis for the corresponding Lie algebra representation $\pi_n$ that diagonalize both  $\pi_n(\boldsymbol{C}) \equiv \hat{J}^2$ and $\pi_n(\boldsymbol{J}_3) \equiv \hat{J}_3$ simultaneously.
		Since $\mathsf{SU}(2)$ is simply connected, the converse must also be true\cite{book:Hall2015}: a complete set of simultaneous eigenfunctions of  $\hat{J}^2$ and $\hat{J}_3$ must conserve symmetry in $\mathsf{SU}(2)$.
		
		The relationship between symmetry breaking and the generator $\hat{\Theta}$ of the time-reversal group $\mathcal{T}$ is much simpler.
		Since $\mathcal{T}$ is cyclic, the character and also eigenvalue of $\hat{\Theta}$ in each one-dimensional irreducible representation is a distinct fourth root of unity (Table~\ref{tab:Tcharactertable}).
		Any state that conserves symmetry in $\hat{\Theta}$ must also be an eigenfunction of $\hat{\Theta}$ that corresponds to one of these eigenvalues, and \textit{vice versa}.
		
		By recognizing that all elements of $\mathsf{SU}(2)$ must take the form\cite{book:Hall2015}
			\begingroup
			\renewcommand\arraystretch{0.65}
			\begin{equation*}
				\boldsymbol{U} = \begin{pmatrix*}[r] U_{11} & U_{12} \\ -U_{12}^* & U_{11}^* \end{pmatrix*}
			\end{equation*}
			\endgroup
		where $|U_{11}|^2 + |U_{12}|^2 = 1$ and by using the definition of $\hat{\Theta}$ in (\ref{eq:optimerev}) together with the property that $\hat{\Theta}$ is antilinear\cite{article:Stedman1980}, we can easily show that $\hat{\Theta}$ commutes with the linear transformation $\Pi_n(\boldsymbol{U})$ in (\ref{eq:SU2rep}) for all $n$ and $\boldsymbol{U}$.
		The group $\mathcal{S} \otimes \mathcal{T}$ is indeed the direct product group of $\mathsf{SU}(2)$ and $\mathcal{T}$, and since there is a one-to-one correspondence between $\mathsf{SU}(2)$ and $\mathfrak{su}(2)$, the above discussion of symmetry breaking and group generators holds for all Fukutome subgroups of $\mathcal{S} \otimes \mathcal{T}$.

	\subsection{\boldmath Commutativity Between $\hat{\Theta}$ and $\hat{J}_3$ for $M_J = 0$}
	\label{appsubsec:commutativitythetaj3}
	
		The effects of $\hat{\Theta}$ and $\hat{J}_3$ on a state $\ket{J, M_J}$ that conserves symmetry in $\mathsf{SU}(2)$ are given by\cite{article:Stedman1980}
			\begin{align}
				\hat{\Theta} \ket{J, M_J} &= (-1)^{J-M_J} \ket{J, -M_J} \label{eq:optimerev} \\
				\hat{J}_3 \ket{J, M_J} &= M_J \ket{J, M_J}
			\end{align}
		which are a generalized version of (\ref{eq:optimerevS}) and (\ref{eq:opS3}) where $J$ is now used in place of $S$ to denote a general form of angular momentum.
		The effect of the commutator $[\hat{\Theta}, \hat{J}_3]$ on this state is therefore
			\ifdefined\twocolumnmode
				\begin{align}
					[\hat{\Theta}, \hat{J}_3] \ket{J, M_J}%
						&= (\hat{\Theta}\hat{J}_3 - \hat{J}_3\hat{\Theta}) \ket{J, M_J} \nonumber \\%
						&= \begin{multlined}[t]%
								\hat{\Theta} M_J \ket{J, M_J} \\ - \hat{J}_3 (-1)^{J-M_J} \ket{J, -M_J}
							 \end{multlined} \nonumber \\%
						&= \begin{multlined}[t]%
								(-1)^{J-M_J} M_J \ket{J, -M_J} \\ + (-1)^{J-M_J} M_J \ket{J, -M_J}
							 \end{multlined} \nonumber \\%
						&= 2 \times (-1)^{J-M_J} M_J \ket{J, -M_J}
						\label{eq:timerevszcommutator}
				\end{align}
			\else
				\begin{align}
					[\hat{\Theta}, \hat{J}_3] \ket{J, M_J}%
						&= (\hat{\Theta}\hat{J}_3 - \hat{J}_3\hat{\Theta}) \ket{J, M_J} \nonumber \\%
						&= \hat{\Theta} M_J \ket{J, M_J} - \hat{J}_3 (-1)^{J-M_J} \ket{J, -M_J} \nonumber \\%
						&= (-1)^{J-M_J} M_J \ket{J, -M_J} + (-1)^{J-M_J} M_J \ket{J, -M_J} \nonumber \\%
						&= 2 \times (-1)^{J-M_J} M_J \ket{J, -M_J}
					\label{eq:timerevszcommutator}
				\end{align}
			\fi
		Now, consider a non-vanishing wavefunction denoted by $\Psi_{M_J}$ that has a well defined $M_J$ but that is symmetry-broken in $\mathsf{SU}(2)$.
		$\Psi_{M_J}$ can be written as a sum over the various contributing symmetry-conserved states,
			\begin{equation*}
				\Psi_{M_J} = \sum_{J} c_J \ket{J, M_J} \quad (c_J \in \mathbb{C})
			\end{equation*}
		From (\ref{eq:timerevszcommutator}), and noting that $\hat{\Theta}$ is antilinear, the effect of $[\hat{\Theta}, \hat{J}_3]$ on this wavefunction is
			\begin{align}
				[\hat{\Theta}, \hat{J}_3] \Psi_{M_J}%
					&= \sum_{J} [\hat{\Theta}, \hat{J}_3] c_J \ket{J, M_J} \nonumber \\
					&= 2M_J \sum_{J} c^*_J (-1)^{J-M_J} \ket{J, -M_J}
			\end{align}
		which vanishes only when $M_J = 0$.	
		The general non-commutativity between $\hat{\Theta}$ and $\hat{J}_3$ versus the commutativity between $\hat{\Theta}$ and $\Pi_n(\boldsymbol{U})$ pointed out in \ref{appsubsec:symbreakingTSgroup} provides a clear example of how Lie group and Lie algebra representations can take on very different physical meanings and behaviors despite their close mathematical relationship.

%% file: apps/symanalysis.tex
\section{Symmetry Analysis}
\label{app:symanalysis}

	The matrix $\boldsymbol{X}$ in (\ref{eq:canonicalorthogonalizationofS}) that ensures that $\boldsymbol{\tilde{S}}$ is of full rank is defined as follows.
	If $N_\mathrm{indept} < N_\mathrm{det}$, \textit{i.e.}, the $N_\mathrm{det}$ degenerate Slater determinants are not all linearly independent, then $\boldsymbol{X}$ is constructed based on Equation~(3.172) of Ref.~\citenum{book:Szabo1996}.
	If $N_\mathrm{indept} = N_\mathrm{det}$, \textit{i.e.}, the $N_\mathrm{det}$ degenerate Slater determinants are already linearly independent, then we set $\boldsymbol{X} = \boldsymbol{I}_{N_\mathrm{det}}$ so as to avoid unnecessary mixing of the degenerate Slater determinants in the symmetry analysis that follows.
	Because of this, $\boldsymbol{\tilde{S}}$ is only guaranteed to be of full rank but not necessarily equal to $\boldsymbol{I}_{N_\mathrm{indept}}$.
	This choice of $\boldsymbol{X}$ does not affect the physics of the problem and still allows us to transform the full NOCI secular equation~(\ref{eq:nocisecular}) into an equivalent one in the linearly independent space,
		\begin{equation}
			\boldsymbol{\tilde{H}\tilde{A}} = \boldsymbol{\tilde{S}\tilde{A}E}
		\end{equation}
	where $\boldsymbol{\tilde{H}} = \boldsymbol{X^{\dagger}HX}$ and $\boldsymbol{A} = \boldsymbol{X\tilde{A}}$.
	The transformed Slater determinants in the linearly independent space are thus
		\begin{equation}
			\label{eq:Slaterdetsetlinindept}
			\left\lbrace
				\prescript{m}{}{\tilde{\Psi}} %
				\ | \ %
				m = 1, 2, \ldots, N_\mathrm{indept} %
			\right\rbrace 
		\end{equation}
	where
		\begin{equation}
			\label{eq:linindeptfromlindep}
			\prescript{m}{}{\tilde{\Psi}} = \sum_w^{N_\mathrm{det}} \prescript{w}{}{\Psi} X_{wm}
		\end{equation}
	and
		\begin{equation}
			\braket{\prescript{m}{}{\tilde{\Psi}}|\prescript{n}{}{\tilde{\Psi}}} = \tilde{S}_{mn}
		\end{equation}

	For a symmetry operation $\hat{R}$, we seek the representation matrix $\boldsymbol{\tilde{D}}(\hat{R})$ in the linearly independent space such that
		\begin{equation}
			\label{eq:repmatlinindeptdef}
			\hat{R} \prescript{m}{}{\tilde{\Psi}} = %
				\sum_n^{N_\mathrm{indept}} %
				\prescript{n}{}{\tilde{\Psi}} \tilde{D}_{nm}(\hat{R})
		\end{equation}
	The non-orthogonal projection operator due to \citeauthor{article:Soriano2014}\cite{article:Soriano2014} in this basis takes the form
		\begin{equation}
			\label{eq:projectionop}
			\hat{P}_m = \sum_n^{N_\mathrm{indept}} %
				\ket{\prescript{m}{}{\tilde{\Psi}}} \tilde{S}^{-1}_{mn} \bra{\prescript{n}{}{\tilde{\Psi}}}
		\end{equation}
	where $\tilde{S}^{-1}_{mn}$ are the matrix elements of $\boldsymbol{\tilde{S}}^{-1}$ which is guaranteed to exist since $\boldsymbol{\tilde{S}}$ contains no zero eigenvalues.
	Noting that the following property always holds\cite{article:Soriano2014},
		\begin{equation}
			\hat{P}_m \ket{\prescript{n}{}{\tilde{\Psi}}} = \delta_{mn} \ket{\prescript{m}{}{\tilde{\Psi}}}
		\end{equation}
	we apply the projection operator to both sides of (\ref{eq:repmatlinindeptdef}), relabeling the indices when necessary, and obtain
		\begin{align}
			\hat{P}_m \ket{\hat{R} \prescript{m'}{}{\tilde{\Psi}}} %
				&= \sum_{n'}^{N_\mathrm{indept}} \hat{P}_m \ket{\prescript{n'}{}{\tilde{\Psi}}} \tilde{D}_{n'm'}(\hat{R}) \nonumber \\%
				&= \sum_{n'}^{N_\mathrm{indept}} \delta_{mn'} \ket{\prescript{m}{}{\tilde{\Psi}}} \tilde{D}_{n'm'}(\hat{R}) \nonumber \\
				&= \ket{\prescript{m}{}{\tilde{\Psi}}} \tilde{D}_{mm'}(\hat{R})
		\end{align}
	Multiplying both sides by $\bra{\prescript{m}{}{\tilde{\Psi}}}$, we get
		\begin{equation}
			\braket{\prescript{m}{}{\tilde{\Psi}} | \hat{P}_m | \hat{R} \prescript{m'}{}{\tilde{\Psi}}} = \tilde{S}_{mm} \tilde{D}_{mm'}(\hat{R})
		\end{equation}
	With the definition of $\hat{P}_m$ in (\ref{eq:projectionop}), the above equation becomes
		\ifdefined\twocolumnmode
			\begin{align}
				\MoveEqLeft[3]{\tilde{D}_{mm'}(\hat{R})} \nonumber \\ %
					={}& \frac{1}{\tilde{S}_{mm}} \sum_n^{N_\mathrm{indept}} %
							\braket{\prescript{m}{}{\tilde{\Psi}} | \prescript{m}{}{\tilde{\Psi}}} %
							\tilde{S}^{-1}_{mn} %
							\braket{\prescript{n}{}{\tilde{\Psi}} | \hat{R} \prescript{m'}{}{\tilde{\Psi}}} \nonumber \\%
					={}& \sum_n^{N_\mathrm{indept}} %
							\tilde{S}^{-1}_{mn} %
							\tilde{T}_{nm'}(\hat{R})
							\label{eq:DtildeRdef}
			\end{align}
		\else
			\begin{align}
				\tilde{D}_{mm'}(\hat{R}) %
					&= \frac{1}{\tilde{S}_{mm}} \sum_n^{N_\mathrm{indept}} %
							\braket{\prescript{m}{}{\tilde{\Psi}} | \prescript{m}{}{\tilde{\Psi}}} %
							\tilde{S}^{-1}_{mn} %
							\braket{\prescript{n}{}{\tilde{\Psi}} | \hat{R} \prescript{m'}{}{\tilde{\Psi}}} \nonumber \\%
					&= \sum_n^{N_\mathrm{indept}} %
							\tilde{S}^{-1}_{mn} %
							\tilde{T}_{nm'}(\hat{R})
							\label{eq:DtildeRdef}
			\end{align}
		\fi
	where we have let $\tilde{T}_{nm'}(\hat{R}) = \braket{\prescript{n}{}{\tilde{\Psi}} | \hat{R} \prescript{m'}{}{\tilde{\Psi}}}$, which, by (\ref{eq:linindeptfromlindep}), can also be written as
		\begin{align}
			\tilde{T}_{nm'}(\hat{R}) %
				&= \sum_{wx}^{N_\mathrm{det}} \braket{\prescript{w}{}{\Psi} X_{wn} | \hat{R} \prescript{x}{}{\Psi} X_{xm'}} \nonumber \\%
				&= \sum_{wx}^{N_\mathrm{det}} X^\dagger_{nw} T_{wx}(\hat{R}) X_{xm'}
				\label{eq:TtildeRdef}
		\end{align}
	with $T_{wx}(\hat{R})$ defined in (\ref{eq:TwxR}).
	By substituting (\ref{eq:TtildeRdef}) into (\ref{eq:DtildeRdef}) and rewriting the result in matrix form, we complete the derivation for (\ref{eq:DtildeR}).
	
	In order to obtain the representation matrix of $\hat{R}$ in the basis of the NOCI wavefunctions given by (\ref{eq:nociwavefunction}), we first consider the NOCI wavefunctions in the linearly independent space,
		\begin{equation}
			\label{eq:nocilinindeptdef}
			\prescript{m}{}{\tilde{\Phi}} = \sum_{n}^{N_\mathrm{indept}} \prescript{n}{}{\tilde{\Psi}} \tilde{A}_{nm}
		\end{equation}
	which satisfy
		\begin{align}
			\tilde{S}^{\mathrm{NOCI}}_{mn} %
				&= \braket{\prescript{m}{}{\tilde{\Phi}} | \prescript{n}{}{\tilde{\Phi}}} \nonumber \\%
				&= \sum_{m'n'}^{N_\mathrm{indept}} \tilde{A}^{\dagger}_{mm'} \braket{\prescript{m'}{}{\tilde{\Psi}} | \prescript{n'}{}{\tilde{\Psi}}} \tilde{A}_{n'n} \nonumber \\%
				&= (\boldsymbol{\tilde{A}^{\dagger}\tilde{S}\tilde{A}})_{mn}
		\end{align}
	We seek $\boldsymbol{\tilde{D}}^{\mathrm{NOCI}}(\hat{R})$ such that
		\begin{equation}
			\label{eq:DtildeNOCIRdef}
			\hat{R} \prescript{m}{}{\tilde{\Phi}} = %
				\sum_{m'}^{N_\mathrm{indept}} \prescript{m'}{}{\tilde{\Phi}} \tilde{D}^{\mathrm{NOCI}}_{m'm}(\hat{R})
		\end{equation}
	To this end, we use the projection operator
		\begin{equation}
			\hat{P}^{\mathrm{NOCI}}_n = \sum_{n'}^{N_\mathrm{indept}} %
			\ket{\prescript{n}{}{\tilde{\Phi}}} (\tilde{S}^{\mathrm{NOCI}})^{-1}_{nn'} \bra{\prescript{n'}{}{\tilde{\Phi}}}
		\end{equation}
	and proceed as above to arrive at
		\begin{equation}
			\tilde{D}^{\mathrm{NOCI}}_{nm}(\hat{R}) = %
				\sum_{n'}^{N_\mathrm{indept}} (\tilde{S}^{\mathrm{NOCI}})^{-1}_{nn'} \braket{\prescript{n'}{}{\tilde{\Phi}} | \hat{R} \prescript{m}{}{\tilde{\Phi}}}
		\end{equation}
	which is equivalent to
		\begin{equation}
			\label{eq:DtildeNOCIR}
			\boldsymbol{\tilde{D}}^{\mathrm{NOCI}}(\hat{R}) = \boldsymbol{\tilde{A}}^{-1} \boldsymbol{\tilde{D}}(\hat{R}) \boldsymbol{\tilde{A}}
		\end{equation}
	by virtue of (\ref{eq:nocilinindeptdef}) and then (\ref{eq:repmatlinindeptdef}).
	
	To obtain $\boldsymbol{D}^{\mathrm{NOCI}}(\hat{R})$ in the basis of the NOCI wavefunctions given by (\ref{eq:nociwavefunction}), we substitute (\ref{eq:nocilinindeptdef}) into (\ref{eq:DtildeNOCIRdef}) to reintroduce the 
	transformed Slater determinants in the linearly independent space and then use (\ref{eq:linindeptfromlindep}) to go back to the original space of linearly dependent Slater determinants.
	This yields
		\begin{equation}
			\hat{R} \prescript{m}{}{\Phi} = \sum_{m'}^{N_\mathrm{indept}} \prescript{m'}{}{\Phi} \tilde{D}^{\mathrm{NOCI}}_{m'm}(\hat{R})
		\end{equation}
	which has to be equivalent to
		\begin{equation}
			\hat{R} \prescript{m}{}{\Phi} = \sum_{m'}^{N_\mathrm{indept}} \prescript{m'}{}{\Phi} D^{\mathrm{NOCI}}_{m'm}(\hat{R})
		\end{equation}
	by definition.
	The linear independence of $\prescript{m}{}{\Phi}$ requires that $\tilde{D}^{\mathrm{NOCI}}_{m'm}(\hat{R})$ and $D^{\mathrm{NOCI}}_{m'm}(\hat{R})$ be identical to each other.
	Thus, by substituting (\ref{eq:DtildeR}) into (\ref{eq:DtildeNOCIR}), we arrive at (\ref{eq:Dnoci}) and complete the derivation.

%% file: apps/symtransformation.tex
\section{Symmetry Transformation}
\label{app:symtrans}

	Let $\hat{R}$ be a spatial symmetry operation which must be unitary (see Section~2.7 of Ref.~\citenum{book:Altmann1986}).
	Applying $\hat{R}$ on $\chi_{i}\left(\boldsymbol{x}_i\right)$ defined in (\ref{eq:spinorb}) gives
		\begin{equation}
			\label{eq:Rchiintermediate}
			\hat{R}\chi_i(\boldsymbol{x}_i) =%
			\omega_{\cdot \delta}(s_i) \hat{R}\left[\varphi_{\cdot \mu}(\boldsymbol{r}_i)\right] G_i^{\delta\mu,\cdot}
		\end{equation}
	since $\hat{R}$ can only affect the spatial basis functions.
	As each spatial basis function is an atomic orbital centred on a particular nucleus in the system, say $A$, that is clamped fixed, we can write $\varphi_{\cdot \mu}(\boldsymbol{r}_i)$ as
		\begin{equation}
			\varphi_{\cdot \mu}(\boldsymbol{r}_i) = \tilde{\varphi}_{\cdot \mu}(\boldsymbol{r}_i - \boldsymbol{R}_A)
		\end{equation}
	where $\boldsymbol{R}_A$ is the position vector of nucleus $A$ in a space-fixed coordinate system and $\tilde{\varphi}$ denotes an atomic orbital.
	To make matters more illuminating, we rewrite the above as a convolution of an atomic orbital centred at the origin with the three-dimensional Dirac delta function shifted to the position of $A$:
		\begin{align}
			\varphi_{\cdot \mu}(\boldsymbol{r}_i) %
				&= \tilde{\varphi}_{\cdot \mu}(\boldsymbol{r}_i) * \delta(\boldsymbol{r}_i - \boldsymbol{R}_A) \nonumber \\
				&= \int %
						\tilde{\varphi}_{\cdot \mu}(\boldsymbol{r}')
						\delta\left[\boldsymbol{r}' - (\boldsymbol{r}_i - \boldsymbol{R}_A)\right]
						\D\boldsymbol{r}'
		\end{align}
	so that
		\begin{align}
			\MoveEqLeft[2] \hat{R} \left[\varphi_{\cdot \mu}(\boldsymbol{r}_i)\right] \nonumber \\%
				={}& \int %
						\hat{R} \tilde{\varphi}_{\cdot \mu}(\boldsymbol{r}')
						\hat{R} \delta\left[\boldsymbol{r}' - (\boldsymbol{r}_i - \boldsymbol{R}_A)\right]
						\hat{R} (\D\boldsymbol{r}') \nonumber \\
				={}& \int %
						\hat{R} \tilde{\varphi}_{\cdot \mu}(\boldsymbol{r}')
						\delta\left[\boldsymbol{r}' - \hat{R}(\boldsymbol{r}_i - \boldsymbol{R}_A)\right]
						\left|J_{\hat{R}}\right| \D\boldsymbol{r}' \label{eq:Rphi2} \\
				={}& \hat{R} \varphi_{\cdot \mu}\left[\hat{R}(\boldsymbol{r}_i - \boldsymbol{R}_A)\right] \label{eq:Rphi3}
		\end{align}
	which is the transformed atomic orbital centered at a possibly different nucleus onto which $A$ is mapped by the action of $\hat{R}$ on the nuclear framework.
	From (\ref{eq:Rphi2}) to (\ref{eq:Rphi3}), we have used the fact that $\hat{R}$ is a unitary transformation and the modulus of the corresponding Jacobian $J_{\hat{R}}$ must be $1$.
	Substituting this result back into (\ref{eq:Rchiintermediate}) gives (\ref{eq:Rchi}) and concludes the derivation.
	
	To determine how each atomic orbital $\tilde{\varphi}$ transforms under $\hat{R}$, we first need to convert any Cartesian Gaussian functions into pure and real spherical harmonics because the rotation-inversion transformation matrices in these bases can be generated with ease using the recursion method of \citeauthor{article:Ivanic1996,*article:Ivanic1998}\cite{article:Ivanic1996,*article:Ivanic1998}.
	Let $\tilde{\chi}_{l_\mathrm{c}}(\boldsymbol{x})$ be a fragment of the overall spin-orbital that is constructed from a single complete set of Cartesian Gaussian basis functions of degree $l_\mathrm{c}$ located on a single atom, $\left\lbrace\left(\tilde{\varphi}_{l_\mathrm{c}}\right)_{\cdot \mu}\right\rbrace$:
		\begin{equation}
		\label{eq:spatialsingleangmom}
		\tilde{\chi}_{l_\mathrm{c}}(\boldsymbol{x}) =
			\omega_{\cdot \delta}(s) %
			\left(\tilde{\varphi}_{l_\mathrm{c}}\right)_{\cdot \mu}(\boldsymbol{r}) %
			G^{\delta\mu,\cdot}_{l_\mathrm{c}}
		\end{equation}
	We cannot simply identify $l_\mathrm{c}$ with the angular momentum of an equivalent set of spherical harmonics because the set $\left\lbrace\left(\tilde{\varphi}_{l_\mathrm{c}}\right)_{\cdot \mu}\right\rbrace$ contains $\frac{1}{2}\left(l_\mathrm{c}+1\right)\left(l_\mathrm{c}+2\right)$ Cartesian Gaussians and therefore spans a larger space than the set of $2l_\mathrm{c}+1$ spherical harmonics of angular momentum $l_\mathrm{c}$ does for all $l_c \geq 2$ \cite{article:Schlegel1995}.
	In fact, since
		\begin{equation}
			\frac{1}{2}\left(l_\mathrm{c}+1\right)\left(l_\mathrm{c}+2\right) %
			= %
			\sum_{\substack{l \leq l_\mathrm{c} \\[3pt] l^P = l_\mathrm{c}^P}} \left(2l+1\right)
		\end{equation}
	where $l^P$ and $l_\mathrm{c}^P$ denote the parities of $l$ and $l_\mathrm{c}$, we must allow $\left(\tilde{\varphi}_{l_\mathrm{c}}\right)_{\cdot \mu}$ to be decomposed into all spherical harmonics of the same parity having angular momenta no greater than $l_\mathrm{c}$:
		\begin{equation}
			\left(\tilde{\varphi}_{l_\mathrm{c}}\right)_{\cdot \mu} = %
				r^{l_\mathrm{c}} \sum_{\substack{l \leq l_\mathrm{c} \\[3pt] l^P = l_\mathrm{c}^P}} \sum_{m} Y_{lm} W^{(ll_c)}_{m\mu}
		\end{equation}
	where $r$ is the radial distance, $Y_{lm}$ a real spherical harmonic function of degree $l$ and order $m$, and $\boldsymbol{W}^{(ll_c)}$ the corresponding transformation matrix with $\left(2l+1\right)$ rows and $\frac{1}{2}\left(l_\mathrm{c}+1\right)\left(l_\mathrm{c}+2\right)$ columns.
	The elements of $\boldsymbol{W}^{(ll_c)}$ and its inverse for any $l$ and $l_\mathrm{c}$ can be easily derived based on earlier work by \citeauthor{article:Schlegel1995} for the specific case $l=l_\mathrm{c}$\cite{article:Schlegel1995}.
	In real spherical harmonic bases, (\ref{eq:spatialsingleangmom}) thus takes the form
		\begin{equation}
			\label{eq:spatialsingleangmomexpanded}
			\tilde{\chi}_{l_\mathrm{c}} = %
			\omega_{\cdot \delta}(s) %
			\ %
			r^{l_\mathrm{c}} \sum_{\substack{l \leq l_\mathrm{c} \\[3pt] l^P = l_\mathrm{c}^P}} \sum_{m} Y_{lm} W^{(ll_c)}_{m\mu} %
			G^{\delta\mu,\cdot}_{l_\mathrm{c}}
		\end{equation}
	Since
		\begin{equation}
			\hat{R}Y_{lm} = \sum_{m'} Y_{lm'} D^{(l)}_{m'm}(\hat{R})
		\end{equation}
	where the representation matrices $\boldsymbol{D}^{(l)}(\hat{R})$ of any integral $l$ for any rotation-inversion operation $\hat{R}$ can be determined recursively\cite{article:Ivanic1996,*article:Ivanic1998}, applying $\hat{R}$ to both sides of (\ref{eq:spatialsingleangmomexpanded}) yields
		\ifdefined\twocolumnmode
			\begin{multline}
				\hat{R} \tilde{\chi}_{l_\mathrm{c}} = \\%
				\omega_{\cdot \delta}(s) %
				\ %
				r^{l_\mathrm{c}} \sum_{\substack{l \leq l_\mathrm{c} \\[3pt] l^P = l_\mathrm{c}^P}} \sum_{m\vphantom{m'}}\sum_{\vphantom{m}m'} Y_{lm'} D^{(l)}_{m'm}(\hat{R}) W^{(ll_c)}_{m\mu} %
				G^{\delta\mu,\cdot}_{l_\mathrm{c}}
			\end{multline}
		\else
			\begin{equation}
				\hat{R} \tilde{\chi}_{l_\mathrm{c}} = %
					\omega_{\cdot \delta}(s) %
					\ %
					r^{l_\mathrm{c}} \sum_{\substack{l \leq l_\mathrm{c} \\[3pt] l^P = l_\mathrm{c}^P}} \sum_{m\vphantom{m'}}\sum_{\vphantom{m}m'} Y_{lm'} D^{(l)}_{m'm}(\hat{R}) W^{(ll_c)}_{m\mu} %
					G^{\delta\mu,\cdot}_{l_\mathrm{c}}
			\end{equation}
		\fi
	A knowledge of $\boldsymbol{D}^{(l)}(\hat{R})$, $\boldsymbol{W}^{(ll_c)}$ and its inverse allows the coefficients of the transformed spin-orbitals expressed in the original basis to be worked out easily.

%% file: apps/nocinatorbs.tex
\section{NOCI Natural Orbitals}
\label{app:nocinatorbs}
	
	To determine the one-particle density matrices of the NOCI wavefunctions in (\ref{eq:nociwavefunction}), we invoke the one-particle density operator
		\begin{equation}
			\hat{\rho}(\boldsymbol{x}) = \sum_{i=1}^{N_\mathrm{e}} \delta (\boldsymbol{x}_i - \boldsymbol{x})
		\end{equation}
	where $\delta (\boldsymbol{x}_i - \boldsymbol{x})$ is the Dirac delta function. The one-particle density of the NOCI wavefunction $\prescript{m}{}{\Phi}$ is therefore
		\begin{align}
			\label{eq:nocidenmattransden}
			\prescript{m}{}{\rho\left(\boldsymbol{x}\right)} %
			&= \sum_{i=1}^{N_\mathrm{e}} \Braket{\prescript{m}{}{\Phi} | \delta (\boldsymbol{x}_i - \boldsymbol{x}) | \prescript{m}{}{\Phi}} \nonumber \\ %
			&= \sum_{wx}^{N_\mathrm{det}} A^{\vphantom{*}}_{xm} A^*_{wm} \prescript{wx}{}{\rho(\boldsymbol{x})}
		\end{align}
	where $\prescript{wx}{}{\rho(\boldsymbol{x})}$ is the transition density given by
		\begin{equation}
		\prescript{wx}{}{\rho(\boldsymbol{x})} = \sum_{i=1}^{N_\mathrm{e}} \Braket{\prescript{w}{}{\Psi} | \delta (\boldsymbol{x}_i - \boldsymbol{x}) | \prescript{x}{}{\Psi}}
		\end{equation}
	Noting that $\prescript{w}{}{\Psi}$ and $\prescript{x}{}{\Psi}$ are single Slater determinants, we can write
		\ifdefined\twocolumnmode
			\begin{multline}
				\label{eq:transdenfull}
				\prescript{wx}{}{\rho(\boldsymbol{x})} = \\%
					\sum_{i=1}^{N_\mathrm{e}} %
					\Braket{ %
						\prescript{w}{}{\Omega} %
						| \delta (\boldsymbol{x}_i - \boldsymbol{x}) | %
						\sum_{\sigma} (-1)^\sigma \hat{\mathscr{P}}_\sigma %
							\left(%
								\prescript{x}{}{\Omega} %
							\right)
					}
			\end{multline}
		\else
			\begin{equation}
				\label{eq:transdenfull}
				\prescript{wx}{}{\rho(\boldsymbol{x})} = %
					\sum_{i=1}^{N_\mathrm{e}} %
					\Braket{ %
						\prescript{w}{}{\Omega} %
						| \delta (\boldsymbol{x}_i - \boldsymbol{x}) | %
						\sum_{\sigma} (-1)^\sigma \hat{\mathscr{P}}_\sigma %
							\left(%
							\prescript{x}{}{\Omega} %
							\right)
					}
			\end{equation}
		\fi
	using the trick in Appendix M of Ref.~\citenum{book:Piela2013}, where
		\begin{align*}
			\prescript{w}{}{\Omega} = \prod_{j}^{N_\mathrm{e}} \prescript{w}{}{\chi}_{j}(\boldsymbol{x}_j), \qquad
			\prescript{x}{}{\Omega} = \prod_{k}^{N_\mathrm{e}} \prescript{x}{}{\chi}_{k}(\boldsymbol{x}_k)
		\end{align*}
	are the corresponding Hartree products of the two Slater determinants.
	Assuming that the two sets of spin-orbitals have been L\"{o}wdin paired\cite{article:Thom2009b,article:Lowdin1962,article:Mayer2010a} to become bi-orthogonal, \textit{i.e.},
		\begin{equation}
			\label{eq:Lowdinpairingoverlap}
			\braket{\prescript{w}{}{\chi}_{j}|\prescript{x}{}{\chi}_{k}} = \prescript{wx}{}{s}_j \delta_{jk} \qquad \textrm{(no sum over $j$)}
		\end{equation}
	which allows (\ref{eq:transdenfull}) to be reduced to a simple form,
		\begin{equation}
			\prescript{wx}{}{\rho(\boldsymbol{x})} = \sum_{i=1}^{N_\mathrm{e}} \; \left(\prod_{j \neq i}^{N_\mathrm{e}} \prescript{wx}{}{s_j}\right) \prescript{x}{}{\chi}_{i} (\boldsymbol{x}) \prescript{w}{}{\chi}^*_{i} (\boldsymbol{x})
		\end{equation}
	(\ref{eq:nocidenmattransden}) thus becomes
		\ifdefined\twocolumnmode
			\begin{multline}
			\prescript{m}{}{\rho(\boldsymbol{x})} = \\%
				\sum_{wx}^{N_\mathrm{det}} \sum_{i=1}^{N_\mathrm{e}} %
				A^{\vphantom{*}}_{xm} A^*_{wm} \; %
				\left(\prod_{j \neq i}^{N_\mathrm{e}} \prescript{wx}{}{s_j}\right)  \prescript{x}{}{\chi}_{i} (\boldsymbol{x}) \prescript{w}{}{\chi}^*_{i} (\boldsymbol{x})
			\end{multline}
		\else
			\begin{equation}
				\prescript{m}{}{\rho(\boldsymbol{x})} = %
					\sum_{wx}^{N_\mathrm{det}} \sum_{i=1}^{N_\mathrm{e}} %
					A^{\vphantom{*}}_{xm} A^*_{wm} \; %
					\left(\prod_{j \neq i}^{N_\mathrm{e}} \prescript{wx}{}{s_j}\right)  \prescript{x}{}{\chi}_{i} (\boldsymbol{x}) \prescript{w}{}{\chi}^*_{i} (\boldsymbol{x})
			\end{equation}
		\fi
	which upon expansion of the spin-orbitals in terms of the basis functions by (\ref{eq:spinorb}) yields the following expression for the one-particle density in the \emph{covariant} basis:
		\begin{equation*}
		\prescript{m}{}{\rho(\boldsymbol{x})} = \prescript{m}{}{P}^{\delta\mu,\varepsilon\nu} %
		\omega_{\cdot \delta}(s) \varphi_{\cdot \mu}(\boldsymbol{r}) %
		\omega^*_{\varepsilon \cdot}(s) \varphi^*_{\nu \cdot }(\boldsymbol{r}) %
		\end{equation*}
	where $\prescript{m}{}{P}^{\delta\mu,\varepsilon\nu}$ is an element of the one-particle density matrix in the \emph{contravariant} representation for the NOCI wavefunction $\prescript{m}{}{\Phi}$:
		\ifdefined\twocolumnmode
			\begin{multline}
				\label{eq:nocidenmat}
				\prescript{m}{}{P}^{\delta\mu,\varepsilon\nu} = \\%
					\sum_{wx}^{N_\mathrm{det}} \sum_{i=1}^{N_\mathrm{e}} %
					A^{\vphantom{*}}_{xm} A^*_{wm} \; %
					\left(\prod_{j \neq i}^{N_\mathrm{e}} \prescript{wx}{}{s_j}\right) %
					\prescript{x}{}{G}_i^{\delta\mu,\cdot} %
					\left(\prescript{w}{}{G}^\dagger_i\right)^{\cdot,\varepsilon\nu} %
			\end{multline}
		\else
			\begin{equation}
				\label{eq:nocidenmat}
				\prescript{m}{}{P}^{\delta\mu,\varepsilon\nu} = \\%
					\sum_{wx}^{N_\mathrm{det}} \sum_{i=1}^{N_\mathrm{e}} %
					A^{\vphantom{*}}_{xm} A^*_{wm} \; %
					\left(\prod_{j \neq i}^{N_\mathrm{e}} \prescript{wx}{}{s_j}\right) %
					\prescript{x}{}{G}_i^{\delta\mu,\cdot} %
					\left(\prescript{w}{}{G}^\dagger_i\right)^{\cdot,\varepsilon\nu} %
			\end{equation}
		\fi
		
	To obtain natural spin-orbital coefficients in the representation consistent with that in (\ref{eq:spinorb}), the density matrix must first be converted to the mixed representation before being diagonalized,
		\begin{equation}
			\label{eq:nocidenmatmixed}
			\prescript{m}{}{P}^{\delta\mu}_{\cdot,\varepsilon\nu} = \prescript{m}{}{P}^{\delta\mu,\zeta\xi}S^{\mathrm{AO}}_{\zeta\xi,\varepsilon\nu}
		\end{equation}
	where $S^{\mathrm{AO}}_{\zeta\xi,\varepsilon\nu}$ denotes the overlap of the underlying basis functions,
		\begin{equation}
			S^{\mathrm{AO}}_{\zeta\xi,\varepsilon\nu} = %
			\braket{\omega_{\zeta \cdot}(s) | \omega_{\cdot \varepsilon}(s)} %
			\braket{\varphi_{\xi \cdot}(\boldsymbol{r}) | \varphi_{\cdot \nu}(\boldsymbol{r})}
		\end{equation}
	The coefficients for the $i^\mathrm{th}$ natural spin-orbital of $\prescript{m}{}{\Phi}$ then satisfy
		\begin{equation}
			\prescript{m}{}{P}^{\delta\mu}_{\cdot,\varepsilon\nu} \prescript{m}{}{G}^{\varepsilon\nu}_{i} %
			= %
			\prescript{m}{}{\lambda}_i \prescript{m}{}{G}^{\delta\mu}_{i}
		\end{equation}
	where $\prescript{m}{}{\lambda}_i$ is the corresponding occupation number.

	By applying appropriate constraints, we can simplify (\ref{eq:nocidenmat}) and obtain expressions for the one-particle density matrix in RHF and UHF.
	Specifically, in UHF, we can write
		\begin{equation*}
			\prescript{m}{}{\rho(\boldsymbol{x})} = %
				\left(\prescript{m}{}{P}^{\mu\nu}_\alpha + \prescript{m}{}{P}^{\mu\nu}_\beta\right) %
				\varphi_{\cdot \mu}(\boldsymbol{r}) %
				\varphi^*_{\nu \cdot}(\boldsymbol{r})
		\end{equation*}
	where
		\ifdefined\twocolumnmode
			\begin{multline*}
				\prescript{m}{}{P}^{\mu\nu}_\alpha = \\%
					\sum_{wx}^{N_\mathrm{det}} \sum_{\substack{i=1 \\ \hphantom{i=N_\alpha+1}}}^{N_{\alpha}} %
					A^{\vphantom{*}}_{xm} A^*_{wm} \; %
					\left(\prod_{j\neq i}^{N_\mathrm{e}} \prescript{wx}{}{s}_j \right) %
					\prescript{x}{}{G}^{1\mu,\cdot}_{i} \left(\prescript{w}{}{G}_i^{\dagger} \right)^{\cdot,1\nu}
			\end{multline*}
		\else
			\begin{equation*}
				\prescript{m}{}{P}^{\mu\nu}_\alpha = %
					\sum_{wx}^{N_\mathrm{det}} \sum_{\substack{i=1 \\ \hphantom{i=N_\alpha+1}}}^{N_{\alpha}} %
					A^{\vphantom{*}}_{xm} A^*_{wm} \; %
					\left(\prod_{j\neq i}^{N_\mathrm{e}} \prescript{wx}{}{s}_j \right) %
					\prescript{x}{}{G}^{1\mu,\cdot}_{i} \left(\prescript{w}{}{G}_i^{\dagger} \right)^{\cdot,1\nu}
			\end{equation*}
		\fi
	and
		\ifdefined\twocolumnmode
			\begin{multline*}
				\prescript{m}{}{P}^{\mu\nu}_\beta = \\%
					\sum_{wx}^{N_\mathrm{det}} \sum_{i=N_\alpha+1}^{N_\beta} %
					A^{\vphantom{*}}_{xm} A^*_{wm} \; %
					\left(\prod_{j\neq i}^{N_\mathrm{e}} \prescript{wx}{}{s}_j \right) %
					\prescript{x}{}{G}^{2\mu,\cdot}_{i} \left(\prescript{w}{}{G}_i^{\dagger} \right)^{\cdot,2\nu}
			\end{multline*}
		\else
			\begin{equation*}
				\prescript{m}{}{P}^{\mu\nu}_\beta = %
					\sum_{wx}^{N_\mathrm{det}} \sum_{i=N_\alpha+1}^{N_\beta} %
					A^{\vphantom{*}}_{xm} A^*_{wm} \; %
					\left(\prod_{j\neq i}^{N_\mathrm{e}} \prescript{wx}{}{s}_j \right) %
					\prescript{x}{}{G}^{2\mu,\cdot}_{i} \left(\prescript{w}{}{G}_i^{\dagger} \right)^{\cdot,2\nu}
			\end{equation*}
		\fi
	which can be diagonalized in a similar manner to obtain the natural orbital coefficients and the corresponding occupation numbers.

%% file: TOC/TOC.tex
\tikzsetexternalprefix{./TOC/tikz/}

\clearpage
\begin{figure*}
	\centering
	% \useexternalfile{scale}{trimleft}{trimright}{name}
	% Figure compiled with tikzexternalize
	% Pre-compiled figure located at ./TOC/tikz/tikzoutput/TOC-output.pdf
	% The compiled figure has been converted to TIFF as per the editor's request.
	% The TIFF print size is 3.02815 in x 1.37444 in which should fit in the required area.
	% This is located at ./TOC/tikz/tikzoutput/TOC-tiff-small.tif.
	\input{\tikzexternal@filenameprefix1.tikz}{0}{0}{TOC}
	\caption*{For Table of Contents Only.}
\end{figure*}

%% file: suppinfo/compdetails/compdetails.tex
% !TeX root = symmultiplescf-suppinfo.tex
\renewcommand*{\codelocation}{suppinfo/compdetails/codesnippets}

\section{Converging into Multiple Solutions}

	Listings~\ref{listing:TiF63-.2.0274.uhf.6-31GSTAR.inp}--\ref{listing:VF63-.MS0.1.9896.uhf.6-31GSTAR.inp} show the \textsc{Q-Chem} input files for preliminary low-convergence metadynamics search of UHF solutions in \ce{[TiF6]^{3-}} and \ce{[VF6]^{3-}}.
	Each of the located solutions is then reconverged into with \qchemremvar{SCF\_CONVERGENCE} set to \texttt{13} to ensure that its DIIS error is smaller than \SI{1e-13}{}.
	
	\newpage
\begin{lstlisting}[
		label={listing:TiF63-.2.0274.uhf.6-31GSTAR.inp},
		language=QChemInput,
		caption={\textsc{Q-Chem} input for preliminary low-convergence metadynamics search of $M_S = \frac{1}{2}$ UHF solutions in \ce{[TiF6]^{3-}}.},
	]
$molecule
    -3 2
    --
    3 2
    Ti   0.0000000   0.0000000   0.0000000
    --
    -1 1
    F    0.0000000   0.0000000   2.0274000
    --
    -1 1
    F    0.0000000   0.0000000  -2.0274000
    --
    -1 1
    F    0.0000000   2.0274000   0.0000000
    --
    -1 1
    F    0.0000000  -2.0274000   0.0000000
    --
    -1 1
    F    2.0274000   0.0000000   0.0000000
    --
    -1 1
    F   -2.0274000   0.0000000   0.0000000
$end
$rem
    BASIS 6-31G*
    EXCHANGE hf
    UNRESTRICTED true
    SCF_ALGORITHM diis
    SCF_GUESS fragmo
    SCF_CONVERGENCE 7
    MAX_SCF_CYCLES 1000
    SYMMETRY off
    DIIS_SEPARATE_ERRVEC true
    SYM_IGNORE true
    MOM_START 1
$end
$rem_frgm
    SCF_CONVERGENCE 5
    THRESH 8
$end
@@@
$molecule
    read
$end
$rem
    BASIS 6-31G*
    EXCHANGE hf
    UNRESTRICTED true
    SCF_GUESS read
    SCF_ALGORITHM diis
    SCF_CONVERGENCE 7
    SCF_SAVEMINIMA 100
    SCF_MAX_CYCLES 1000000
    MOM_START 1
    PRINT_ORBITALS true
    SCF_MINFIND_INITNORM 00010
    SCF_MINFIND_INCREASEFACTOR 10800
    SCF_MINFIND_WELLTHRESH 6
    SCF_MINFIND_RANDOMMIXING -15708
    SCF_MINFIND_NRANDOMMIXES 2
    SCF_MINFIND_MIXMETHOD 1
    SCF_MINFIND_MIXENERGY 00200
    SCF_MINFIND_RESTARTSTEPS 1000
    SYMMETRY off
    DIIS_SEPARATE_ERRVEC true
    SYM_IGNORE true
$end
\end{lstlisting}
	
	\newpage
\begin{lstlisting}[
		label={listing:VF63-.MS1.1.9896.uhf.6-31GSTAR.inp},
		language=QChemInput,
		caption={\textsc{Q-Chem} input for preliminary low-convergence metadynamics search of $M_S = 1$ UHF solutions in \ce{[VF6]^{3-}}.},
	]
$molecule
    -3 3
    --
    3 3
    V   0.0000000    0.0000000   0.0000000
    --
    -1 1
    F   0.0000000    0.0000000   1.9896000
    --
    -1 1
    F   0.0000000    0.0000000  -1.9896000
    --
    -1 1
    F   0.0000000    1.9896000   0.0000000
    --
    -1 1
    F   0.0000000   -1.9896000   0.0000000
    --
    -1 1
    F   1.9896000    0.0000000   0.0000000
    --
    -1 1
    F  -1.9896000    0.0000000   0.0000000
$end
$rem
    BASIS 6-31G*
    EXCHANGE hf
    UNRESTRICTED true
    SCF_ALGORITHM diis
    SCF_GUESS fragmo
    SCF_CONVERGENCE 7
    MAX_SCF_CYCLES 1000
    SYMMETRY off
    DIIS_SEPARATE_ERRVEC true
    SYM_IGNORE true
    MOM_START 1
$end
$rem_frgm
    SCF_CONVERGENCE 3
    THRESH 6
$end
@@@
$molecule
    read
$end
$rem
    BASIS 6-31G*
    EXCHANGE hf
    UNRESTRICTED true
    SCF_GUESS read
    SCF_ALGORITHM diis
    SCF_CONVERGENCE 7
    SCF_SAVEMINIMA 700
    SCF_MAX_CYCLES 10000000
    MOM_START 1
    PRINT_ORBITALS true
    SCF_MINFIND_INITNORM 00010
    SCF_MINFIND_INCREASEFACTOR 10800
    SCF_MINFIND_WELLTHRESH 6
    SCF_MINFIND_RANDOMMIXING -15708
    SCF_MINFIND_NRANDOMMIXES 2
    SCF_MINFIND_MIXMETHOD 1
    SCF_MINFIND_MIXENERGY 00200
    SCF_MINFIND_RESTARTSTEPS 1000
    SYMMETRY off
    DIIS_SEPARATE_ERRVEC true
    SYM_IGNORE true
$end
\end{lstlisting}
	
	\newpage
\begin{lstlisting}[
		label={listing:VF63-.MS0.1.9896.uhf.6-31GSTAR.inp},
		language=QChemInput,
		caption={\textsc{Q-Chem} input for preliminary low-convergence metadynamics search of $M_S = 0$ UHF solutions in \ce{[VF6]^{3-}}.},
	]
$molecule
    -3 1
    --
    3 1
    V   0.0000000    0.0000000   0.0000000
    --
    -1 1
    F   0.0000000    0.0000000   1.9896000
    --
    -1 1
    F   0.0000000    0.0000000  -1.9896000
    --
    -1 1
    F   0.0000000    1.9896000   0.0000000
    --
    -1 1
    F   0.0000000   -1.9896000   0.0000000
    --
    -1 1
    F   1.9896000    0.0000000   0.0000000
    --
    -1 1
    F  -1.9896000    0.0000000   0.0000000
$end
$rem
    BASIS 6-31G*
    EXCHANGE hf
    UNRESTRICTED true
    SCF_ALGORITHM diis
    SCF_GUESS fragmo
    SCF_CONVERGENCE 7
    MAX_SCF_CYCLES 100000
    SCF_MINFIND_RESTARTSTEPS 10000
    SYMMETRY off
    DIIS_SEPARATE_ERRVEC true
    SYM_IGNORE true
    MOM_START 1
$end
$rem_frgm
    SCF_CONVERGENCE 6
    THRESH 9
$end
@@@
$molecule
    read
$end
$rem
    BASIS 6-31G*
    EXCHANGE hf
    UNRESTRICTED true
    SCF_GUESS read
    SCF_ALGORITHM diis
    SCF_CONVERGENCE 7
    SCF_SAVEMINIMA 500
    SCF_MAX_CYCLES 100000000
    MOM_START 1
    PRINT_ORBITALS true
    SCF_MINFIND_INITNORM 00010
    SCF_MINFIND_INCREASEFACTOR 10800
    SCF_MINFIND_WELLTHRESH 6
    SCF_MINFIND_RANDOMMIXING -15708
    SCF_MINFIND_NRANDOMMIXES 2
    SCF_MINFIND_MIXMETHOD 1
    SCF_MINFIND_MIXENERGY 00180
    SCF_MINFIND_RESTARTSTEPS 3000
    SYMMETRY off
    DIIS_SEPARATE_ERRVEC true
    SYM_IGNORE true
$end
\end{lstlisting}
	
	\clearpage

%% file: suppinfo/results-d1/results-d1-supp.tex
% !TeX root = symmultiplescf-suppinfo.tex

\section{\ce{[TiF6]^{3-}}: Detailed Results}

	\begin{table*}
		\centering
		\caption{Overlap matrices for degenerate, linearly independent UHF solutions in \ce{[TiF6]^{3-}}.}
		\label{tab:TiF63-ovlap}
		\footnotesize
		\begingroup
		\renewcommand\arraystretch{1.35}
		\scalebox{1}{
		\begin{subtable}[t]{\textwidth}
			\centering
			\caption{$\mathrm{A}_{\frac{1}{2}}$ solutions in \ce{[TiF6]^{3-}}.}
			\begin{tabular}[t]{%
					r |%
					*{6}{S[table-format=2.4, table-alignment=right]}
				}
				\toprule
				& $\mathrm{A}_{\frac{1}{2},1}$ & $\mathrm{A}_{\frac{1}{2},2}$ & $\mathrm{A}_{\frac{1}{2},3}$ & $\mathrm{A}_{\frac{1}{2},4}$ & $\mathrm{A}_{\frac{1}{2},5}$ & $\mathrm{A}_{\frac{1}{2},6}$ \\
				\midrule
				$\mathrm{A}_{\frac{1}{2},1}$ &  1.0000 & -0.4991 &  0.4991 & -0.4991 & -0.4991 &  0.0000 \\
				$\mathrm{A}_{\frac{1}{2},2}$ & -0.4991 &  1.0000 &  0.0000 & -0.4991 &  0.4991 &  0.4991 \\
				$\mathrm{A}_{\frac{1}{2},3}$ &  0.4991 &  0.0000 &  1.0000 & -0.4991 &  0.0000 & -0.4991 \\
				$\mathrm{A}_{\frac{1}{2},4}$ & -0.4991 & -0.4991 & -0.4991 &  1.0000 &  0.0000 & -0.4991 \\
				$\mathrm{A}_{\frac{1}{2},5}$ & -0.4991 &  0.4991 &  0.4991 &  0.0000 & 1.0000 & -0.4991 \\
				$\mathrm{A}_{\frac{1}{2},6}$ &  0.0000 &  0.4991 & -0.4991 & -0.4991 & -0.4991 & 1.0000 \\
				\bottomrule
			\end{tabular}
		\end{subtable}
		}
	
		\vspace{1cm}
		\scalebox{1}{
		\begin{subtable}[t]{.48\textwidth}
			\centering
			\caption{$\mathrm{A}'_{\frac{1}{2}}$ solutions in \ce{[TiF6]^{3-}}.}
			\begin{tabular}[t]{%
					r |%
					*{3}{S[table-format=2.4, table-alignment=right]}
				}
				\toprule
				& $\mathrm{A}'_{\frac{1}{2},1}$ & $\mathrm{A}'_{\frac{1}{2},2}$ & $\mathrm{A}'_{\frac{1}{2},3}$ \\
				\midrule
				$\mathrm{A}'_{\frac{1}{2},1}$ &  1.0000 &  0.0000 &  0.0000 \\
				$\mathrm{A}'_{\frac{1}{2},2}$ &  0.0000 &  1.0000 &  0.0000 \\
				$\mathrm{A}'_{\frac{1}{2},3}$ &  0.0000 &  0.0000 &  1.0000 \\
				\bottomrule
			\end{tabular}
		\end{subtable}
		}
		\hfill
		\scalebox{1}{
		\begin{subtable}[t]{.48\textwidth}
			\centering
			\caption{$\mathrm{B}_{\frac{1}{2}}$ solutions in \ce{[TiF6]^{3-}}.}
			\begin{tabular}[t]{%
					r |%
					*{3}{S[table-format=2.4, table-alignment=right]}
				}
				\toprule
				& $\mathrm{B}_{\frac{1}{2},1}$ & $\mathrm{B}_{\frac{1}{2},2}$ & $\mathrm{B}_{\frac{1}{2},3}$ \\
				\midrule
				$\mathrm{B}_{\frac{1}{2},1}$ &  1.0000 &  0.4988 & -0.4988 \\
				$\mathrm{B}_{\frac{1}{2},2}$ &  0.4988 &  1.0000 &  0.4988 \\
				$\mathrm{B}_{\frac{1}{2},3}$ & -0.4988 &  0.4988 &  1.0000 \\
				\bottomrule
			\end{tabular}
		\end{subtable}
		}
		\endgroup
	\end{table*}

	\begin{table}
		\centering
		\caption{
			Energies and $\langle \hat{S}^2 \rangle$ values of $M_S = \frac{1}{2}$ UHF solutions in \ce{[TiF6]^{3-}}.
		}
		\label{tab:d1_uhfenergy}
		\footnotesize
		\begingroup
		\renewcommand\arraystretch{1.35}
		\begin{tabular}[t]{%
			>{\raggedright\arraybackslash}m{0.9cm}%
			S[table-format=4.7, table-alignment=left]%
			S[table-format=1.4, table-alignment=left]%
		}
			\toprule
			$\Psi_\mathrm{UHF}$ &	{Energy/\si{\hartree}} & {$\langle \hat{S}^2 \rangle$} \\
			\midrule
			$\mathrm{A}_{\frac{1}{2}}$ & -1444.9306822 & 0.7522 \\
			$\mathrm{A}_{\frac{1}{2}}'$ & -1444.9306706 & 0.7522 \\
			$\mathrm{B}_{\frac{1}{2}}$ & -1444.8666826 & 0.7527 \\
			\bottomrule
		\end{tabular}
		\endgroup
	\end{table}

	\begin{table}
		\centering
		\caption{
			Energies and $\langle \hat{S}^2 \rangle$ values of $M_S = \frac{1}{2}$ NOCI wavefunctions in \ce{[TiF6]^{3-}}.
		}
		\label{tab:d1_nocienergy}
		\footnotesize
		\begingroup
		\renewcommand\arraystretch{1.35}
		\begin{tabular}[t]{
				>{\raggedright\arraybackslash}m{2.7cm}%
				S[table-format=4.7, table-alignment=left]%
				S[table-format=1.4, table-alignment=left]}
			\toprule
			$\Phi$ &	{Energy/\si{\hartree}} & {$\langle \hat{S}^2 \rangle$} \\
			\midrule
			$\prescript{2}{}{T}_{2g}[\mathrm{A}_{\frac{1}{2}}]$ & -1444.9326372 & 0.7512 \\
			$\prescript{\varnothing}{}{T}_{1g}[\mathrm{A}_{\frac{1}{2}}]$ & -1442.6252833 & 1.8403 \\
			\midrule
			$\prescript{2}{}{T}_{2g}[\mathrm{A}_{\frac{1}{2}}\oplus\mathrm{A}_{\frac{1}{2}}']$ & -1444.9327765 & 0.7512 \\
			$\prescript{\varnothing}{}{T}_{2g}[\mathrm{A}_{\frac{1}{2}}\oplus \mathrm{A}_{\frac{1}{2}}']$ & -1442.6931022 & 1.5549 \\
			$\prescript{\varnothing}{}{T}_{1g}[\mathrm{A}_{\frac{1}{2}}\oplus \mathrm{A}_{\frac{1}{2}}']$ & -1442.6252833 & 1.8403 \\
			\midrule
			$\prescript{2}{}{E}_g[\mathrm{B}_{\frac{1}{2}}]$ & -1444.8682433 & 0.7517 \\
			$\prescript{\varnothing}{}{A}_{2g}[\mathrm{B}_{\frac{1}{2}}]$ & -1442.9308067 & 2.0321 \\
			\bottomrule
		\end{tabular}
		\endgroup
	\end{table}

	\clearpage

%% file: suppinfo/results-d2/results-d2-supp.tex
% !TeX root = symmultiplescf-suppinfo.tex

\section{\ce{[(B^{3+})(q^-)6]}: Spin-Orbitals}

	%% 2e5d isosurface plots
	\begin{table*}
		\centering
		\caption{
			Representative isosurface plots for the Pipek--Mezey-localized spatial parts of the occupied spin-orbitals of the UHF solutions in the toy system \ce{[(B^{3+})(q^-)6]}.
			The point charges (not shown here) are located octahedrally on the Cartesian axes.
			In (\subref{subtab:2e5d_MS1}), both spin-orbitals $\chi_{1}$ and $\chi_{2}$ have $m_s = \frac{1}{2}$.
			In (\subref{subtab:2e5d_MS0}), $\chi_{1}$ has $m_s = \frac{1}{2}$ whereas $\bar{\chi}_{2}$ has $m_s = -\frac{1}{2}$ as indicated by the bar.
			The spin-orbitals of all solutions within one set have similar forms and are related by the symmetry operations of $\mathcal{O}_h$.
			Axis triad: red\textendash $x$; green\textendash $y$; blue\textendash $z$.
		}
		\label{tab:2e5d}
		\footnotesize
		\begin{subtable}[h]{.48\textwidth}
			\centering
			\caption{$M_S = 1$}
			\label{subtab:2e5d_MS1}
			\begingroup
			\renewcommand\arraystretch{0.8}
			\begin{tabular}[t]{>{\raggedright\arraybackslash}m{0.9cm} >{\centering\arraybackslash}m{1.9cm} >{\centering\arraybackslash}m{1.9cm}}
				\toprule
				$\Psi_\mathrm{UHF}$ & $\chi_{1}$ & $\chi_{2}$ \\
				\midrule
				$\mathrm{a}_1$ & \includegraphics[trim=0 5 0 5, clip=true]{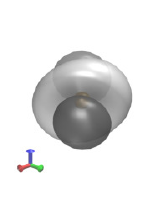} & \includegraphics[trim=0 5 0 5, clip=true]{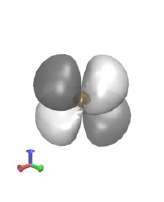} \\
				$\mathrm{a}'_1$ & \includegraphics[trim=0 5 0 5, clip=true]{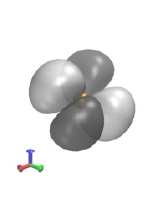} & \includegraphics[trim=0 5 0 5, clip=true]{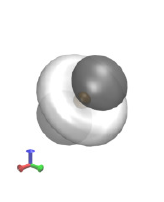} \\
				$\mathrm{b}_1$ & \includegraphics[trim=0 5 0 5, clip=true]{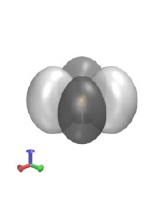} & \includegraphics[trim=0 5 0 5, clip=true]{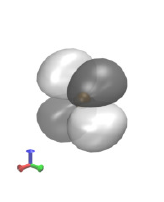} \\
				$\mathrm{c}_1$ & \includegraphics[trim=0 5 0 5, clip=true]{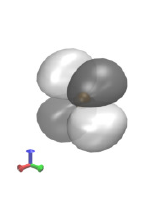} & \includegraphics[trim=0 5 0 5, clip=true]{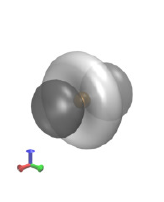} \\
				$\mathrm{d}_1$ & \includegraphics[trim=0 5 0 5, clip=true]{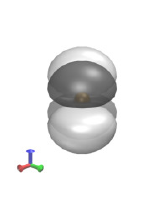} & \includegraphics[trim=0 5 0 5, clip=true]{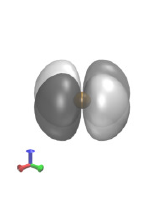} \\
				$\mathrm{e}_1$ & \includegraphics[trim=0 5 0 5, clip=true]{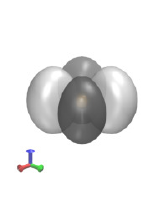} & \includegraphics[trim=0 5 0 5, clip=true]{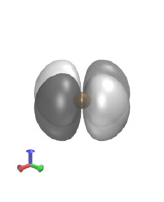} \\
				$\mathrm{f}_1$ & \includegraphics[trim=0 5 0 5, clip=true]{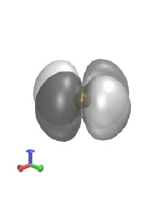} & \includegraphics[trim=0 5 0 5, clip=true]{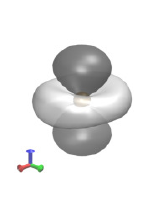} \\
				\bottomrule
			\end{tabular}
			\endgroup
		\end{subtable}
		\hfill
		\begin{subtable}[h]{.48\textwidth}
			\centering
			\caption{$M_S = 0$}
			\label{subtab:2e5d_MS0}
			\begingroup
			\renewcommand\arraystretch{0.8}
			\begin{tabular}[t]{>{\raggedright\arraybackslash}m{0.9cm} >{\centering\arraybackslash}m{1.9cm} >{\centering\arraybackslash}m{1.9cm}}
				\toprule
				$\Psi_\mathrm{UHF}$ & $\chi_{1}$ & $\bar{\chi}_{2}$ \\
				\midrule
				$\mathrm{a}^*_0$ & \includegraphics[trim=0 5 0 5, clip=true]{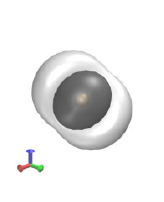} & \includegraphics[trim=0 5 0 5, clip=true]{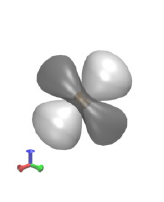} \\
				$\mathrm{a}'_0$ & \includegraphics[trim=0 5 0 5, clip=true]{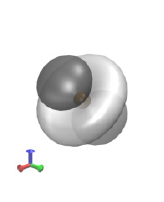} & \includegraphics[trim=0 5 0 5, clip=true]{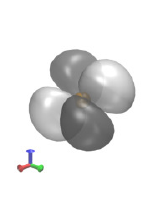} \\
				$\mathrm{a}''_0$ & \includegraphics[trim=0 5 0 5, clip=true]{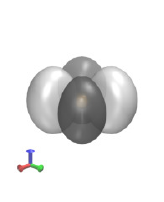} & \includegraphics[trim=0 5 0 5, clip=true]{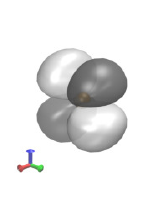} \\
				$\mathrm{b}_0$ & \includegraphics[trim=0 5 0 5, clip=true]{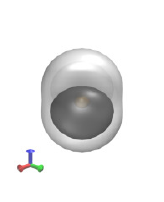} & \includegraphics[trim=0 5 0 5, clip=true]{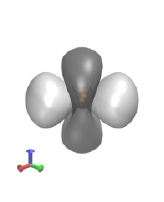} \\
				$\mathrm{c}_0$ & \includegraphics[trim=0 5 0 5, clip=true]{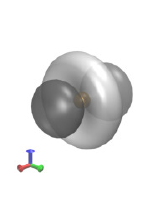} & \includegraphics[trim=0 5 0 5, clip=true]{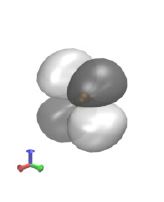} \\
				$\mathrm{d}_0$ & \includegraphics[trim=0 5 0 5, clip=true]{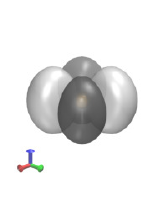} & \includegraphics[trim=0 5 0 5, clip=true]{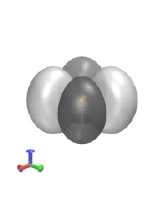} \\
				$\mathrm{e}_0$ & \includegraphics[trim=0 5 0 5, clip=true]{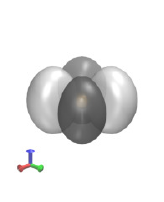} & \includegraphics[trim=0 5 0 5, clip=true]{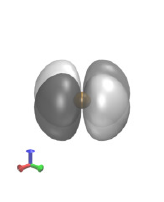} \\
				\bottomrule
			\end{tabular}
			\endgroup
		\end{subtable}
	\end{table*}

\clearpage

\section{\ce{[VF6]^{3-}}: Detailed Results}

	%%%%%%%%%%%%%%%%%%%%%%%%%%%%%%%%%%%%%%%%%
	% Overlap matrices
	%%%%%%%%%%%%%%%%%%%%%%%%%%%%%%%%%%%%%%%%%

	\begin{table*}
		\centering
		\caption{Overlap matrices for degenerate, linearly independent UHF solutions in \ce{[VF6]^{3-}}.}
		\label{tab:VF63-ovlap}
		\footnotesize
		\begingroup
		\renewcommand\arraystretch{1.35}
		\scalebox{1}{
			\begin{subtable}[t]{\textwidth}
				\centering
				\caption{$\mathrm{A}_1$ solutions in \ce{[VF6]^{3-}}.}
				\begin{tabular}[t]{%
						r |%
						*{4}{S[table-format=2.4, table-alignment=right]}
					}
					\toprule
					& $\mathrm{A}_{1,1}$ & $\mathrm{A}_{1,2}$ & $\mathrm{A}_{1,3}$ & $\mathrm{A}_{1,4}$ \\
					\midrule
					$\mathrm{A}_{1,1}$ &  1.0000 & -0.3318 & -0.3318 & 0.3318 \\
					$\mathrm{A}_{1,2}$ & -0.3318 &  1.0000 & -0.3318 & 0.3318 \\
					$\mathrm{A}_{1,3}$ & -0.3318 & -0.3318 &  1.0000 & 0.3318 \\
					$\mathrm{A}_{1,4}$ &  0.3318 &  0.3318 &  0.3318 & 1.0000 \\
					\bottomrule
				\end{tabular}
			\end{subtable}
		}
	
		\vspace{1cm}
		\scalebox{1}{
			\begin{subtable}[t]{1\textwidth}
				\centering
				\caption{$\mathrm{A}'_1$ solutions in \ce{[VF6]^{3-}}.}
				\begin{tabular}[t]{%
						r |%
						*{6}{S[table-format=2.4, table-alignment=right]}
					}
					\toprule
					& $\mathrm{A}'_{1,1}$ & $\mathrm{A}'_{1,2}$ & $\mathrm{A}'_{1,3}$ & $\mathrm{A}'_{1,4}$ & $\mathrm{A}'_{1,5}$ & $\mathrm{A}'_{1,6}$ \\
					\midrule
					$\mathrm{A}'_{1,1}$ &  1.0000 & -0.4851 & -0.4851 & -0.4851 & -0.4851 &  0.0000 \\
					$\mathrm{A}'_{1,2}$ & -0.4851 &  1.0000 &  0.0000 &  0.4851 & -0.4851 & -0.4851 \\
					$\mathrm{A}'_{1,3}$ & -0.4851 &  0.0000 &  1.0000 & -0.4851 &  0.4851 & -0.4851 \\
					$\mathrm{A}'_{1,4}$ & -0.4851 &  0.4851 & -0.4851 &  1.0000 &  0.0000 &  0.4851 \\
					$\mathrm{A}'_{1,5}$ & -0.4851 & -0.4851 &  0.4851 &  0.0000 &  1.0000 &  0.4851 \\
					$\mathrm{A}'_{1,6}$ &  0.0000 & -0.4851 & -0.4851 &  0.4851 &  0.4851 &  1.0000 \\
					\bottomrule
				\end{tabular}
			\end{subtable}
		}
		
		\vspace{1cm}
		\scalebox{1}{
			\begin{subtable}[t]{.48\textwidth}
				\centering
				\caption{$\mathrm{B}_{1}$ solutions in \ce{[VF6]^{3-}}.}
				\begin{tabular}[t]{%
						r |%
						*{3}{S[table-format=2.4, table-alignment=right]}
					}
					\toprule
					& $\mathrm{B}_{1,1}$ & $\mathrm{B}_{1,2}$ & $\mathrm{B}_{1,3}$ \\
					\midrule
					$\mathrm{B}_{1,1}$ &  1.0000 &  0.0000 &  0.0000 \\
					$\mathrm{B}_{1,2}$ &  0.0000 &  1.0000 &  0.0000 \\
					$\mathrm{B}_{1,3}$ &  0.0000 &  0.0000 &  1.0000 \\
					\bottomrule
				\end{tabular}
			\end{subtable}
		}
		\hfill
		\scalebox{1}{
			\begin{subtable}[t]{.48\textwidth}
				\centering
				\caption{$\mathrm{C}_{1}$ solutions in \ce{[VF6]^{3-}}.}
				\begin{tabular}[t]{%
						r |%
						*{3}{S[table-format=2.4, table-alignment=right]}
					}
					\toprule
					& $\mathrm{C}_{1,1}$ & $\mathrm{C}_{1,2}$ & $\mathrm{C}_{1,3}$ \\
					\midrule
					$\mathrm{C}_{1,1}$ &  1.0000 &  0.0000 &  0.0000 \\
					$\mathrm{C}_{1,2}$ &  0.0000 &  1.0000 &  0.0000 \\
					$\mathrm{C}_{1,3}$ &  0.0000 &  0.0000 &  1.0000 \\
					\bottomrule
				\end{tabular}
			\end{subtable}
		}
		\endgroup
	\end{table*}
		
	\begin{table*}
		\ContinuedFloat
		\footnotesize
		\begingroup
		\renewcommand\arraystretch{1.35}
		\scalebox{1}{
			\begin{subtable}[t]{1\textwidth}
				\centering
				\caption{$\mathrm{D}_1$ solutions in \ce{[VF6]^{3-}}.}
				\begin{tabular}[t]{%
						r |%
						*{6}{S[table-format=2.4, table-alignment=right]}
					}
					\toprule
					& $\mathrm{D}_{1,1}$ & $\mathrm{D}_{1,2}$ & $\mathrm{D}_{1,3}$ & $\mathrm{D}_{1,4}$ & $\mathrm{D}_{1,5}$ & $\mathrm{D}_{1,6}$ \\
					\midrule
					$\mathrm{D}_{1,1}$ &  1.0000 &  0.0000 &  0.2492 & -0.2492 & -0.2492 &  0.2492 \\
					$\mathrm{D}_{1,2}$ &  0.0000 &  1.0000 &  0.2492 & -0.2492 &  0.2492 & -0.2492 \\
					$\mathrm{D}_{1,3}$ &  0.2492 &  0.2492 &  1.0000 &  0.0000 & -0.2492 & -0.2492 \\
					$\mathrm{D}_{1,4}$ & -0.2492 & -0.2492 &  0.0000 &  1.0000 & -0.2492 & -0.2492 \\
					$\mathrm{D}_{1,5}$ & -0.2492 &  0.2492 &  0.2492 & -0.2492 &  1.0000 &  0.0000 \\
					$\mathrm{D}_{1,6}$ &  0.2492 & -0.2492 & -0.2492 & -0.2492 &  0.0000 &  1.0000 \\
					\bottomrule
				\end{tabular}
			\end{subtable}
		}
	
		\vspace{1cm}
		\scalebox{1}{
			\begin{subtable}[t]{.48\textwidth}
				\centering
				\caption{$\mathrm{E}_{1}$ solutions in \ce{[VF6]^{3-}}.}
				\begin{tabular}[t]{%
						r |%
						*{3}{S[table-format=2.4, table-alignment=right]}
					}
					\toprule
					& $\mathrm{E}_{1,1}$ & $\mathrm{E}_{1,2}$ & $\mathrm{E}_{1,3}$ \\
					\midrule
					$\mathrm{E}_{1,1}$ &  1.0000 &  0.0000 &  0.0000 \\
					$\mathrm{E}_{1,2}$ &  0.0000 &  1.0000 &  0.0000 \\
					$\mathrm{E}_{1,3}$ &  0.0000 &  0.0000 &  1.0000 \\
					\bottomrule
				\end{tabular}
			\end{subtable}
		}
		\scalebox{1}{
			\begin{subtable}[t]{.48\textwidth}
				\centering
				\caption{$\mathrm{F}_{1}$ solutions in \ce{[VF6]^{3-}}.}
				\begin{tabular}[t]{%
						r |%
						*{1}{S[table-format=2.4, table-alignment=right]}
					}
					\toprule
					& $\mathrm{F}_{1,1}$ \\
					\midrule
					$\mathrm{F}_{1,1}$ &  1.0000 \\
					\bottomrule
				\end{tabular}
			\end{subtable}
		}
		\endgroup
	\end{table*}

	\begin{landscape}
		\begin{table*}
			\ContinuedFloat
			\footnotesize
			\begingroup
			\renewcommand\arraystretch{1.35}
			\scalebox{0.6}{
				\begin{subtable}[t]{2.3\textwidth}
					\centering
					\caption{$\mathrm{A}_0$ solutions in \ce{[VF6]^{3-}}.}
					\begin{tabular}[t]{%
							r |%
							*{24}{S[table-format=2.4, table-alignment=right]}
						}
						\toprule
						& $\mathrm{A}_{0,1}$ & $\mathrm{A}_{0,2}$ & $\mathrm{A}_{0,3}$ & $\mathrm{A}_{0,4}$ & $\mathrm{A}_{0,5}$ & $\mathrm{A}_{0,6}$ & $\mathrm{A}_{0,7}$ & $\mathrm{A}_{0,8}$ & $\mathrm{A}_{0,9}$ & $\mathrm{A}_{0,10}$ & $\mathrm{A}_{0,11}$ & $\mathrm{A}_{0,12}$ & $\mathrm{A}_{0,13}$ & $\mathrm{A}_{0,14}$ & $\mathrm{A}_{0,15}$ & $\mathrm{A}_{0,16}$ & $\mathrm{A}_{0,17}$ & $\mathrm{A}_{0,18}$ & $\mathrm{A}_{0,19}$ & $\mathrm{A}_{0,20}$ & $\mathrm{A}_{0,21}$ & $\mathrm{A}_{0,22}$ & $\mathrm{A}_{0,23}$ & $\mathrm{A}_{0,24}$ \\
						\midrule
						$\mathrm{A}_{0,1}$ & 1.0000  & -0.1601 & 0.0000  & -0.5036 & -0.7469 & 0.2936  & -0.1601 & 0.2936  & -0.7469 & -0.0709 & 0.3182  & 0.3182  & 0.0156  & -0.2471 & -0.3182 & 0.3182  & -0.0156 & -0.2471 & 0.1221  & -0.0390 & -0.0390 & -0.1221 & 0.0390  & 0.0390  \\
						$\mathrm{A}_{0,2}$ & -0.1601 & 1.0000  & -0.1221 & -0.1221 & 0.0390  & 0.0390  & 0.0709  & -0.7469 & 0.2936  & 0.1601  & -0.0156 & -0.2471 & -0.3182 & 0.3182  & -0.3182 & 0.3182  & -0.2471 & -0.0156 & 0.5036  & -0.2936 & -0.0390 & 0.0000  & -0.7469 & 0.0390  \\
						$\mathrm{A}_{0,3}$ & 0.0000  & -0.1221 & 1.0000  & 0.0709  & -0.2471 & -0.0156 & -0.1221 & -0.0156 & -0.2471 & 0.5036  & 0.0390  & 0.0390  & -0.2936 & -0.7469 & -0.0390 & 0.0390  & 0.2936  & -0.7469 & 0.1601  & -0.3182 & -0.3182 & -0.1601 & 0.3182  & 0.3182  \\
						$\mathrm{A}_{0,4}$ & -0.5036 & -0.1221 & 0.0709  & 1.0000  & 0.3182  & 0.3182  & -0.1221 & 0.3182  & 0.3182  & 0.0000  & -0.7469 & 0.2936  & -0.0390 & 0.0390  & -0.2936 & -0.7469 & 0.0390  & 0.0390  & 0.1601  & 0.2471  & 0.0156  & -0.1601 & -0.0156 & -0.2471 \\
						$\mathrm{A}_{0,5}$ & -0.7469 & 0.0390  & -0.2471 & 0.3182  & 1.0000  & -0.1601 & 0.2936  & -0.3182 & 0.2471  & -0.0390 & -0.1221 & -0.5036 & 0.1221  & 0.0000  & 0.0390  & -0.2936 & -0.0390 & 0.7469  & 0.0156  & 0.1601  & 0.3182  & 0.3182  & 0.0709  & 0.0156  \\
						$\mathrm{A}_{0,6}$ & 0.2936  & 0.0390  & -0.0156 & 0.3182  & -0.1601 & 1.0000  & -0.7469 & 0.0156  & -0.3182 & -0.0390 & -0.5036 & -0.1221 & 0.0000  & -0.1221 & -0.7469 & -0.0390 & -0.2936 & -0.0390 & 0.2471  & -0.0709 & -0.2471 & 0.3182  & -0.1601 & -0.3182 \\
						$\mathrm{A}_{0,7}$ & -0.1601 & 0.0709  & -0.1221 & -0.1221 & 0.2936  & -0.7469 & 1.0000  & 0.0390  & 0.0390  & 0.1601  & 0.3182  & 0.3182  & 0.2471  & -0.0156 & 0.2471  & -0.0156 & 0.3182  & 0.3182  & 0.0000  & -0.0390 & 0.7469  & -0.5036 & 0.0390  & 0.2936  \\
						$\mathrm{A}_{0,8}$ & 0.2936  & -0.7469 & -0.0156 & 0.3182  & -0.3182 & 0.0156  & 0.0390  & 1.0000  & -0.1601 & -0.0390 & -0.0390 & 0.7469  & 0.2936  & -0.0390 & 0.1221  & -0.5036 & 0.0000  & -0.1221 & -0.3182 & 0.3182  & 0.1601  & -0.2471 & 0.2471  & 0.0709  \\
						$\mathrm{A}_{0,9}$ & -0.7469 & 0.2936  & -0.2471 & 0.3182  & 0.2471  & -0.3182 & 0.0390  & -0.1601 & 1.0000  & -0.0390 & -0.2936 & -0.0390 & 0.0390  & 0.7469  & 0.5036  & -0.1221 & -0.1221 & 0.0000  & -0.3182 & -0.0156 & -0.0709 & -0.0156 & -0.3182 & -0.1601 \\
						$\mathrm{A}_{0,10}$ & -0.0709 & 0.1601  & 0.5036  & 0.0000  & -0.0390 & -0.0390 & 0.1601  & -0.0390 & -0.0390 & 1.0000  & 0.2471  & 0.0156  & 0.3182  & -0.3182 & -0.0156 & 0.2471  & -0.3182 & -0.3182 & -0.1221 & -0.7469 & 0.2936  & 0.1221  & -0.2936 & 0.7469  \\
						$\mathrm{A}_{0,11}$ & 0.3182  & -0.0156 & 0.0390  & -0.7469 & -0.1221 & -0.5036 & 0.3182  & -0.0390 & -0.2936 & 0.2471  & 1.0000  & -0.1601 & -0.0709 & -0.1601 & 0.3182  & 0.2471  & -0.3182 & 0.0156  & -0.0390 & 0.0000  & 0.0390  & 0.2936  & -0.1221 & 0.7469  \\
						$\mathrm{A}_{0,12}$ & 0.3182  & -0.2471 & 0.0390  & 0.2936  & -0.5036 & -0.1221 & 0.3182  & 0.7469  & -0.0390 & 0.0156  & -0.1601 & 1.0000  & 0.1601  & 0.0709  & -0.0156 & -0.3182 & 0.2471  & -0.3182 & -0.0390 & 0.1221  & 0.2936  & -0.7469 & 0.0000  & -0.0390 \\
						$\mathrm{A}_{0,13}$ & 0.0156  & -0.3182 & -0.2936 & -0.0390 & 0.1221  & 0.0000  & 0.2471  & 0.2936  & 0.0390  & 0.3182  & -0.0709 & 0.1601  & 1.0000  & 0.1601  & 0.2471  & 0.3182  & -0.0156 & 0.3182  & -0.7469 & -0.5036 & 0.7469  & -0.0390 & 0.1221  & 0.0390  \\
						$\mathrm{A}_{0,14}$ & -0.2471 & 0.3182  & -0.7469 & 0.0390  & 0.0000  & -0.1221 & -0.0156 & -0.0390 & 0.7469  & -0.3182 & -0.1601 & 0.0709  & 0.1601  & 1.0000  & 0.3182  & 0.0156  & -0.3182 & 0.2471  & -0.2936 & 0.1221  & 0.0390  & 0.0390  & -0.5036 & -0.2936 \\
						$\mathrm{A}_{0,15}$ & -0.3182 & -0.3182 & -0.0390 & -0.2936 & 0.0390  & -0.7469 & 0.2471  & 0.1221  & 0.5036  & -0.0156 & 0.3182  & -0.0156 & 0.2471  & 0.3182  & 1.0000  & 0.1601  & 0.1601  & -0.0709 & -0.7469 & -0.0390 & 0.0000  & -0.0390 & 0.2936  & 0.1221  \\
						$\mathrm{A}_{0,16}$ & 0.3182  & 0.3182  & 0.0390  & -0.7469 & -0.2936 & -0.0390 & -0.0156 & -0.5036 & -0.1221 & 0.2471  & 0.2471  & -0.3182 & 0.3182  & 0.0156  & 0.1601  & 1.0000  & 0.0709  & -0.1601 & -0.2936 & -0.7469 & 0.1221  & 0.0390  & -0.0390 & 0.0000  \\
						$\mathrm{A}_{0,17}$ & -0.0156 & -0.2471 & 0.2936  & 0.0390  & -0.0390 & -0.2936 & 0.3182  & 0.0000  & -0.1221 & -0.3182 & -0.3182 & 0.2471  & -0.0156 & -0.3182 & 0.1601  & 0.0709  & 1.0000  & -0.1601 & -0.0390 & 0.0390  & 0.1221  & -0.7469 & 0.7469  & -0.5036 \\
						$\mathrm{A}_{0,18}$ & -0.2471 & -0.0156 & -0.7469 & 0.0390  & 0.7469  & -0.0390 & 0.3182  & -0.1221 & 0.0000  & -0.3182 & 0.0156  & -0.3182 & 0.3182  & 0.2471  & -0.0709 & -0.1601 & -0.1601 & 1.0000  & -0.0390 & 0.2936  & 0.5036  & 0.2936  & -0.0390 & -0.1221 \\
						$\mathrm{A}_{0,19}$ & 0.1221  & 0.5036  & 0.1601  & 0.1601  & 0.0156  & 0.2471  & 0.0000  & -0.3182 & -0.3182 & -0.1221 & -0.0390 & -0.0390 & -0.7469 & -0.2936 & -0.7469 & -0.2936 & -0.0390 & -0.0390 & 1.0000  & 0.3182  & -0.2471 & -0.0709 & -0.3182 & 0.0156  \\
						$\mathrm{A}_{0,20}$ & -0.0390 & -0.2936 & -0.3182 & 0.2471  & 0.1601  & -0.0709 & -0.0390 & 0.3182  & -0.0156 & -0.7469 & 0.0000  & 0.1221  & -0.5036 & 0.1221  & -0.0390 & -0.7469 & 0.0390  & 0.2936  & 0.3182  & 1.0000  & -0.3182 & 0.0156  & 0.1601  & -0.2471 \\
						$\mathrm{A}_{0,21}$ & -0.0390 & -0.0390 & -0.3182 & 0.0156  & 0.3182  & -0.2471 & 0.7469  & 0.1601  & -0.0709 & 0.2936  & 0.0390  & 0.2936  & 0.7469  & 0.0390  & 0.0000  & 0.1221  & 0.1221  & 0.5036  & -0.2471 & -0.3182 & 1.0000  & -0.3182 & -0.0156 & 0.1601  \\
						$\mathrm{A}_{0,22}$ & -0.1221 & 0.0000  & -0.1601 & -0.1601 & 0.3182  & 0.3182  & -0.5036 & -0.2471 & -0.0156 & 0.1221  & 0.2936  & -0.7469 & -0.0390 & 0.0390  & -0.0390 & 0.0390  & -0.7469 & 0.2936  & -0.0709 & 0.0156  & -0.3182 & 1.0000  & -0.2471 & 0.3182  \\
						$\mathrm{A}_{0,23}$ & 0.0390  & -0.7469 & 0.3182  & -0.0156 & 0.0709  & -0.1601 & 0.0390  & 0.2471  & -0.3182 & -0.2936 & -0.1221 & 0.0000  & 0.1221  & -0.5036 & 0.2936  & -0.0390 & 0.7469  & -0.0390 & -0.3182 & 0.1601  & -0.0156 & -0.2471 & 1.0000  & -0.3182 \\
						$\mathrm{A}_{0,24}$ & 0.0390  & 0.0390  & 0.3182  & -0.2471 & 0.0156  & -0.3182 & 0.2936  & 0.0709  & -0.1601 & 0.7469  & 0.7469  & -0.0390 & 0.0390  & -0.2936 & 0.1221  & 0.0000  & -0.5036 & -0.1221 & 0.0156  & -0.2471 & 0.1601  & 0.3182  & -0.3182 & 1.0000 \\
						\bottomrule
					\end{tabular}
				\end{subtable}
			}
		
			\vspace{1cm}
			\scalebox{0.6}{
				\begin{subtable}[t]{1.25\textwidth}
					\centering
					\caption{$\mathrm{A}'_0$ solutions in \ce{[VF6]^{3-}}.}
					\begin{tabular}[t]{%
							r |%
							*{12}{S[table-format=2.4, table-alignment=right]}
						}
						\toprule
						& $\mathrm{A}'_{0,1}$ & $\mathrm{A}'_{0,2}$ & $\mathrm{A}'_{0,3}$ & $\mathrm{A}'_{0,4}$ & $\mathrm{A}'_{0,5}$ & $\mathrm{A}'_{0,6}$ & $\mathrm{A}'_{0,7}$ & $\mathrm{A}'_{0,8}$ & $\mathrm{A}'_{0,9}$ & $\mathrm{A}'_{0,10}$ & $\mathrm{A}'_{0,11}$ & $\mathrm{A}'_{0,12}$ \\
						\midrule
						$\mathrm{A}'_{0,1}$ & 1.0000  & -0.0074 & -0.0074 & 0.0074  & -0.0074 & 0.0000  & 0.0000  & -0.4844 & 0.4844  & 0.4844  & -0.4844 & 0.0000  \\
						$\mathrm{A}'_{0,2}$ & -0.0074 & 1.0000  & 0.0000  & -0.0074 & -0.0074 & 0.0074  & -0.4844 & 0.0000  & 0.0000  & -0.4844 & -0.4844 & 0.4844  \\
						$\mathrm{A}'_{0,3}$ & -0.0074 & 0.0000  & 1.0000  & 0.0074  & 0.0074  & 0.0074  & -0.4844 & 0.0000  & 0.0000  & 0.4844  & 0.4844  & 0.4844  \\
						$\mathrm{A}'_{0,4}$ & 0.0074  & -0.0074 & 0.0074  & 1.0000  & 0.0000  & 0.0074  & 0.4844  & -0.4844 & -0.4844 & 0.0000  & 0.0000  & 0.4844  \\
						$\mathrm{A}'_{0,5}$ & -0.0074 & -0.0074 & 0.0074  & 0.0000  & 1.0000  & -0.0074 & -0.4844 & -0.4844 & -0.4844 & 0.0000  & 0.0000  & -0.4844 \\
						$\mathrm{A}'_{0,6}$ & 0.0000  & 0.0074  & 0.0074  & 0.0074  & -0.0074 & 1.0000  & 0.0000  & 0.4844  & -0.4844 & 0.4844  & -0.4844 & 0.0000  \\
						$\mathrm{A}'_{0,7}$ & 0.0000  & -0.4844 & -0.4844 & 0.4844  & -0.4844 & 0.0000  & 1.0000  & -0.0074 & 0.0074  & 0.0074  & -0.0074 & 0.0000  \\
						$\mathrm{A}'_{0,8}$ & -0.4844 & 0.0000  & 0.0000  & -0.4844 & -0.4844 & 0.4844  & -0.0074 & 1.0000  & 0.0000  & -0.0074 & -0.0074 & 0.0074  \\
						$\mathrm{A}'_{0,9}$ & 0.4844  & 0.0000  & 0.0000  & -0.4844 & -0.4844 & -0.4844 & 0.0074  & 0.0000  & 1.0000  & -0.0074 & -0.0074 & -0.0074 \\
						$\mathrm{A}'_{0,10}$ & 0.4844  & -0.4844 & 0.4844  & 0.0000  & 0.0000  & 0.4844  & 0.0074  & -0.0074 & -0.0074 & 1.0000  & 0.0000  & 0.0074  \\
						$\mathrm{A}'_{0,11}$ & -0.4844 & -0.4844 & 0.4844  & 0.0000  & 0.0000  & -0.4844 & -0.0074 & -0.0074 & -0.0074 & 0.0000  & 1.0000  & -0.0074 \\
						$\mathrm{A}'_{0,12}$ & 0.0000  & 0.4844  & 0.4844  & 0.4844  & -0.4844 & 0.0000  & 0.0000  & 0.0074  & -0.0074 & 0.0074  & -0.0074 & 1.0000  \\
						\bottomrule
					\end{tabular}
				\end{subtable}
			}
			\endgroup
		\end{table*}
	\end{landscape}

	\begin{table*}
		\ContinuedFloat
		\footnotesize
		\begingroup
		\renewcommand\arraystretch{1.35}
		\scalebox{1}{
			\begin{subtable}[t]{1\textwidth}
				\centering
				\caption{$\mathrm{A}''_0$ solutions in \ce{[VF6]^{3-}}.}
				\begin{tabular}[t]{%
						r |%
						*{6}{S[table-format=2.4, table-alignment=right]}
					}
					\toprule
					& $\mathrm{A}''_{0,1}$ & $\mathrm{A}''_{0,2}$ & $\mathrm{A}''_{0,3}$ & $\mathrm{A}''_{0,4}$ & $\mathrm{A}''_{0,5}$ & $\mathrm{A}''_{0,6}$ \\
					\midrule
					$\mathrm{A}''_{0,1}$ &  1.0000 &  0.0000 &  0.0000 &  0.0000 &   0.0000 &  0.0000 \\
					$\mathrm{A}''_{0,2}$ &  0.0000 &  1.0000 &  0.0000 &  0.0000 &   0.0000 &  0.0000 \\
					$\mathrm{A}''_{0,3}$ &  0.0000 &  0.0000 &  1.0000 &  0.0000 &  0.0000 &  0.0000 \\
					$\mathrm{A}''_{0,4}$ &  0.0000 &  0.0000 &  0.0000 &  1.0000 &  0.0000 &  0.0000 \\
					$\mathrm{A}''_{0,5}$ &  0.0000 &  0.0000 &  0.0000 &  0.0000 &  1.0000 &  0.0000 \\
					$\mathrm{A}''_{0,6}$ &  0.0000 &  0.0000 &  0.0000 &  0.0000 &  0.0000 &  1.0000 \\
					\bottomrule
				\end{tabular}
			\end{subtable}
		}
		
		\vspace{1cm}
		\scalebox{1}{
			\begin{subtable}[t]{1\textwidth}
				\centering
				\caption{$\mathrm{B}_{1}$ solutions in \ce{[VF6]^{3-}}.}
				\begin{tabular}[t]{%
						r |%
						*{6}{S[table-format=2.4, table-alignment=right]}
					}
					\toprule
					& $\mathrm{B}_{0,1}$ & $\mathrm{B}_{0,2}$ & $\mathrm{B}_{0,3}$ & $\mathrm{B}_{0,4}$ & $\mathrm{B}_{0,5}$ & $\mathrm{B}_{0,6}$ \\
					\midrule
					$\mathrm{B}_{0,1}$ &  1.0000 &  0.0196 & -0.0196 &  0.0196 & -0.0196 &  0.1927 \\
					$\mathrm{B}_{0,2}$ &  0.0196 &  1.0000 & -0.1927 &  0.0196 & -0.0196 &  0.0196 \\
					$\mathrm{B}_{0,3}$ & -0.0196 & -0.1927 &  1.0000 & -0.0196 &  0.0196 & -0.0196 \\
					$\mathrm{B}_{0,4}$ &  0.0196 &  0.0196 & -0.0196 &  1.0000 & -0.1927 &  0.0196 \\
					$\mathrm{B}_{0,5}$ & -0.0196 & -0.0196 &  0.0196 & -0.1927 &  1.0000 & -0.0196 \\
					$\mathrm{B}_{0,6}$ &  0.1927 &  0.0196 & -0.0196 &  0.0196 & -0.0196 &  1.0000 \\
					\bottomrule
				\end{tabular}
			\end{subtable}
		}
	
		\vspace{1cm}
		\scalebox{1}{
			\begin{subtable}[t]{1\textwidth}
				\centering
				\caption{$\mathrm{C}_0$ solutions in \ce{[VF6]^{3-}}.}
				\begin{tabular}[t]{%
						r |%
						*{6}{S[table-format=2.4, table-alignment=right]}
					}
					\toprule
					& $\mathrm{C}_{0,1}$ & $\mathrm{C}_{0,2}$ & $\mathrm{C}_{0,3}$ & $\mathrm{C}_{0,4}$ & $\mathrm{C}_{0,5}$ & $\mathrm{C}_{0,6}$ \\
					\midrule
					$\mathrm{C}_{0,1}$ &  1.0000 &  0.0000 &  0.0000 &  0.0000 &   0.0000 &  0.0000 \\
					$\mathrm{C}_{0,2}$ &  0.0000 &  1.0000 &  0.0000 &  0.0000 &   0.0000 &  0.0000 \\
					$\mathrm{C}_{0,3}$ &  0.0000 &  0.0000 &  1.0000 &  0.0000 &  0.0000 &  0.0000 \\
					$\mathrm{C}_{0,4}$ &  0.0000 &  0.0000 &  0.0000 &  1.0000 &  0.0000 &  0.0000 \\
					$\mathrm{C}_{0,5}$ &  0.0000 &  0.0000 &  0.0000 &  0.0000 &  1.0000 &  0.0000 \\
					$\mathrm{C}_{0,6}$ &  0.0000 &  0.0000 &  0.0000 &  0.0000 &  0.0000 &  1.0000 \\
					\bottomrule
				\end{tabular}
			\end{subtable}
		}
	
		\vspace{1cm}
		\scalebox{1}{
			\begin{subtable}[t]{1\textwidth}
				\centering
				\caption{$\mathrm{D}_{0}$ solutions in \ce{[VF6]^{3-}}.}
				\begin{tabular}[t]{%
						r |%
						*{3}{S[table-format=2.4, table-alignment=right]}
					}
					\toprule
					& $\mathrm{D}_{0,1}$ & $\mathrm{D}_{0,2}$ & $\mathrm{D}_{0,3}$ \\
					\midrule
					$\mathrm{D}_{0,1}$ &  1.0000 &  0.0000 &  0.0000 \\
					$\mathrm{D}_{0,2}$ &  0.0000 &  1.0000 &  0.0000 \\
					$\mathrm{D}_{0,3}$ &  0.0000 &  0.0000 &  1.0000 \\
					\bottomrule
				\end{tabular}
			\end{subtable}
		}
		\endgroup
	\end{table*}

	\begin{table*}
		\ContinuedFloat
		\footnotesize
		\begingroup
		\renewcommand\arraystretch{1.35}
		\scalebox{1}{
			\begin{subtable}[t]{1\textwidth}
				\centering
				\caption{$\mathrm{E}_0$ solutions in \ce{[VF6]^{3-}}.}
				\begin{tabular}[t]{%
						r |%
						*{6}{S[table-format=2.4, table-alignment=right]}
					}
					\toprule
					& $\mathrm{E}_{0,1}$ & $\mathrm{E}_{0,2}$ & $\mathrm{E}_{0,3}$ & $\mathrm{E}_{0,4}$ & $\mathrm{E}_{0,5}$ & $\mathrm{E}_{0,6}$ \\
					\midrule
					$\mathrm{E}_{0,1}$ &  1.0000 &  0.0000 &  0.0000 &  0.0000 &   0.0000 &  0.0000 \\
					$\mathrm{E}_{0,2}$ &  0.0000 &  1.0000 &  0.0000 &  0.0000 &   0.0000 &  0.0000 \\
					$\mathrm{E}_{0,3}$ &  0.0000 &  0.0000 &  1.0000 &  0.0000 &  0.0000 &  0.0000 \\
					$\mathrm{E}_{0,4}$ &  0.0000 &  0.0000 &  0.0000 &  1.0000 &  0.0000 &  0.0000 \\
					$\mathrm{E}_{0,5}$ &  0.0000 &  0.0000 &  0.0000 &  0.0000 &  1.0000 &  0.0000 \\
					$\mathrm{E}_{0,6}$ &  0.0000 &  0.0000 &  0.0000 &  0.0000 &  0.0000 &  1.0000 \\
					\bottomrule
				\end{tabular}
			\end{subtable}
		}
		\endgroup
	\end{table*}

	%%%%%%%%%%%%%%%%%%%%%%%%%%%%%%%%%%%%%%%%%
	% Energies and other properties
	%%%%%%%%%%%%%%%%%%%%%%%%%%%%%%%%%%%%%%%%%

	\begin{table*}
		\centering
		\caption{
			Energies and $\langle \hat{S}^2 \rangle$ values of UHF solutions in \ce{[VF6]^{3-}}.
		}
		\label{tab:d2_uhfenergy}
		\footnotesize
		\begingroup
		\renewcommand\arraystretch{1.35}
		\begin{subtable}[h]{.48\textwidth}
			\centering
			\caption{$M_S = 1$}
			\label{subtab:d2_uhfenergy_MS1}
			\begin{tabular}[t]{%
					>{\raggedright\arraybackslash}m{0.9cm}%
					S[table-format=4.10, table-alignment=left]%
					S[table-format=1.4, table-alignment=left]%
				}
				\toprule
				$\Psi_\mathrm{UHF}$ &	{Energy/\si{\hartree}} & {$\langle \hat{S}^2 \rangle$} \\
				\midrule
				$\mathrm{A}_1$ & -1539.3424268 & 2.0040 \\
				$\mathrm{A}'_1$ & -1539.3416345 & 2.0040 \\
				$\mathrm{B}_1$ & -1539.3360346 & 2.0045 \\
				$\mathrm{C}_1$ & -1539.2883111 & 2.0035 \\
				$\mathrm{D}_1$ & -1539.2750572 & 2.0039 \\
				$\mathrm{E}_1$ & -1539.2354939 & 2.0049 \\
				$\mathrm{F}_1$ & -1539.2277383 & 2.0044 \\
				\bottomrule
			\end{tabular}
		\end{subtable}
		\hfill
		\begin{subtable}[h]{.48\textwidth}
			\centering
			\caption{$M_S = 0$}
			\label{subtab:d2_uhfenergy_MS0}
			\begin{tabular}[t]{%
					>{\raggedright\arraybackslash}m{0.9cm}%
					S[table-format=4.10, table-alignment=left]%
					S[table-format=1.4, table-alignment=left]%
				}
				\toprule
				$\Psi_\mathrm{UHF}$ &	{Energy/\si{\hartree}} & {$\langle \hat{S}^2 \rangle$} \\
				\midrule
				$\mathrm{A}_0$ & -1539.3086155 & 1.0025 \\
				$\mathrm{A}'_0$ & -1539.3080469 & 1.0029 \\
				$\mathrm{A}''_0$ & -1539.3062026 & 1.0037 \\
				$\mathrm{B}_0$ & -1539.2659974 & 0.8090 \\
				$\mathrm{C}_0$ & -1539.2528868 & 1.0034 \\
				$\mathrm{D}_0$ & -1539.2440258 & 0.0000 \\
				$\mathrm{E}_0$ & -1539.2204473 & 1.0046 \\
				\bottomrule
			\end{tabular}
		\end{subtable}
		\endgroup
	\end{table*}

	\begin{table*}
		\centering
		\caption{
			Energies and $\langle \hat{S}^2 \rangle$ values of NOCI wavefunctions constructed from single sets of degenerate UHF solutions in \ce{[VF6]^{3-}}.
		}
		\label{tab:d2_singlenocienergy}
		\begin{subtable}[t]{\textwidth}
			\centering
			\caption{
				$M_S = 1$
			}
			\footnotesize
			\begingroup
			\renewcommand\arraystretch{1.35}
			\begin{tabular}[t]{
					>{\raggedright\arraybackslash}m{1.9cm}%
					S[table-format=4.7, table-alignment=left]%
					S[table-format=1.4, table-alignment=left]}
				\toprule
				$\Phi$ &	{Energy/\si{\hartree}} & {$\langle \hat{S}^2 \rangle$} \\
				\midrule
				$\prescript{3}{}{T}_{1g}[\mathrm{A}_1]$ & -1539.3435881 & 2.0035 \\
				$\prescript{\varnothing}{}{A}_{2g}[\mathrm{A}_1]$ & -1538.3251392 & 2.4185 \\
				\midrule
				$\prescript{3}{}{T}_{1g}[\mathrm{A}'_1]$ & -1539.3442214 & 2.0033 \\
				$\prescript{\varnothing}{}{T}_{2g}[\mathrm{A}'_1]$ & -1539.1704720 & 2.0483 \\
				\midrule
				$\prescript{3}{}{T}_{2g}[\mathrm{D}_1]$ & -1539.2912895 & 2.0025 \\
				$\prescript{3}{}{T}_{1g}[\mathrm{D}_1]$ & -1539.2265782 & 2.0079 \\
				\bottomrule
			\end{tabular}
			\endgroup
		\end{subtable}
	
		\vspace{0.5cm}
		\begin{subtable}[t]{\textwidth}
			\centering
			\caption{
				$M_S = 0$
			}
			\footnotesize
			\begingroup
			\renewcommand\arraystretch{1.35}
			\begin{tabular}[t]{
					>{\raggedright\arraybackslash}m{1.9cm}%
					S[table-format=4.7, table-alignment=left]%
					S[table-format=1.4, table-alignment=left]%
					>{\centering\arraybackslash}m{0.5cm}%%
					>{\raggedright\arraybackslash}m{1.9cm}%
					S[table-format=4.7, table-alignment=left]%
					S[table-format=1.4, table-alignment=left]
				}
				\toprule
				$\Phi$ &	{Energy/\si{\hartree}} & {$\langle \hat{S}^2 \rangle$} && $\Phi$ &	{Energy/\si{\hartree}} & {$\langle \hat{S}^2 \rangle$}\\
				\midrule
				$\prescript{3}{}{T}_{1g}[\mathrm{A}_0]$ & -1539.3445628 & 2.0000 && %
				$\prescript{1}{}{T}_{2g}[\mathrm{A}_0]$ & -1539.2803723 & 0.0038 \\
				$\prescript{3}{}{A}_{2g}[\mathrm{A}_0]$ & -1537.8222025 & 2.0006 && %
				$\prescript{1}{}{E}_{g}[\mathrm{A}_0]$ & -1539.2770923 & 0.0032 \\
				$\prescript{3}{}{T}_{1g}[\mathrm{A}_0]$ & -1536.9596631 & 2.0016 && %
				$\prescript{\varnothing}{}{T}_{1g}[\mathrm{A}_0]$ & -1539.0836504 & 0.1003 \\
				$\prescript{3}{}{T}_{2g}[\mathrm{A}_0]$ & -1536.9406332 & 2.0008 && %
				$\prescript{\varnothing}{}{A}_{1g}[\mathrm{A}_0]$ & -1539.0545711 & 0.0455 \\
				$\prescript{3}{}{E}_{g}[\mathrm{A}_0]$ & -1536.3383087 & 2.0011 && %
				$\prescript{\varnothing}{}{T}_{2g}[\mathrm{A}_0]$ & -1539.0014173 & 0.1574 \\
				\midrule
				$\prescript{3}{}{T}_{1g}[\mathrm{A}'_0]$ & -1539.3433740 & 2.0000 && %
				$\prescript{1}{}{T}_{2g}[\mathrm{A}'_0]$ & -1539.2791715 & 0.0042 \\
				$\prescript{3}{}{T}_{2g}[\mathrm{A}'_0]$ & -1539.0187820 & 2.0002 && %
				$\prescript{\varnothing}{}{T}_{1g}[\mathrm{A}'_0]$ & -1539.1142797 & 0.0750 \\
				\midrule
				$\prescript{3}{}{T}_{1g}[\mathrm{A}''_0]$ & -1539.3372071 & 2.0000 && %
				$\prescript{1}{}{T}_{2g}[\mathrm{A}''_0]$ & -1539.2751980 & 0.0073 \\
				\midrule
				$\prescript{3}{}{T}_{2g}[\mathrm{B}_0]$ & -1539.2847593 & 2.0000 && %
				$\prescript{1}{}{E}_{g}[\mathrm{B}_0]$ & -1539.2735656 & 0.0029 \\
				&&&& %
				$\prescript{1}{}{A}_{1g}[\mathrm{B}_0]$ & -1539.2165102 & 0.0027 \\
				\midrule
				$\prescript{3}{}{T}_{2g}[\mathrm{C}_0]$ & -1539.2868717 & 2.0000 && %
				$\prescript{1}{}{T}_{2g}[\mathrm{C}_0]$ & -1539.2189019 & 0.0067 \\
				\midrule
				&&&& %
				$\prescript{1}{}{E}_{g}[\mathrm{D}_0]$ & -1539.2742054 & 0.0000 \\
				&&&& %
				$\prescript{1}{}{A}_{1g}[\mathrm{D}_0]$ & -1539.1836669 & 0.0000 \\
				\midrule
				$\prescript{3}{}{T}_{1g}[\mathrm{E}_0]$ & -1539.2374281 & 2.0000 && %
				$\prescript{1}{}{T}_{1g}[\mathrm{E}_0]$ & -1539.2034665 & 0.0092 \\
				\bottomrule
			\end{tabular}
			\endgroup
		\end{subtable}
	\end{table*}

	%%
	%% All noci - MS = 1
	\begin{table*}
		\centering
		\caption{
			Energies and $\langle \hat{S}^2 \rangle$ values of $M_S = 1$ NOCI wavefunctions constructed from all interacting low-lying UHF solutions in \ce{[VF6]^{3-}}.
		}
		\label{tab:d2_allnocienergy_MS1}
		\begin{subtable}[t]{.48\textwidth}
			\centering
			\caption{
				$\prescript{3}{}{T}_{1g}[\mathrm{A}_1, \mathrm{A}'_1, \mathrm{B}_1\ (\mathrm{D}_1, \mathrm{E}_1)]$
			}
			\footnotesize
			\begingroup
			\renewcommand\arraystretch{1.35}
			\scalebox{.8}{
				\begin{tabular}[t]{
						>{\raggedright\arraybackslash}m{5.0cm}%
						S[table-format=4.7, table-alignment=left]%
						S[table-format=1.4, table-alignment=left]}
					\toprule
					$\Phi$ &	{Energy/\si{\hartree}} & {$\langle \hat{S}^2 \rangle$} \\
					\midrule
					$\prescript{3}{}{T}_{1g}[\mathrm{A}_1]$ & -1539.3435881 & 2.0035 \\
					$\prescript{3}{}{T}_{1g}[\mathrm{A}_1 \oplus \mathrm{D}_1]$ & -1539.3436373 & 2.0035 \\
					$\prescript{3}{}{T}_{1g}[\mathrm{A}_1 \oplus \mathrm{D}_1 \oplus \mathrm{E}_1]$ & -1539.3436454 & 2.0035 \\
					$\prescript{3}{}{T}_{1g}[\mathrm{A}_1 \oplus \mathrm{E}_1]$ & -1539.3436451 & 2.0035 \\
					\midrule
					$\prescript{3}{}{T}_{1g}[\mathrm{A}_1 \oplus \mathrm{A}'_1]$ & -1539.3442685 & 2.0033 \\
					$\prescript{3}{}{T}_{1g}[\mathrm{A}_1 \oplus \mathrm{A}'_1 \oplus \mathrm{D}_1]$ & -1539.3445505 & 2.0033 \\
					$\prescript{3}{}{T}_{1g}[\mathrm{A}_1 \oplus \mathrm{A}'_1 \oplus \mathrm{D}_1 \oplus \mathrm{E}_1]$ & -1539.3447079 & 2.0033 \\
					$\prescript{3}{}{T}_{1g}[\mathrm{A}_1 \oplus \mathrm{A}'_1 \oplus \mathrm{E}_1]$ & -1539.3446971 & 2.0033 \\
					\midrule
					$\prescript{3}{}{T}_{1g}[\mathrm{A}_1 \oplus \mathrm{A}'_1 \oplus \mathrm{B}_1]$ & -1539.3443200 & 2.0033 \\
					$\prescript{3}{}{T}_{1g}[\mathrm{A}_1 \oplus \mathrm{A}'_1 \oplus \mathrm{B}_1 \oplus \mathrm{D}_1]$ & -1539.3446913 & 2.0034 \\
					$\prescript{3}{}{T}_{1g}[\mathrm{A}_1 \oplus \mathrm{A}'_1 \oplus \mathrm{B}_1 \oplus \mathrm{D}_1 \oplus \mathrm{E}_1]$ & -1539.3448266 & 2.0034 \\
					$\prescript{3}{}{T}_{1g}[\mathrm{A}_1 \oplus \mathrm{A}'_1 \oplus \mathrm{B}_1 \oplus \mathrm{E}_1]$ & -1539.3448110 & 2.0034 \\
					\midrule
					$\prescript{3}{}{T}_{1g}[\mathrm{A}_1 \oplus \mathrm{B}_1]$ & -1539.3439815 & 2.0034 \\
					$\prescript{3}{}{T}_{1g}[\mathrm{A}_1 \oplus \mathrm{B}_1 \oplus \mathrm{D}_1]$ & -1539.3446908 & 2.0034 \\
					$\prescript{3}{}{T}_{1g}[\mathrm{A}_1 \oplus \mathrm{B}_1 \oplus \mathrm{D}_1 \oplus \mathrm{E}_1]$ & -1539.3448244 & 2.0034 \\
					$\prescript{3}{}{T}_{1g}[\mathrm{A}_1 \oplus \mathrm{B}_1 \oplus \mathrm{E}_1]$ & -1539.3447885 & 2.0034 \\
					\midrule
					$\prescript{3}{}{T}_{1g}[\mathrm{A}'_1]$ & -1539.3442214 & 2.0033 \\
					$\prescript{3}{}{T}_{1g}[\mathrm{A}'_1 \oplus \mathrm{D}_1]$ & -1539.3443994 & 2.0033 \\
					$\prescript{3}{}{T}_{1g}[\mathrm{A}'_1 \oplus \mathrm{D}_1 \oplus \mathrm{E}_1]$ & -1539.3444664 & 2.0033 \\
					$\prescript{3}{}{T}_{1g}[\mathrm{A}'_1 \oplus \mathrm{E}_1]$ & -1539.3444642 & 2.0033 \\
					\midrule
					$\prescript{3}{}{T}_{1g}[\mathrm{A}'_1 \oplus \mathrm{B}_1]$ & -1539.3442249 & 2.0033 \\
					$\prescript{3}{}{T}_{1g}[\mathrm{A}'_1 \oplus \mathrm{B}_1 \oplus \mathrm{D}_1]$ & -1539.3446631 & 2.0033 \\
					$\prescript{3}{}{T}_{1g}[\mathrm{A}'_1 \oplus \mathrm{B}_1 \oplus \mathrm{D}_1 \oplus \mathrm{E}_1]$ & -1539.3448189 & 2.0033 \\
					$\prescript{3}{}{T}_{1g}[\mathrm{A}'_1 \oplus \mathrm{B}_1 \oplus \mathrm{E}_1]$ & -1539.3448107 & 2.0034 \\
					\midrule
					$\prescript{3}{}{T}_{1g}[\mathrm{B}_1 \oplus \mathrm{D}_1]$ & -1539.3424715 & 2.0044 \\
					$\prescript{3}{}{T}_{1g}[\mathrm{B}_1 \oplus \mathrm{D}_1 \oplus \mathrm{E}_1]$ & -1539.3430467 & 2.0044 \\
					$\prescript{3}{}{T}_{1g}[\mathrm{B}_1 \oplus \mathrm{E}_1]$ & -1539.3428730 & 2.0046 \\
					\bottomrule
				\end{tabular}
			}
			\endgroup
		\end{subtable}
		\hfill
		\begin{subtable}[t]{.48\textwidth}
			\centering
			\caption{
				$\prescript{3}{}{T}_{1g}[\mathrm{D}_1, \mathrm{E}_1\ (\mathrm{A}_1, \mathrm{A}'_1, \mathrm{B}_1)]$
			}
			\footnotesize
			\begingroup
			\renewcommand\arraystretch{1.35}
			\scalebox{.8}{
				\begin{tabular}[t]{
						>{\raggedright\arraybackslash}m{5.0cm}%
						S[table-format=4.7, table-alignment=left]%
						S[table-format=1.4, table-alignment=left]}
					\toprule
					$\Phi$ &	{Energy/\si{\hartree}} & {$\langle \hat{S}^2 \rangle$} \\
					\midrule
					$\prescript{3}{}{T}_{1g}[\mathrm{D}_1]$ & -1539.2265782 & 2.0079 \\
					$\prescript{3}{}{T}_{1g}[\mathrm{A}_1 \oplus \mathrm{D}_1]$ & -1539.2187002 & 2.0086 \\
					$\prescript{3}{}{T}_{1g}[\mathrm{A}_1 \oplus \mathrm{A}'_1 \oplus \mathrm{D}_1]$ & -1539.2208846 & 2.0077 \\
					$\prescript{3}{}{T}_{1g}[\mathrm{A}_1 \oplus \mathrm{A}'_1 \oplus \mathrm{B}_1 \oplus \mathrm{D}_1]$ & -1539.2209930 & 2.0076 \\
					$\prescript{3}{}{T}_{1g}[\mathrm{A}_1 \oplus \mathrm{B}_1 \oplus \mathrm{D}_1]$ & -1539.2209551 & 2.0076 \\
					$\prescript{3}{}{T}_{1g}[\mathrm{A}'_1 \oplus \mathrm{D}_1]$ & -1539.2191020 & 2.0085 \\
					$\prescript{3}{}{T}_{1g}[\mathrm{A}'_1 \oplus \mathrm{B}_1 \oplus \mathrm{D}_1]$ & -1539.2209142 & 2.0076 \\
					$\prescript{3}{}{T}_{1g}[\mathrm{B}_1 \oplus \mathrm{D}_1]$ & -1539.2201542 & 2.0080 \\
					\midrule
					$\prescript{3}{}{T}_{1g}[\mathrm{D}_1 \oplus \mathrm{E}_1]$ & -1539.2375245 & 2.0033 \\
					$\prescript{3}{}{T}_{1g}[\mathrm{A}_1 \oplus \mathrm{D}_1 \oplus \mathrm{E}_1]$ & -1539.2302007 & 2.0036 \\
					$\prescript{3}{}{T}_{1g}[\mathrm{A}_1 \oplus \mathrm{A}'_1 \oplus \mathrm{D}_1 \oplus \mathrm{E}_1]$ & -1539.2303581 & 2.0035 \\
					$\prescript{3}{}{T}_{1g}[\mathrm{A}_1 \oplus \mathrm{A}'_1 \oplus \mathrm{B}_1 \oplus \mathrm{D}_1 \oplus \mathrm{E}_1]$ & -1539.2306619 & 2.0035 \\
					$\prescript{3}{}{T}_{1g}[\mathrm{A}_1 \oplus \mathrm{B}_1 \oplus \mathrm{D}_1 \oplus \mathrm{E}_1]$ & -1539.2305215 & 2.0034 \\
					$\prescript{3}{}{T}_{1g}[\mathrm{A}'_1 \oplus \mathrm{D}_1 \oplus \mathrm{E}_1]$ & -1539.2302913 & 2.0035 \\
					$\prescript{3}{}{T}_{1g}[\mathrm{A}'_1 \oplus \mathrm{B}_1 \oplus \mathrm{D}_1 \oplus \mathrm{E}_1]$ & -1539.2305196 & 2.0034 \\
					$\prescript{3}{}{T}_{1g}[\mathrm{B}_1 \oplus \mathrm{D}_1 \oplus \mathrm{E}_1]$ & -1539.2305195 & 2.0034 \\
					\midrule
					$\prescript{3}{}{T}_{1g}[\mathrm{A}_1 \oplus \mathrm{E}_1]$ & -1539.2280548 & 2.0053 \\
					$\prescript{3}{}{T}_{1g}[\mathrm{A}_1 \oplus \mathrm{A}'_1 \oplus \mathrm{E}_1]$ & -1539.2281044 & 2.0052 \\
					$\prescript{3}{}{T}_{1g}[\mathrm{A}_1 \oplus \mathrm{A}'_1 \oplus \mathrm{B}_1 \oplus \mathrm{E}_1]$ & -1539.2299323 & 2.0043 \\
					$\prescript{3}{}{T}_{1g}[\mathrm{A}_1 \oplus \mathrm{B}_1 \oplus \mathrm{E}_1]$ & -1539.2286892 & 2.0048 \\
					$\prescript{3}{}{T}_{1g}[\mathrm{A}'_1 \oplus \mathrm{E}_1]$ & -1539.2280974 & 2.0052 \\
					$\prescript{3}{}{T}_{1g}[\mathrm{A}'_1 \oplus \mathrm{B}_1 \oplus \mathrm{E}_1]$ & -1539.2287562 & 2.0048 \\
					$\prescript{3}{}{T}_{1g}[\mathrm{B}_1 \oplus \mathrm{E}_1]$ & -1539.2286554 & 2.0049 \\
					\bottomrule
				\end{tabular}
			}
			\endgroup
		\end{subtable}

		\vspace{1cm}
		\begin{subtable}[t]{.48\textwidth}
			\centering
			\caption{
				$\prescript{3}{}{T}_{2g}[\mathrm{D}_1\ (\mathrm{C}_1)]$
			}
			\footnotesize
			\begingroup
			\renewcommand\arraystretch{1.35}
			\scalebox{.8}{
				\begin{tabular}[t]{
						>{\raggedright\arraybackslash}m{2.5cm}%
						S[table-format=4.7, table-alignment=left]%
						S[table-format=1.4, table-alignment=left]}
					\toprule
					$\Phi$ &	{Energy/\si{\hartree}} & {$\langle \hat{S}^2 \rangle$} \\
					\midrule
					$\prescript{3}{}{T}_{1g}[\mathrm{C}_1 \oplus \mathrm{D}_1]$ & -1539.2914711 & 2.0024 \\
					$\prescript{3}{}{T}_{2g}[\mathrm{D}_1]$ & -1539.2912895 & 2.0025 \\
					\bottomrule
				\end{tabular}
			}
			\endgroup
		\end{subtable}
	\end{table*}

	%%
	%% All noci - MS = 0
	\begin{table*}
		\centering
		\caption{
			Energies and $\langle \hat{S}^2 \rangle$ values of $M_S = 0$ NOCI wavefunctions constructed from all interacting low-lying UHF solutions in \ce{[VF6]^{3-}}.
		}
		\label{tab:d2_allnocienergy_MS0}
		\begin{subtable}[t]{.48\textwidth}
			\centering
			\caption{
				$\prescript{3}{}{T}_{1g}[\mathrm{A}_0, \mathrm{A}'_0, \mathrm{A}''_0\ (\mathrm{E}_0)]$
			}
			\footnotesize
			\begingroup
			\renewcommand\arraystretch{1.35}
			\scalebox{.8}{
				\begin{tabular}[t]{
						>{\raggedright\arraybackslash}m{4.3cm}%
						S[table-format=4.7, table-alignment=left]%
						S[table-format=1.4, table-alignment=left]}
					\toprule
					$\Phi$ &	{Energy/\si{\hartree}} & {$\langle \hat{S}^2 \rangle$} \\
					\midrule
					$\prescript{3}{}{T}_{1g}[\mathrm{A}_0]$ & -1539.3445628 & 2.0000 \\
					$\prescript{3}{}{T}_{1g}[\mathrm{A}_0 \oplus \mathrm{E}_0]$ & -1539.3451477 & 2.0000 \\
					\midrule
					$\prescript{3}{}{T}_{1g}[\mathrm{A}_0 \oplus \mathrm{A}'_0]$ & -1539.3448695 & 2.0000 \\
					$\prescript{3}{}{T}_{1g}[\mathrm{A}_0 \oplus \mathrm{A}'_0 \oplus \mathrm{E}_0]$ & -1539.3460102 & 2.0000 \\
					\midrule
					$\prescript{3}{}{T}_{1g}[\mathrm{A}_0 \oplus \mathrm{A}'_0 \oplus \mathrm{A}''_0]$ & -1539.3448808 & 2.0000 \\
					$\prescript{3}{}{T}_{1g}[\mathrm{A}_0 \oplus \mathrm{A}'_0 \oplus \mathrm{A}''_0 \oplus \mathrm{E}_0]$ & -1539.3466382 & 2.0000 \\
					\midrule
					$\prescript{3}{}{T}_{1g}[\mathrm{A}_0 \oplus \mathrm{A}''_0]$ & -1539.3447589 & 2.0000 \\
					$\prescript{3}{}{T}_{1g}[\mathrm{A}_0 \oplus \mathrm{A}''_0 \oplus \mathrm{E}_0]$ & -1539.3465501 & 2.0000 \\
					\midrule
					$\prescript{3}{}{T}_{1g}[\mathrm{A}'_0]$ & -1539.3433740 & 2.0000 \\
					$\prescript{3}{}{T}_{1g}[\mathrm{A}'_0 \oplus \mathrm{E}_0]$ & -1539.3444053 & 2.0000 \\
					\midrule
					$\prescript{3}{}{T}_{1g}[\mathrm{A}'_0 \oplus \mathrm{A}''_0]$ & -1539.3434856 & 2.0000 \\
					$\prescript{3}{}{T}_{1g}[\mathrm{A}'_0 \oplus \mathrm{A}''_0 \oplus \mathrm{E}_0]$ & -1539.3446437 & 2.0000 \\
					\midrule
					$\prescript{3}{}{T}_{1g}[\mathrm{A}''_0]$ & -1539.3372071 & 2.0000 \\
					$\prescript{3}{}{T}_{1g}[\mathrm{A}''_0 \oplus \mathrm{E}_0]$ & -1539.3435151 & 2.0000 \\
					\bottomrule
				\end{tabular}
			}
			\endgroup
		\end{subtable}
		\hfill
		\begin{subtable}[t]{.48\textwidth}
			\centering
			\caption{
				$\prescript{3}{}{T}_{1g}[\mathrm{E}_0\ (\mathrm{A}_0, \mathrm{A}'_0, \mathrm{A}''_0)]$
			}
			\footnotesize
			\begingroup
			\renewcommand\arraystretch{1.35}
			\scalebox{.8}{
				\begin{tabular}[t]{
						>{\raggedright\arraybackslash}m{4.3cm}%
						S[table-format=4.7, table-alignment=left]%
						S[table-format=1.4, table-alignment=left]}
					\toprule
					$\Phi$ &	{Energy/\si{\hartree}} & {$\langle \hat{S}^2 \rangle$} \\
					\midrule
					$\prescript{3}{}{T}_{1g}[\mathrm{E}_0]$ & -1539.2374281 & 2.0000 \\
					$\prescript{3}{}{T}_{1g}[\mathrm{A}_0 \oplus \mathrm{E}_0]$ & -1539.2307151 & 2.0000 \\
					$\prescript{3}{}{T}_{1g}[\mathrm{A}_0 \oplus \mathrm{A}'_0 \oplus \mathrm{E}_0]$ & -1539.2313706 & 2.0000 \\
					$\prescript{3}{}{T}_{1g}[\mathrm{A}_0 \oplus \mathrm{A}'_0 \oplus \mathrm{A}''_0 \oplus \mathrm{E}_0]$ & -1539.2315576 & 2.0000 \\
					$\prescript{3}{}{T}_{1g}[\mathrm{A}_0 \oplus \mathrm{A}''_0 \oplus \mathrm{E}_0]$ & -1539.2315567 & 2.0000 \\
					$\prescript{3}{}{T}_{1g}[\mathrm{A}'_0 \oplus \mathrm{E}_0]$ & -1539.2307352 & 2.0000 \\
					$\prescript{3}{}{T}_{1g}[\mathrm{A}'_0 \oplus \mathrm{A}''_0 \oplus \mathrm{E}_0]$ & -1539.2312293 & 2.0000 \\
					$\prescript{3}{}{T}_{1g}[\mathrm{A}''_0 \oplus \mathrm{E}_0]$ & -1539.2311201 & 2.0000 \\
					\bottomrule
				\end{tabular}
			}
			\endgroup
		\end{subtable}
		
		\vspace{1cm}
		\begin{subtable}[t]{.48\textwidth}
			\centering
			\caption{
				$\prescript{3}{}{T}_{2g}[\mathrm{B}_0, \mathrm{C}_0]$
			}
			\footnotesize
			\begingroup
			\renewcommand\arraystretch{1.35}
			\scalebox{.8}{
				\begin{tabular}[t]{
						>{\raggedright\arraybackslash}m{2.5cm}%
						S[table-format=4.7, table-alignment=left]%
						S[table-format=1.4, table-alignment=left]}
					\toprule
					$\Phi$ &	{Energy/\si{\hartree}} & {$\langle \hat{S}^2 \rangle$} \\
					\midrule
					$\prescript{3}{}{T}_{1g}[\mathrm{B}_0]$ & -1539.2847593 & 2.0000 \\
					$\prescript{3}{}{T}_{2g}[\mathrm{B}_0 \oplus \mathrm{C}_0]$ & -1539.2890629 & 2.0000 \\
					$\prescript{3}{}{T}_{1g}[\mathrm{C}_0]$ & -1539.2868717 & 2.0000 \\
					\bottomrule
				\end{tabular}
			}
			\endgroup
		\end{subtable}
	\end{table*}

	%%
	%% All noci - MS = 0 (continued)
	\begin{table*}
		\ContinuedFloat
		\begin{subtable}[t]{.48\textwidth}
			\centering
			\caption{
				$\prescript{1}{}{T}_{2g}[\mathrm{A}_0, \mathrm{A}'_0, \mathrm{A}''_0\ (\mathrm{C}_0)]$
			}
			\footnotesize
			\begingroup
			\renewcommand\arraystretch{1.35}
			\scalebox{.8}{
				\begin{tabular}[t]{
						>{\raggedright\arraybackslash}m{4.3cm}%
						S[table-format=4.7, table-alignment=left]%
						S[table-format=1.4, table-alignment=left]}
					\toprule
					$\Phi$ &	{Energy/\si{\hartree}} & {$\langle \hat{S}^2 \rangle$} \\
					\midrule
					$\prescript{1}{}{T}_{2g}[\mathrm{A}_0]$ & -1539.2803723 & 0.0038 \\
					$\prescript{1}{}{T}_{2g}[\mathrm{A}_0 \oplus \mathrm{C}_0]$ & -1539.2816639 & 0.0040 \\
					\midrule
					$\prescript{1}{}{T}_{2g}[\mathrm{A}_0 \oplus \mathrm{A}'_0]$ & -1539.2803724 & 0.0038 \\
					$\prescript{1}{}{T}_{2g}[\mathrm{A}_0 \oplus \mathrm{A}'_0 \oplus \mathrm{C}_0]$ & -1539.2816643 & 0.0040 \\
					\midrule
					$\prescript{1}{}{T}_{2g}[\mathrm{A}_0 \oplus \mathrm{A}'_0 \oplus \mathrm{A}''_0]$ & -1539.2804618 & 0.0038 \\
					$\prescript{1}{}{T}_{2g}[\mathrm{A}_0 \oplus \mathrm{A}'_0 \oplus \mathrm{A}''_0 \oplus \mathrm{C}_0]$ & -1539.2818519 & 0.0041 \\
					\midrule
					$\prescript{1}{}{T}_{2g}[\mathrm{A}_0 \oplus \mathrm{A}''_0]$ & -1539.2804345 & 0.0038 \\
					$\prescript{1}{}{T}_{2g}[\mathrm{A}_0 \oplus \mathrm{A}''_0 \oplus \mathrm{C}_0]$ & -1539.2817958 & 0.0040 \\
					\midrule
					$\prescript{1}{}{T}_{2g}[\mathrm{A}'_0]$ & -1539.2791715 & 0.0042 \\
					$\prescript{1}{}{T}_{2g}[\mathrm{A}'_0 \oplus \mathrm{C}_0]$ & -1539.2810148 & 0.0042 \\
					\midrule
					$\prescript{1}{}{T}_{2g}[\mathrm{A}'_0 \oplus \mathrm{A}''_0]$ & -1539.2792889 & 0.0039 \\
					$\prescript{1}{}{T}_{2g}[\mathrm{A}'_0 \oplus \mathrm{A}''_0 \oplus \mathrm{C}_0]$ & -1539.2810824 & 0.0045 \\
					\midrule
					$\prescript{1}{}{T}_{2g}[\mathrm{A}''_0]$ & -1539.2751980 & 0.0073 \\
					$\prescript{1}{}{T}_{2g}[\mathrm{A}''_0 \oplus \mathrm{C}_0]$ & -1539.2796099 & 0.0072 \\
					\bottomrule
				\end{tabular}
			}
			\endgroup
		\end{subtable}
		\hfill
		\begin{subtable}[t]{.48\textwidth}
			\centering
			\caption{
				$\prescript{1}{}{T}_{2g}[\mathrm{C}_0\ (\mathrm{A}_0, \mathrm{A}'_0, \mathrm{A}''_0)]$
			}
			\footnotesize
			\begingroup
			\renewcommand\arraystretch{1.35}
			\scalebox{.8}{
				\begin{tabular}[t]{
						>{\raggedright\arraybackslash}m{4.3cm}%
						S[table-format=4.7, table-alignment=left]%
						S[table-format=1.4, table-alignment=left]}
					\toprule
					$\Phi$ &	{Energy/\si{\hartree}} & {$\langle \hat{S}^2 \rangle$} \\
					\midrule
					$\prescript{1}{}{T}_{2g}[\mathrm{C}_0]$ & -1539.2189019 & 0.0067 \\
					$\prescript{1}{}{T}_{2g}[\mathrm{A}_0 \oplus \mathrm{C}_0]$ & -1539.2146041 & 0.0069 \\
					$\prescript{1}{}{T}_{2g}[\mathrm{A}_0 \oplus \mathrm{A}'_0 \oplus \mathrm{C}_0]$ & -1539.2147239 & 0.0067 \\
					$\prescript{1}{}{T}_{2g}[\mathrm{A}_0 \oplus \mathrm{A}'_0 \oplus \mathrm{A}''_0 \oplus \mathrm{C}_0]$ & -1539.2147699 & 0.0068 \\
					$\prescript{1}{}{T}_{2g}[\mathrm{A}_0 \oplus \mathrm{A}''_0 \oplus \mathrm{C}_0]$ & -1539.2146049 & 0.0069 \\
					$\prescript{1}{}{T}_{2g}[\mathrm{A}'_0 \oplus \mathrm{C}_0]$ & -1539.2146977 & 0.0068 \\
					$\prescript{1}{}{T}_{2g}[\mathrm{A}'_0 \oplus \mathrm{A}''_0 \oplus \mathrm{C}_0]$ & -1539.2147039 & 0.0068 \\
					$\prescript{1}{}{T}_{2g}[\mathrm{A}''_0 \oplus \mathrm{C}_0]$ & -1539.2144900 & 0.0067 \\
					\bottomrule
				\end{tabular}
			}
			\endgroup
		\end{subtable}
	
		\vspace{1cm}
		\begin{subtable}[t]{.48\textwidth}
			\centering
			\caption{
				$\prescript{1}{}{E}_{g}[\mathrm{A}_0, \mathrm{B}_0, \mathrm{D}_0]$
			}
			\footnotesize
			\begingroup
			\renewcommand\arraystretch{1.35}
			\scalebox{.8}{
				\begin{tabular}[t]{
						>{\raggedright\arraybackslash}m{3.4cm}%
						S[table-format=4.7, table-alignment=left]%
						S[table-format=1.4, table-alignment=left]}
					\toprule
					$\Phi$ &	{Energy/\si{\hartree}} & {$\langle \hat{S}^2 \rangle$} \\
					\midrule
					$\prescript{1}{}{E}_{g}[\mathrm{A}_0]$ & -1539.2770923 & 0.0032 \\
					$\prescript{1}{}{E}_{g}[\mathrm{A}_0 \oplus \mathrm{B}_0]$ & -1539.2816837 & 0.0016 \\
					$\prescript{1}{}{E}_{g}[\mathrm{A}_0 \oplus \mathrm{B}_0 \oplus \mathrm{D}_0]$ & -1539.2816837 & 0.0016 \\
					$\prescript{1}{}{E}_{g}[\mathrm{A}_0 \oplus \mathrm{D}_0]$ & -1539.2785202 & 0.0013 \\
					$\prescript{1}{}{E}_{g}[\mathrm{B}_0]$ & -1539.2735656 & 0.0029 \\
					$\prescript{1}{}{E}_{g}[\mathrm{B}_0 \oplus \mathrm{D}_0]$ & -1539.2781317 & 0.0070 \\
					$\prescript{1}{}{E}_{g}[\mathrm{D}_0]$ & -1539.2742054 & 0.0000\\
					\bottomrule
				\end{tabular}
			}
			\endgroup
		\end{subtable}
		\hfill
		\begin{subtable}[t]{.48\textwidth}
			\centering
			\caption{
				$\prescript{1}{}{A}_{1g}[\mathrm{B}_0\ (\mathrm{D}_0)]$
			}
			\footnotesize
			\begingroup
			\renewcommand\arraystretch{1.35}
			\scalebox{.8}{
				\begin{tabular}[t]{
						>{\raggedright\arraybackslash}m{2.7cm}%
						S[table-format=4.7, table-alignment=left]%
						S[table-format=1.4, table-alignment=left]}
					\toprule
					$\Phi$ &	{Energy/\si{\hartree}} & {$\langle \hat{S}^2 \rangle$} \\
					\midrule
					$\prescript{1}{}{A}_{1g}[\mathrm{B}_0]$ & -1539.2165102 & 0.0027 \\
					$\prescript{1}{}{A}_{1g}[\mathrm{B}_0 \oplus \mathrm{D}_0]$ & -1539.2165109 & 0.0027 \\
					\bottomrule
				\end{tabular}
			}
			\endgroup
		\end{subtable}
	\end{table*}

	\clearpage

%% file: symmultiplescf-arxiv.bbl
\begin{mcitethebibliography}{82}
\bibitem[Not()]{Note-1}
Repeated representations are not a problem. For example, if both $\left\lbrace
  \prescript{m}{}{\Phi} \right\rbrace$ and $\left\lbrace \prescript{n}{}{\Phi}
  \right\rbrace$ form two bases for the same irreducible representation
  $\Gamma$, NOCI must already take into account their symmetry-allowed mutual
  interaction and lift their degeneracy with respect to each other so that the
  $\left\lbrace \prescript{m}{}{\Phi} \right\rbrace$ wavefunctions only
  transform amongst themselves under any symmetry operation of the group, as do
  the $\left\lbrace \prescript{n}{}{\Phi} \right\rbrace$ wavefunctions.\relax
\mciteBstWouldAddEndPunctfalse
\mciteSetBstMidEndSepPunct{\mcitedefaultmidpunct}
{}{\mcitedefaultseppunct}\relax
\EndOfBibitem
\bibitem[Not()]{Note-3}
The D shell in vanadium's 6-31G* is commonly constructed with six Cartesian
  functions of degree $2$ for the angular part. There is, however, an $s$
  component contained in this (see Appendix~\ref{app:symtrans}) which is of no
  interest to the model system where the two electrons are constrained to
  occupy only the five $d$ orbitals. We thus use the pure (and real) $d$ form
  instead for the angular part of the basis functions in the D shell to
  eliminate the unnecessary totally symmetric $s$ component.\relax
\mciteBstWouldAddEndPunctfalse
\mciteSetBstMidEndSepPunct{\mcitedefaultmidpunct}
{}{\mcitedefaultseppunct}\relax
\EndOfBibitem
\end{mcitethebibliography}
